\documentclass[twocolumn, secnumarabic, amssymb, nobibnotes, pre, superscriptaddress]{revtex4-2}

\usepackage{comment}

\usepackage{graphicx}
\usepackage{amsmath}
\usepackage{float}
\usepackage{color}
\usepackage[english]{babel}
\usepackage[T1]{fontenc}
\usepackage[usenames,dvipsnames]{xcolor}
\usepackage{setspace}
\usepackage{hyperref}
\usepackage{bm}
\usepackage{lipsum}
\usepackage{ifthen}
\usepackage{calc} 
\usepackage{tikz}
\usetikzlibrary{patterns}
\usetikzlibrary{arrows.meta}
\usetikzlibrary{decorations.pathreplacing}
\usetikzlibrary{decorations.pathmorphing}
\usetikzlibrary{decorations.markings}

\usetikzlibrary{math}
\usetikzlibrary{tikzmark}
\usepackage{stmaryrd}

\definecolor{color1}{rgb}{0.122,0.467,0.706}

\definecolor{ColorSP}{rgb}{0.925, 0.110, 0.141}
\definecolor{ColorSM}{rgb}{0.259, 0.396, 0.686}

\definecolor{ColorSL}{rgb}{0.639, 0.784, 0.639}

\definecolor{colorFerro}{rgb}{0.957,0.843,0.890}
\definecolor{colorAntiFerro}{rgb}{1.0,0.965,0.835}
\definecolor{colorMixed}{rgb}{1.0,0.902,0.835}
\definecolor{colorPreisach}{rgb}{0.949,0.949,0.949}

\definecolor{ColorBG}{rgb}{0.973, 0.922, 0.859}

\definecolor{ColorDarkGreen}{rgb}{0.0, 0.267, 0.106}

\makeatletter
\renewcommand*{\fnum@figure}{{\normalfont \small{FIG.}~\thefigure}}
\makeatother

\makeatletter
\def\@seccntformat#1{\csname the#1\endcsname\quad}
\renewcommand\thesection{\arabic{section}}
\renewcommand\thesubsection{\thesection.\arabic{subsection}}

\renewcommand\paragraph{\theparagraph.\arabic{paragraph}}
\makeatother

\pgfdeclarepatternformonly{my crosshatch dots}{\pgfqpoint{-1pt}{-1pt}}{\pgfqpoint{5pt}{5pt}}{\pgfqpoint{6pt}{6pt}}%
{
    \pgfpathcircle{\pgfqpoint{0pt}{0pt}}{.5pt}
    \pgfpathcircle{\pgfqpoint{3pt}{3pt}}{.5pt}
    \pgfusepath{fill}
}

\begin{document}

\title{Dynamic self-loops in networks of passive and active binary elements}

\author{Paul Baconnier}
\affiliation{AMOLF, 1098 XG Amsterdam, The Netherlands.}
\author{Margot H. Teunisse}
\affiliation{AMOLF, 1098 XG Amsterdam, The Netherlands.}
\affiliation{Huygens-Kamerlingh Onnes Laboratory, Leiden University, 2300 RA Leiden, The Netherlands.}
\author{Martin van Hecke}
\affiliation{AMOLF, 1098 XG Amsterdam, The Netherlands.}
\affiliation{Huygens-Kamerlingh Onnes Laboratory, Leiden University, 2300 RA Leiden, The Netherlands.}

\begin{abstract}
Models of coupled binary elements capture 
memory effects in complex dissipative materials,
such as transient responses or 
sequential computing, when their interactions are chosen appropriately.
However, for random interactions, self-loops—cyclic transition sequences incompatible with dissipative dynamics—dominate the response and undermine statistical approaches.
Here we reveal that self-loops originate from energy injection and limit cycles in the underlying physical system.
We furthermore introduce interaction ensembles that strongly suppress or completely eliminate self-loops, allowing statistical studies of memory in large dissipative systems.
Our work opens a route towards a unified description of passive and active multistable materials using hysteron models.
\end{abstract}

\maketitle

\begin{figure}[tb]
\centering
\begin{tikzpicture}

\node[draw, rounded corners=10pt, minimum width=1cm, minimum height=0.75cm, align=center, fill=gray!30] (A) at (0.0,-2.15) {$S^{0}$};
\node[draw, rounded corners=10pt, minimum width=1cm, minimum height=0.75cm, align=center, fill=gray!30] (B) at (-1.4,-0.75) {$S^{1}$};
\node[draw, rounded corners=10pt, minimum width=1cm, minimum height=0.75cm, align=center, fill=gray!30] (C) at (0.0,0.65) {$S^{2}$};
\node[draw, rounded corners=10pt, minimum width=1cm, minimum height=0.75cm, align=center, fill=gray!30] (D) at (1.4,-0.75) {$S^{3}$};

\draw[->, thick, bend left=30, >={Triangle[length=1.4mm, width=1.4mm]}, color=ColorSP] (A) to (B);
\draw[->, thick, bend left=30, >={Triangle[length=1.4mm, width=1.4mm]}, color=ColorSP] (B) to (C);
\draw[->, thick, bend left=30, >={Triangle[length=1.4mm, width=1.4mm]}, color=ColorSM] (C) to (D);
\draw[->, thick, bend left=30, >={Triangle[length=1.4mm, width=1.4mm]}, color=ColorSM] (D) to (A);

\node[] at (-1.75,0.8) {\small (d)};

\node[rotate=0, anchor=south west] at (-5.9,1.2) {\includegraphics[height=2.1cm]{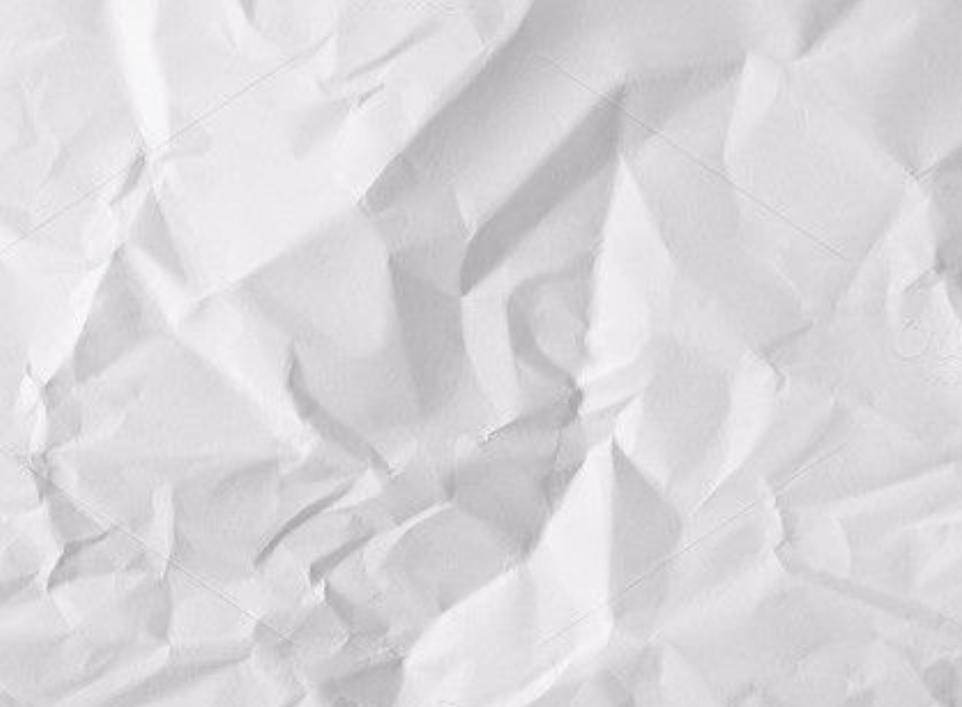}};

\draw[] (-5.78,1.32) rectangle (-2.92,3.42);

\draw[->, >=Triangle, thick] (-6.1,2.37) -- (-5.78,2.37);
\draw[->, >=Triangle, thick] (-2.6,2.37) -- (-2.92,2.37);

\node[] at (-2.6,2.67) {\small $H$};

\node[] at (-6.12,3.15) {\small (a)};

\draw[] (-4.8,1.7) circle (0.15);
\draw[] (-4.4,1.9) circle (0.2);
\node[] at (-5.1,1.7) {\small $i$};
\node[] at (-4.05,1.9) {\small $j$};

\node[] at (-6.12,0.8) {\small (b)};

\node[rotate=0, anchor=south west] at (-6.25,-2.4) {\includegraphics[height=3.0cm]{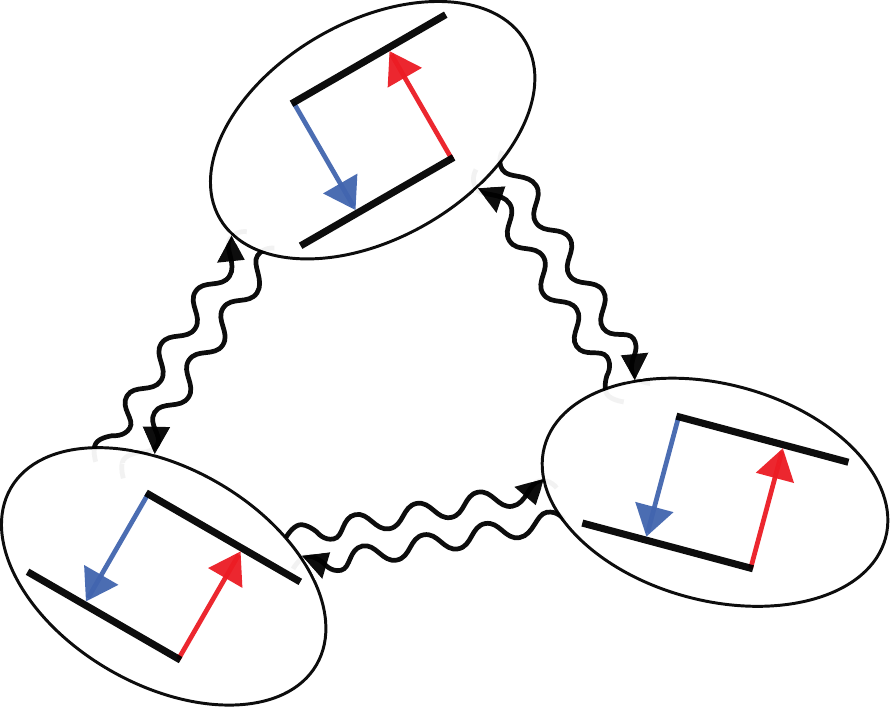}};

\node[] at (-1.77,3.15) {\small (c)};

\node[] at (-1.88,2.01) {\small $i$};
\node[] at (1.9,1.9) {\small $j$};

\draw[] (0.95,2.4) ellipse [x radius=1.0, y radius=0.65];

\draw[->, >={Triangle[length=1.5mm, width=1.5mm]}, thick, color=ColorSM] (0.65,2.7) -- (0.65,2.1);
\draw[<-, >={Triangle[length=1.5mm, width=1.5mm]}, thick, color=ColorSP] (1.25,2.7) -- (1.25,2.1);

\draw[densely dashed, line width=0.01cm] (0.65,2.1) -- (0.65,1.7);
\node[] at (0.65,1.45) {\small $h_j^-$};
\draw[densely dashed, line width=0.01cm] (1.25,2.1) -- (1.25,1.7);
\node[] at (1.25,1.45) {\small $h_j^+$};

\draw[very thick] (0.25,2.1) -- (1.25,2.1);
\draw[very thick] (0.65,2.7) -- (1.65,2.7);

\draw[] (-1.15,2.4) ellipse [x radius=0.7, y radius=0.5];

\draw[<-, >={Triangle[length=1.5mm, width=1.5mm]}, thick, color=ColorSP] (-1.0,2.55) -- (-1.0,2.25);
\draw[->, >={Triangle[length=1.5mm, width=1.5mm]}, thick, color=ColorSM] (-1.3,2.55) -- (-1.3,2.25);

\draw[very thick] (-1.0,2.25) -- (-1.65,2.25);
\draw[very thick] (-1.3,2.55) -- (-0.65,2.55);

\draw[<-, >={stealth[length=1.2mm, width=1.2mm]}, very thick, color=black] (-0.65,2.8) to[out=40, in=140] (0.05,2.8);

\draw[->, >={stealth[length=1.2mm, width=1.2mm]}, color=black] (-0.65,2.0) to[out=-40, in=220] (0.05,2.0);

\node[] at (-0.3,3.15) {\small $c_{ij}$};
\node[] at (-0.3,1.65) {\small $c_{ji}$};

\node[] at (1.5,2.5) {\footnotesize $+1$};
\node[] at (0.36,2.26) {\footnotesize $-1$};

\node[] at (-0.75,2.35) {\footnotesize $+1$};
\node[] at (-1.59,2.41) {\footnotesize $-1$};

\end{tikzpicture}
\vspace*{-0.5cm}
\caption{(a) Crumpled sheet with bistable  elements (ridges $i$ and $j$)
\cite{shohat2022memory}. (b) Abstract hysteron models employ interacting binary elements. (c) Variations in strength of physical elements
map to asymmetric hysteron interactions ($|c_{ij}|>|c_{ji}|$). (d) Self-loop in a (partial) transition graph. When the switching thresholds satisfy $H^{+}(S^{1})\!<\!H^{+}(S^{0})$, $H^{-}(S^{2})\!>\!H^{+}(S^{0})$ and $H^{-}(S^{3})\!>\!H^{+}(S^{0})$ and 
the system is in state $S^{0}$, 
increasing the driving $H$ above its 
threshold $H^{+}(S^{0})$ triggers a self-loop as all states are unstable (Supplemental Material).}
\label{fig:loops_or_not}
\end{figure}

Sequences of transitions between metastable states govern the hysteresis \cite{preisach1935magnetische}, memory \cite{keim2019memory, paulsen2019minimal, lindeman2021multiple, keim2021multiperiodic, jules2022delicate, shohat2022memory, keim2022mechanical, paulsen2024review}, emergent computing \cite{bense2021complex, kwakernaak2023counting, liu2024controlled}, sequential shape-morphing \cite{melancon2022inflatable, meeussen2023multistable}, and adaptive behavior \cite{kamp2024reprogrammable, veenstra2025adaptive} of driven dissipative materials, such as crumpled sheets, disordered media, and metamaterials \cite{paulsen2024review} (Fig.~\ref{fig:loops_or_not}a). 
As these states are often composed of local, two-state elements, with or without hysteresis, the response is commonly described
by models of interacting hysterons or binary spins {at zero-temperature} (Fig.~\ref{fig:loops_or_not}b).
While models without interactions are well understood \cite{preisach1935magnetische, mungan2019networks, mungan2019structure, terzi2020state}, interactions are crucial for capturing  complex responses such as avalanches, transient responses, and multiperiodic cycles \cite{van2021profusion, keim2021multiperiodic}.
Pairwise interactions are captured by a matrix \( c_{ij} \) which represents how element \( i \)'s flipping thresholds are influenced by element \( j \). (Fig.~\ref{fig:loops_or_not}c).
In some cases $c_{ij}$ can be measured \cite{bense2021complex,shohat2022memory}, and for networks of physical bistable elements it can be modeled \cite{puglisi2000mechanics,nicolas2018deformation,kumar2022mapping,kumar2024self,liu2024controlled,shohat2025geometric}. Hysteron models with appropriate interactions then
enable accurate predictions of the systems response and memory effects  \cite{bense2021complex,ding2022sequential,liu2024controlled,keim2021multiperiodic, keim2022mechanical,paulsen2024mechanical,shohat2025geometric}. 

Yet, the precise interactions are  often unknown, 
or the goal is to understand system classes rather than a single experiment.
Strikingly, assigning 
random interaction coefficients \cite{van2021profusion,keim2020global,keim2021multiperiodic,lindeman2021multiple,teunisse2024transition,lindeman2025generalizing} 
overwhelmingly produces self-loops (Fig.~\ref{fig:loops_or_not}d): avalanches that get trapped in a repeating sequence of states which never settle \cite{gutfreund1988nature, nutzel1993subtle, eissfeller1994mean, hwang2019number, van2021profusion}. 
Such dynamic loops---prohibited in dissipative systems---can be avoided by using symmetric interactions ($c_{ij} = c_{ji}$), as commonly done in spin glasses where asymmetry implies energy input and oscillations \cite{
kirkpatrick1978infinite,eissfeller1994mean, 
hopfield1982neural,crisanti1987dynamics,gutfreund1988nature, nutzel1993subtle, eissfeller1994mean,panchenko2013sherrington,hwang2019number,fruchart2021non,lorenzana2024non}.
However, for hysteretic binary elements, whose strong nonlinearity invalidates Maxwell-Betti reciprocity~\cite{coulais2017static},
asymmetric interactions are compatible with dissipation.
For example, a small ridge \( i \) may weakly affect a larger one \( j \) but not vice versa: \( |c_{ij}| < |c_{ji}| \)
(Fig.~\ref{fig:loops_or_not}c).
Such asymmetric interactions are experimentally observed and theoretically modeled~\cite{bense2021complex,liu2024controlled,shohat2025geometric}, and are crucial for explaining memory effects~\cite{keim2021multiperiodic, lindeman2025generalizing}. 
A conundrum thus arises: though physical and essential, random asymmetric interactions yield unphysical self-loops.

Here we uncover the mechanism that generates self-loops and show that their probability approaches one in large systems of asymmetrically coupled hysterons.
We discuss physical networks with {\em active} bistable elements, map them to interacting hysterons, and show that these produce self-loops linked to limit cycles in the networks: rather than being unphysical, self-loops reveal that coupled hysteron models span passive and active physical systems. Surprisingly, active elements may give rise to symmetric hysteron interactions.
Finally, we introduce weakly asymmetric interaction ensembles to suppress self-loops, develop strict ensembles to eliminate them, and use these to study avalanche statistics and cyclic response in large hysteron systems.
Our work uncovers subtle relations between the symmetry of hysteron interactions, self-loops, and physical energy injection. Moreover, it shows that hysteron models unify the description of passive and active multistable materials, connects the self-loop sector to limit cycles in active systems, and enables statistical studies of memory effects and pathways in dissipative and mixed systems.

{\em Model.---} We consider $N$ binary elements, $s_i \!=\! \pm 1$, which form collective states $S\!=\!(s_1,s_2,\dots)$. The system is driven by a global field $H$, and the stability range of each element $i$ in state $S$ is given by switching thresholds $H_i^\pm(S)$. For pairwise interactions:
\begin{equation} \label{eq:coupled_hysterons_model}
 H_{i}^{\pm}(S) = h_i^{\pm} - \sum_{j \neq i} c_{ij} s_j~, \end{equation}
where $h_i^\pm$ are the bare switching thresholds of element $i$. To model spins, we take $h_i^+ = h_i^-$, whereas for hysterons, $h_i^+ > h_i^-$ \cite{preisach1935magnetische,mungan2019structure,terzi2020state}. The matrix $c_{ij}$, with $c_{ii}=0$, encodes cooperative ($c_{ij} > 0$) or frustrated ($c_{ij} < 0$) interactions that may be asymmetric ($c_{ij} \neq c_{ji}$) \cite{liu2024controlled,keim2021multiperiodic,paulsen2024mechanical,shohat2025geometric}.

In this model, each state $S$ has a range of stability, encoded in state switching thresholds $H^\pm(S)$ which follow from the extrema of $H_i^\pm(S)$: $H^{+}(S)\!:=\!\min_{i^-} ( H_i^{+}(S))$ and $H^{-}(S) \!:=\! \max_{i^+} ( H_i^{-}(S) )$, where $i^\pm$ are the indices where $s_i=\pm 1$. When the system is in state $S^0$ and $H$ is increased above $H^{+}(S^0)$ or decreased below $H^{-}(S^0)$, state $S^0$ loses stability and
its unstable hysteron flips. Depending on $n_u$, the number of unstable hysterons in the resulting state $S^1$, three different scenarios arise \cite{van2021profusion,teunisse2024transition}.
When $n_u=0$, state $S^1$ is stable; when $n_u=1$, state $S^1$ is unstable and its unstable hysteron flips; when $n_u>1$, multiple hysterons are unstable. The latter case, which is abundant in large systems (Supplemental Material), leads to a race condition, and requires a dynamical rule to specify the next step in the transition
\cite{van2021profusion,bense2021complex,teunisse2024transition,liu2024controlled,paulsen2024mechanical}. In the remainder, we flip the most unstable element first \cite{keim2021multiperiodic}; this rule is physically plausible and corresponds to the zero-temperature limit of the Glauber dynamics \cite{parisi2003statistical, jesi2016spin} (for other rules see Supplemental Material).

\begin{figure}[t!]
\centering
\hspace*{-0.25cm}
\begin{tikzpicture}

\node[rotate=0] at (0.0,-12.9) {\includegraphics[height=4.3cm]{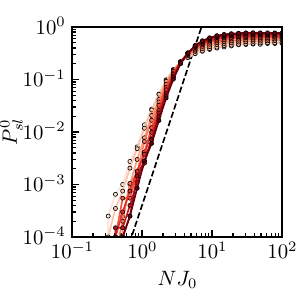}};

\node[rotate=0] at (4.4,-12.9) {\includegraphics[height=4.3cm]{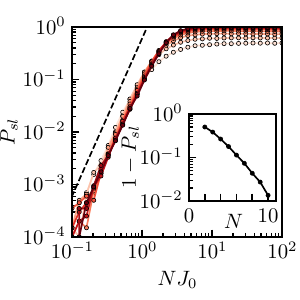}};

\node[rotate=0] at (-0.03,-8.6) {\includegraphics[height=4.3cm]{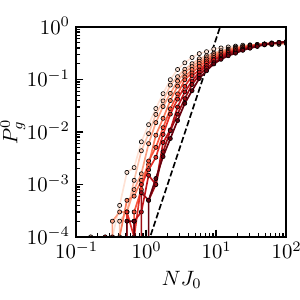}};

\node[rotate=0] at (4.4,-8.6) {\includegraphics[height=4.3cm]{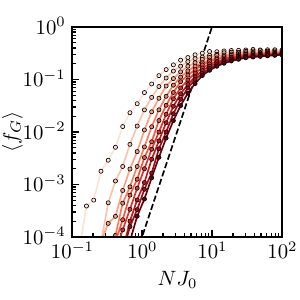}};

\node[rotate=0] at (-0.75,-11.45) {\small (c)};
\node[rotate=0] at (3.65,-11.45) {\small (d)};

\node[rotate=0] at (-0.75,-7.15) {\small (a)};
\node[rotate=0] at (3.65,-7.15) {\small (b)};

\draw[->] (-0.1,-7.5) -- (0.55,-8.7);
\node[rotate=0] at (0.7,-8.95) {\small $N$};

\end{tikzpicture}
\vspace{-0.8cm}
\caption{Statistical measures for gaps and self-loops scale when plotted as function of $N J_0$
($10^5$ samples; color from light to dark as $N$ increases from $2$ to $10$).
(a) Probability $P_g^0$ of finding a gap at $H = 0$  (dashed line indicates slope $4$).
(b) Averaged fraction of gaps $f_G$, where $f_G$ is defined as the ratio of the size of intervals where no stable states exist divided by 
$H^{-}(++\dots) - H^{+}(--\dots)$
(dashed line indicates slope $4$).
(c) Probability $P_{sl}^0$ of finding at least one self-loops at $H = 0$  (dashed line indicates slope $4$).
(d) Probability $P_{sl}$ of finding at least one self-loop for any value of $H$ (dashed line indicates slope $3$).
Inset: The probability to be self-loop free,
$1-P_{sl}$, decays to zero exponentially with $N$ for large couplings ($NJ_0 = 10^2$).}
\label{fig:hysteron_problem}
\end{figure}

{\em {Hysterons with symmetric interactions}.---}
Although hysteron interactions are not expected to be {symmetric}, numerical sampling reveals that symmetric interactions consistently avoid self-loops. This can be proven by showing that the
quantity \( V(S) \), written as:
\begin{equation} \label{eq:energy_like_quantity}
 V(S) = - \sum_{i^{\mp}} \left[ s_i(H - h_{i}^{\pm}) + \frac{1}{2} \sum_{j \neq i} c_{ij} s_i s_j  \right],
\end{equation}
is a Lyapunov-like function of the system, which guarantees that the latter always converges toward a stable state, and thus prohibits self-loops (App. \ref{app:demo_reciprocal}). 

{\em Random asymmetric coupling: gaps and self-loops.---} 
We now exploit random interactions, where $c_{ij}$ and $c_{ji}$ are sampled independently. For any value of $H$, each isolated element has one stable phase, or two within the hysteretic range. Without interactions, stable states can thus be easily formed by combining stable elements. However, interactions make the switching thresholds dependent on the collective state, effectively randomizing stability ranges and creating {\em gaps}, ranges of $H$ where no state is stable.

We find two related scenarios where interactions lead to self-loops. The length of a self-loop, i.e. the number of unstable states visited within the cycle, is denoted $L$. 
We consider state $S$ going unstable by
the driving crossing a critical value $H^c$. In the first scenario, 
$H^c$ lies at the edge of a {gap}, and the system gets inevitably trapped in a self-loop as there are no stable states. 
In the second, a set of $L$ distinct unstable states transition toward each other, forming a periodic attractor, and the 
system gets trapped, despite the presence of other stable states at $H=H^c$.

To investigate the statistics of gaps, self-loops, and the self-loop size distribution (Supplemental Material), we sample the model 
using an event-driven algorithm \cite{keim2021multiperiodic}. We consider 
collections of hysterons with thresholds in a compact range \cite{keim2021multiperiodic, keim2022mechanical, lindeman2025generalizing}: we flatly sample the midpoints of the bare switching thresholds $h_{i}^c = (h^{+}_{i} + h^{-}_{i})/2$ from the interval $\left[-1, 1\right]$ and the interaction coefficients $c_{ij}$ from $\left[-J_0 , J_0\right]$. Unless noted otherwise, we flatly sample the spans $\sigma_i = h^{+}_{i} - h^{-}_{i} > 0$ from $\left[0, 0.5\right]$.

We find that the probability \( P_g^0 \) of a gap, meaning the absence of a stable state at \( H = 0 \), and the fraction of gaps \( f_G \), both increase as power laws when \( NJ_0 \ll 1 \) (Figs.~\ref{fig:hysteron_problem}a-b). For \( NJ_0 \gg 1 \), they saturate at significant values. 
The probability that states have a finite stability range decreases exponentially with \( N \): for large $N$,
stable states become rare, prohibiting statistical studies of transitions (Supplemental Material).

Gaps imply self-loops, but self-loops can also occur outside of gaps.  Therefore, the probability of a self-loop occurring starting from a random state at $H=0$, $P_{sl}^0$,
is larger than the corresponding gap probability $P_g^0$ (Fig. \ref{fig:hysteron_problem}c and Supplemental Material). Similarly, we calculated the total probability of observing a self-loop at any value of $H$, $P_{sl}$, by starting from every stable state, in/decreasing $H$, and checking whether the ensuing transitions yield at least one self-loop. We find that $P_{sl}$ approaches one in large, strongly coupled systems (Fig. \ref{fig:hysteron_problem}d; Supplemental Material). This dominance of self-loops is robust; hysterons with fixed spans $\sigma_i=0.5$ and binary spins where $\sigma_i=0$ also have $P_{sl} \rightarrow 1$ in the large coupling limit (Supplemental Material)
\cite{gutfreund1988nature, nutzel1993subtle, eissfeller1994mean, hwang2019number}. 
Hence, self-loops, incompatible with the dissipative systems we aim to model, are unavoidable for random interactions, and for large systems completely overwhelm the response.

{\em Networks of Active Elements.---}
To gain further insights into the emergence of self-loops in physical systems, we consider networks of bilinear, hysteretic springs: 
\begin{equation}
 f_i (x_i) = x_i - d_i s_i, ~\Delta E_i:= 2d_i \sigma_i^0
\end{equation}
where transitions occur at $x_i = x_i^{\pm}$,
$\sigma_i^0\!:=\!x_i^+\!-\!x_i^-$ is 
taken positive, and $\Delta E_i$ is the dissipated energy per hysteresis loop (Figs. \ref{fig:self_loop_size_4}a-b).
Once the geometry of the network and the parameters of all elements are specified, the network can be mapped to an interacting hysteron model \cite{liu2024controlled,shohat2025geometric}.
This mapping never yields hysteron parameters that produce
self-loops \emph{as long as the physical elements dissipate energy}, i.e., when $d_i >0$.

We find that networks that include active elements ($d_i\!<\!0$),  enabled by, e.g., force-generating components activated when \( s_i \!=\! 1 \) (Fig.~\ref{fig:self_loop_size_4}b), are mapped to hysteron models which can feature self-loops. 
For simplicity we consider a network of two serially-coupled elements,
where the mapping yields
$c_{ij}=-d_j$ and  
$\sigma_i = 2 (\sigma_i^0 + d_{i})$
\cite{liu2024controlled,shohat2025geometric}, producing  four distinct behaviors (Fig.~\ref{fig:self_loop_size_4}c-d). 
{\em(i)} When both elements are passive, the hysteron spans $\sigma_i$ are positive,
the interactions coefficients are both negative, and no self-loops occur
\cite{liu2024controlled}.
{\em(ii)} For strongly active elements, 
i.e., $d_i<-\sigma_i^0$, the hysteron spans $\sigma_i$ become negative (Fig. \ref{fig:self_loop_size_4}e). 
Here we expect the physical system to feature dynamic limit cycles \cite{xia2021nonlinear,sarkar2022organic,belleri2024unravelling},
(partly) captured by an extended hysteron model.
{\em(iii)} Weakly active elements map to hysterons with $\sigma_i > 0$ and
interaction coefficients not accessible with passive elements. Strikingly, active elements with $d_1 = d_2 <0$
map to hysteron models with symmetric interactions (Figs. \ref{fig:self_loop_size_4}d and f, orange line): there is no simple relation between the symmetry of $c_{ij}$ and energy input.
{\em(iv)} Crucially, for appropriately chosen activity, the physical system maps to the hysteron model with an $L=4$ self-loop. For details, see Supplemental Material.

Hence, active elements extend the range of realizable hysteron parameters, may produce negative hysteron spans, and allow to access the hysteron sector with self-loops. Similar mechanisms also arise in fully dynamic models combining activity and bistability \cite{veenstra2024non}.
While larger, more complex networks remain to be understood, our example shows that self-loops can be physically interpreted as arising from energy injection in networks of two-state elements.
The hysteron model features self-loops because {\em it encompasses both passive and active systems}.

\begin{figure}[t!]
\hspace*{-0.33cm}
\begin{tikzpicture}

\node[] at (-5.65,2.41) {\includegraphics[width=2.15cm]{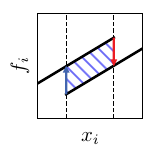}};

\node[] at (-3.55,2.41) {\includegraphics[width=2.15cm]{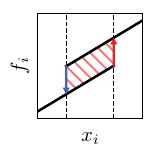}};

\node[] at (-6.48,3.45) {\small (a)};
\node[] at (-4.38,3.45) {\small (b)};

\node[rotate=0] at (-5.2,2.01) {\small \color{blue!50} $d_i > 0$};
\node[rotate=0] at (-3.1,2.01) {\small \color{red!50} $d_i < 0$};

\node[] at (-5.7,3.55) {\small $x_i^-$};
\node[] at (-5.02,3.55) {\small $x_i^+$};

\node[] at (-3.31,3.68) {\small $\sigma_i^0$};
\draw[<->] (-3.64,3.45) -- (-3.02,3.45);

\node[rotate=30] at (-4.90,2.46) {\scriptsize $+1$};
\node[rotate=30] at (-6.0,2.56) {\scriptsize $-1$};

\node[rotate=30] at (-2.80,2.86) {\scriptsize $+1$};
\node[rotate=30] at (-3.91,2.15) {\scriptsize $-1$};

\node[] at (-4.85,-0.7) {\includegraphics[width=4.3cm]{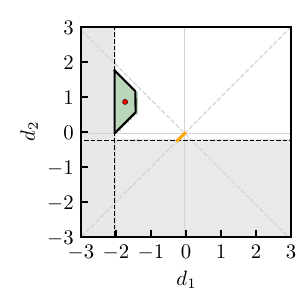}};

\node[rotate=0] at (-0.35,-0.7) {\includegraphics[height=4.3cm]{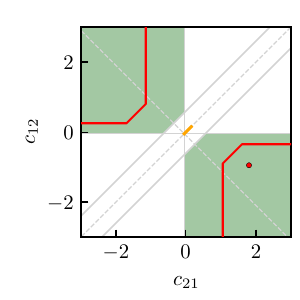}};

\def\xmin{-4.33}
\def\xmax{-2.84}
\def\ymin{-0.44}
\def\ymax{1.06}

\def\spacing{0.1}       
\def\linecolor{blue}
\def\linewidth{0.4pt}
\def\lineopacity{0.6}

\pgfmathsetmacro{\cmin}{\xmin + \ymin}
\pgfmathsetmacro{\cmax}{\xmax + \ymax}
\pgfmathsetmacro{\cstep}{\spacing * sqrt(2)}

\pgfmathsetmacro{\nlines}{int(ceil((\cmax - \cmin)/\cstep))}

\begin{scope}
\clip (\xmin,\ymin) rectangle (\xmax,\ymax);

\foreach \i in {0,...,\nlines} {
    \pgfmathsetmacro{\c}{\cmin + \i * \cstep}


    \pgfmathsetmacro{\yleft}{\c - \xmin}
    \pgfmathsetmacro{\yright}{\c - \xmax}
    \pgfmathsetmacro{\xbottom}{\c - \ymin}
    \pgfmathsetmacro{\xtop}{\c - \ymax}

    \def\pA{}
    \def\pB{}

    \ifdim \yleft pt > \ymin pt
      \ifdim \yleft pt < \ymax pt
        \ifx\pA\empty
          \xdef\pA{(\xmin,\yleft)}
        \else
          \xdef\pB{(\xmin,\yleft)}
        \fi
      \fi
    \fi

    \ifdim \yright pt > \ymin pt
      \ifdim \yright pt < \ymax pt
        \ifx\pA\empty
          \xdef\pA{(\xmax,\yright)}
        \else
          \xdef\pB{(\xmax,\yright)}
        \fi
      \fi
    \fi

    \ifdim \xbottom pt > \xmin pt
      \ifdim \xbottom pt < \xmax pt
        \ifx\pA\empty
          \xdef\pA{(\xbottom,\ymin)}
        \else
          \xdef\pB{(\xbottom,\ymin)}
        \fi
      \fi
    \fi

    \ifdim \xtop pt > \xmin pt
      \ifdim \xtop pt < \xmax pt
        \ifx\pA\empty
          \xdef\pA{(\xtop,\ymax)}
        \else
          \xdef\pB{(\xtop,\ymax)}
        \fi
      \fi
    \fi

    \ifx\pA\empty
    \else
      \ifx\pB\empty
      \else
        \draw[line width=\linewidth, draw=\linecolor, opacity=\lineopacity] \pA -- \pB;
      \fi
    \fi
}

\end{scope}

\node[rotate=0] at (-6.55,0.72) {\small (d)};

\node[rotate=0] at (-3.6,0.32) {\footnotesize (i)};

\node[rotate=0] at (-4.35,-1.19) {\footnotesize (ii)};
\node[rotate=0] at (-5.6,0.32) {\footnotesize (ii)};
\node[rotate=0] at (-5.6,-1.19) {\footnotesize (ii)};

\node[rotate=0] at (-4.85,0.8) {\footnotesize (iii)};

\node[rotate=0] at (-4.67,0.23) {\footnotesize (iv)};
\draw[->] (-4.87,0.20) -- (-5.05,0.15);

\node[rotate=0] at (-2.0,0.72) {\small (f)};

\def\xminA{-1.33}
\def\xmaxA{0.16}
\def\yminA{-1.95}
\def\ymaxA{-0.45}

\def\spacing{0.1}       
\def\linecolor{blue}
\def\linewidth{0.4pt}
\def\lineopacity{0.6}

\pgfmathsetmacro{\cminA}{\xminA + \yminA}
\pgfmathsetmacro{\cmaxA}{\xmaxA + \ymaxA}
\pgfmathsetmacro{\cstep}{\spacing * sqrt(2)}

\pgfmathsetmacro{\nlines}{int(ceil((\cmaxA - \cminA)/\cstep))}

\begin{scope}
\clip (\xminA,\yminA) rectangle (\xmaxA,\ymaxA);

\foreach \i in {0,...,\nlines} {
    \pgfmathsetmacro{\c}{\cminA + \i * \cstep}


    \pgfmathsetmacro{\yleft}{\c - \xminA}
    \pgfmathsetmacro{\yright}{\c - \xmaxA}
    \pgfmathsetmacro{\xbottom}{\c - \yminA}
    \pgfmathsetmacro{\xtop}{\c - \ymaxA}

    \def\pA{}
    \def\pB{}

    \ifdim \yleft pt > \yminA pt
      \ifdim \yleft pt < \ymaxA pt
        \ifx\pA\empty
          \xdef\pA{(\xminA,\yleft)}
        \else
          \xdef\pB{(\xminA,\yleft)}
        \fi
      \fi
    \fi

    \ifdim \yright pt > \yminA pt
      \ifdim \yright pt < \ymaxA pt
        \ifx\pA\empty
          \xdef\pA{(\xmaxA,\yright)}
        \else
          \xdef\pB{(\xmaxA,\yright)}
        \fi
      \fi
    \fi

    \ifdim \xbottom pt > \xminA pt
      \ifdim \xbottom pt < \xmaxA pt
        \ifx\pA\empty
          \xdef\pA{(\xbottom,\yminA)}
        \else
          \xdef\pB{(\xbottom,\yminA)}
        \fi
      \fi
    \fi

    \ifdim \xtop pt > \xminA pt
      \ifdim \xtop pt < \xmaxA pt
        \ifx\pA\empty
          \xdef\pA{(\xtop,\ymaxA)}
        \else
          \xdef\pB{(\xtop,\ymaxA)}
        \fi
      \fi
    \fi

    \ifx\pA\empty
    \else
      \ifx\pB\empty
      \else
        \draw[line width=\linewidth, draw=\linecolor, opacity=\lineopacity] \pA -- \pB;
      \fi
    \fi
}

\end{scope}

\node[] at (-2.15,3.45) {\small (c)};

\foreach \y in {-0.3,-0.2,...,0.3} {
    \draw[] (-2.3,\y+2.45) -- (-2.1,\y+2.65);
}
\draw[thick] (-2.1,2.15) -- (-2.1,2.95);

\draw[thick] (-2.1,2.55) -- (-1.8,2.55);

\fill[color=ColorBG] (-1.8,1.8) rectangle (-0.35,3.3);
\draw[thick] (-1.8,1.8) rectangle (-0.35,3.3);

\draw[black, thick, domain=-0.6:0.2, samples=100]
    plot (\x - 1.1, {1.0*(\x) + 2.82});
\draw[black, thick, domain=-0.2:0.6, samples=100]
    plot (\x - 1.1, {1.0*(\x) + 2.32});
\draw[ColorSP, ->, >=stealth] (-0.9,3.02) -- (-0.9,2.52);
\draw[ColorSM, ->, >=stealth] (-1.3,2.12) -- (-1.3,2.62);

\node[rotate=45] at (-0.6,2.56) {\scriptsize $+1$};
\node[rotate=45] at (-1.6,2.54) {\scriptsize $-1$};

\draw[thick] (-0.35,2.55) -- (-0.1,2.55);

\fill[color=ColorBG] (-0.1,1.8) rectangle (1.35,3.3);
\draw[thick] (-0.1,1.8) rectangle (1.35,3.3);

\draw[black, thick, domain=-0.53:0.2, samples=100]
    plot (\x + 0.6, {1.0*(\x) + 2.42});
\draw[black, thick, domain=-0.2:0.53, samples=100]
    plot (\x + 0.6, {1.0*(\x) + 2.67});
\draw[ColorSP, <-, >=stealth] (0.8,2.85) -- (0.8,2.61);
\draw[ColorSM, <-, >=stealth] (0.4,2.24) -- (0.4,2.48);

\node[rotate=45] at (1.08,2.89) {\scriptsize $+1$};
\node[rotate=45] at (0.12,2.16) {\scriptsize $-1$};

\draw[thick, ->] (1.35,2.55) -- (1.7,2.55);
\node[] at (1.63,2.22) {\footnotesize $U$};

\node[] at (-0.51,1.96) {\scriptsize $1$};
\node[] at (1.2,1.96) {\scriptsize $2$};

\node[rotate=0] at (-1.0,-2.95) {\small (g)};

\fill[ColorSL] (-0.65,-4.6) rectangle (1.7,-3.0);
\draw[] (-0.65,-4.6) rectangle (1.7,-3.0);
\draw[] (0.515,-4.6) -- (0.515,-3.0);

\draw[->] (-0.2,-3.0) -- (-5.29,-0.17);
\draw[->] (0.4,-3.0) -- (1.03,-1.02);

\draw[->] (0.8,-3.0) -- (-0.7,0.1);

\node[rotate=0] at (-0.08,-3.8) {\includegraphics[height=2.1cm]{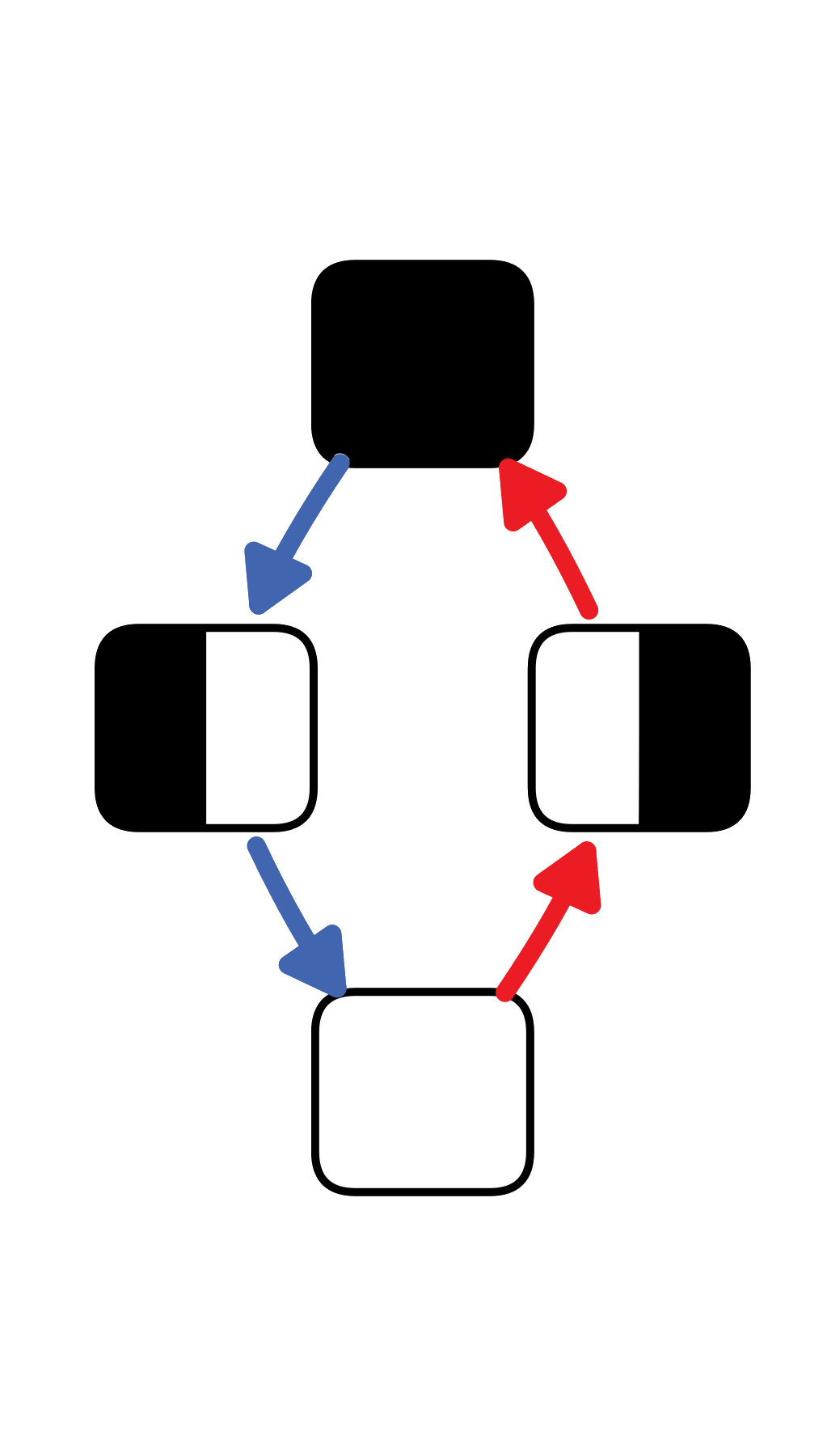}};

\node[rotate=0] at (-0.065,-4.325) {\small \scalebox{0.45}{$\boldsymbol{--}$}};
\node[rotate=0] at (-0.065,-3.275) {\small \scalebox{0.45}{$\color{white}\boldsymbol{++}$}};
\node[rotate=0] at (-0.38,-3.795) {\small \scalebox{0.45}{$\color{white}\boldsymbol{+}\color{black}\boldsymbol{-}$}};
\node[rotate=0] at (0.255,-3.795) {\small \scalebox{0.45}{$\color{black}\boldsymbol{-}\color{white}\boldsymbol{+}$}};

\node[rotate=0] at (1.088,-3.8) {\includegraphics[height=2.1cm]{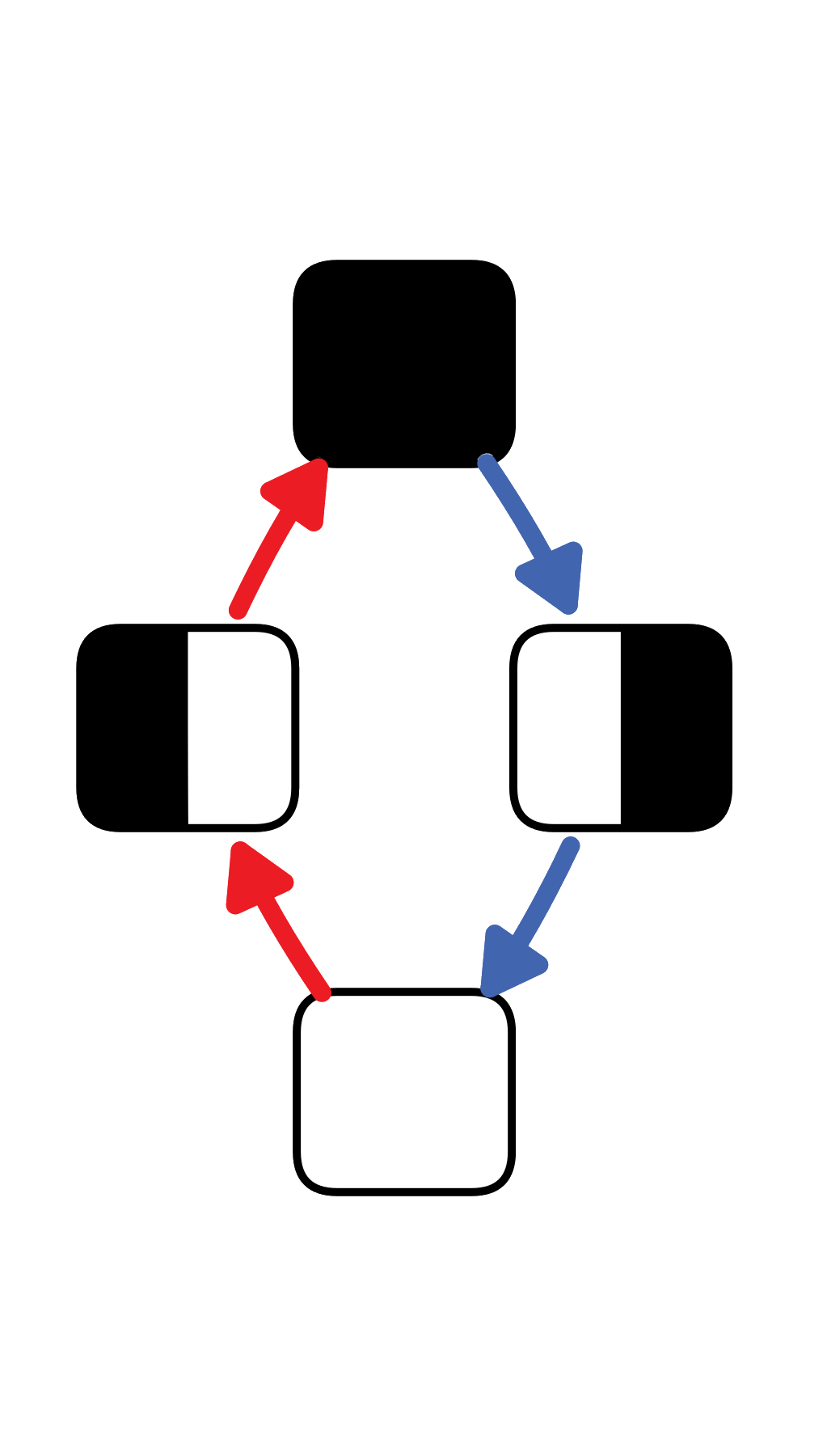}};

\node[rotate=0] at (1.085,-4.325) {\small \scalebox{0.45}{$\boldsymbol{--}$}};
\node[rotate=0] at (1.085,-3.275) {\small \scalebox{0.45}{$\color{white}\boldsymbol{++}$}};
\node[rotate=0] at (0.77,-3.795) {\small \scalebox{0.45}{$\color{white}\boldsymbol{+}\color{black}\boldsymbol{-}$}};
\node[rotate=0] at (1.405,-3.795) {\small \scalebox{0.45}{$\color{black}\boldsymbol{-}\color{white}\boldsymbol{+}$}};

\node[rotate=0] at (-6.65,-2.95) {\small (e)};

\node[rotate=0] at (-5.9,-3.94) {\includegraphics[height=2.26cm]{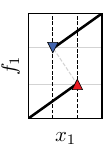}};

\node[rotate=0] at (-4.1,-3.94) {\includegraphics[height=2.26cm]{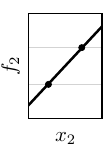}};

\node[rotate=0] at (-2.3,-3.94) {\includegraphics[height=2.26cm]{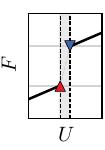}};

\node[rotate=0] at (-4.82,-4.4) {\small $\boldsymbol{+}$};
\node[rotate=0] at (-3.01,-4.4) {\small $\boldsymbol{=}$};

\draw[] (-4.37,-4.5) -- (-4.08,-4.5) -- (-4.08,-4.2) -- cycle;
\node[rotate=0] at (-3.9,-4.35) {\small $k$};

\node[rotate=37] at (-5.36,-3.37) {\scriptsize $+1$};
\node[rotate=37] at (-6.09,-4.30) {\scriptsize $-1$};

\node[rotate=23] at (-1.79,-3.58) {\scriptsize $+1$};
\node[rotate=23] at (-2.47,-4.08) {\scriptsize $-1$};

\end{tikzpicture}
\vspace*{-0.9cm}
\caption{\small{Self-loops in networks containing active elements. (a-b) Force-displacement curves for a dissipative element with $d_i>0$ and $\Delta E_i<0$ (a, blue), and for an active element with $d_i < 0$ and $\Delta E_i>0$ (b, red).
(c) Two serially coupled elements, with total displacement $U = x_1 + x_2$.
(d) Four classes of behaviors for two serially-coupled elements as function of $d_1$ and $d_2$; here 
$(x_1^-, x_1^+) = (-1,1)$, $(x_2^-, x_2^+) = (0.2,0.4)$.
(e) Serially coupling one active element and a linear spring (representing one of the elastic branches of the second element) can produce a gap without stable states ($U^+(\triangle) < U^-(\triangledown)$) if $d_1$ is strongly negative (Supplemental Material). (f) Regions where self-loops emerge for two coupled hysterons; for fixed $\sigma_1 = \sigma_2 = 0$ (green areas), and for $(\sigma_1,\sigma_2)$ determined for the physical system represented by the red dot in (d) (red boundaries). The blue hatched regions in (d,f) represent purely dissipative systems.
Orange lines in (d,f) indicate active elements that lead to symmetric hysteron interactions.
(g) Two $L=4$ self-loops observed in (d,f).}}
\label{fig:self_loop_size_4}
\end{figure}

{\em Proliferation of self-loops.---}
We now return to the hysteron model, and seek to understand the parameters for which self-loops occur. Each
self-loop is associated with
a set of linear inequalities and occurs in a polytope in parameter space $(h_i^\pm,c_{ij})$
\cite{van2021profusion,keim2021multiperiodic,teunisse2024transition,lindeman2025generalizing}.
We first consider $L\!=\!4$ loops (Fig. \ref{fig:self_loop_size_4}g). For two spins, we find that a gap of size $|\Delta c| - |\Delta h^c|$ opens up when  $c_{12}c_{21} < 0$ and $|\Delta c| > |\Delta h^c|$, where $\Delta c := c_{12} - c_{21}$ and $\Delta h^c := h_2^c - h_1^c$. These conditions are sufficient and necessary for the emergence of a self-loop (Fig.~\ref{fig:self_loop_size_4}f). For two hysterons, the
range in parameter space where these self-loops can occur shrinks (Fig.~\ref{fig:self_loop_size_4}f, Supplemental Material).
Finally, we can extend these conditions to  arbitrary $N$; as only two elements $i$ and $j$ are involved for $L=4$ self-loops, such short self-loops are prohibited by requiring $c_{ij}c_{ji} \ge 0$ for all pairs $(i,j)$ (Supplemental Material).

By identifying all potential self-loops with $L \leq 2^N$, one could, in principle, determine all corresponding polytopes; the complement of their union is then free of self-loops.
We have investigated the number of distinct self-loops, each corresponding to a different polytope in parameter space, with $L$.
While our approach allows an efficient classification of the self-loop structures, we find that their number grows extremely rapidly with $N$  (App. \ref{app:fundamental_loop}; Supplemental Material). This
proliferation of the number, length, and complexity of self-loops makes deriving explicit and sharp conditions that identify all self-loops unfeasible. In the remainder, we first introduce a lenient strategy that removes the shortest and suppresses longer self-loops, followed by strict ensembles that fully eliminate all self-loops but are overly restrictive.

\begin{figure}[t!]
\hspace*{-0.3cm}
\begin{tikzpicture}

\node[rotate=0] at (-0.08,0.0) {\includegraphics[height=4.3cm]{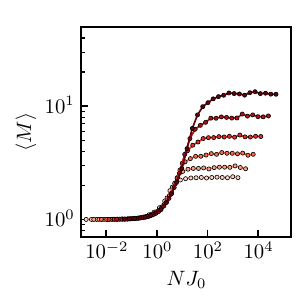}};

\node[rotate=0] at (-4.58,0.0) {\includegraphics[height=4.3cm]{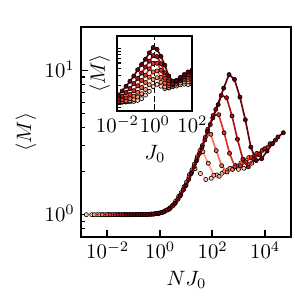}};

\node[rotate=0] at (-0.7,1.42) {\small (b)};\node[rotate=0] at (-2.92,1.42) {\small (a)};

\fill[white] (-5.45,0.93) rectangle (-5.14,1.26);
\node[rotate=90] at (-5.30,1.09) {\small $A$};
\fill[white] (-6.51,0.1) rectangle (-6.2,0.43);
\node[rotate=90] at (-6.36,0.26) {\small $A$};
\fill[white] (-2.03,0.1) rectangle (-1.72,0.43);
\node[rotate=90] at (-1.88,0.26) {\small $A$};

\draw[->] (0.9,-0.6) -- (1.3,1.15);
\node[rotate=0] at (1.35,1.35) {\small $N$};

\draw[->] (-3.95,-0.5) -- (-3.03,0.53);
\node[rotate=0] at (-2.82,0.69) {\small $N$};

\node[rotate=0] at (0.0,-4.2) {\includegraphics[height=4.3cm]{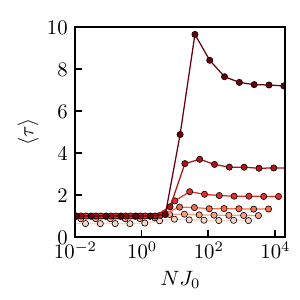}};

\node[rotate=0] at (-4.5,-4.2) {\includegraphics[height=4.3cm]{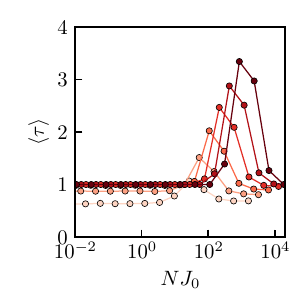}};

\node[rotate=0] at (-0.7,-2.78) {\small (d)};
\node[rotate=0] at (-5.2,-2.78) {\small (c)};

\draw[->] (0.6,-5.3) -- (1.3,-3.0);
\node[rotate=0] at (1.38,-2.8) {\small $N$};

\draw[->] (-3.6,-5.1) -- (-2.9,-2.9);
\node[rotate=0] at (-2.8,-2.67) {\small $N$};

\end{tikzpicture}
\vspace*{-0.8cm}
\caption{\small{Simulations of large systems of coupled hysterons in the constant-columns (left) and {symmetric} (right) ensembles
($N\!=\!16, 32, \dots, 512 $ for increasingly dark 
colors). (a-b) Ensemble averaged avalanche size $\langle A \rangle$. To determine these, we initialize the system at a 
stable state $S^0$ at $H=0$, increase $H$, and measure the number of flips before the system settles on a stable state. (c-d) Ensemble averaged transient $\langle \tau \rangle$, where $\tau$ is the number of cyclic drive cycles after which the system reaches a periodic orbit (Supplemental Material).}}
\label{fig:large_N}
\end{figure}

{\em Lenient strategy: weak {asymmetry}.---}
We introduce the notion of Weak Asymmetry (WA): $c_{ij} c_{ji} \geq 0$ for all pairs $(i,j)$. 
Not only does WA eliminates $L=4$ self-loops, but it also suppresses the number of longer self-loops that are realizable (App. \ref{app:fundamental_loop}, Supplemental Material). Statistical sampling reveals that WA is an effective strategy to suppress self-loops. In particular, {$P_{sl}^0 \rightarrow 0$} for large $N$, allowing to sample individual transitions, although $P_{sl}$ slowly grows with $N$: the combinatorial possibilities of finding a self-loop dominates in large systems (Supplemental Material). Nevertheless, for intermediate $N$, WA strongly suppresses self-loops, e.g., $P_{sl} \approx 14\%$ 
for large couplings and $N=10$.
Hence, {WA} strictly prohibits short self-loops and suppresses longer self-loops. 

{\em Strict ensembles.---}
We now present ensembles of asymmetric interactions which strictly prohibit self-loops. First, if all interactions are positive ($c_{ij}\ge 0$), avalanches exhibit monotonic evolution of the magnetization {$m:=\Sigma s_i$}, thus prohibiting self-loops (App. \ref{app:demo_positiveCouplings}).
If all interactions are negative, and either $c_{ik}=-d_{k}$ (constant-columns) or $c_{ki}=-d_{k}$ (constant-rows), where $d_k \ge 0$,
self-loops are also prohibited. In the former case (which corresponds to the hashed blue region in Fig. \ref{fig:self_loop_size_4}d), the interactions prohibit scrambling \cite{liu2024controlled}, which in turn prohibits self-loops (App. \ref{app:demo_serial}); in the latter case, the interactions only allow avalanches of length two, too short to form a self-loop when $\sigma_i \geq 0$ (App. \ref{app:demo_transposedserial}, Supplemental Material).
We note that the statistics of, e.g., avalanches and self-loops drastically depends on the dynamical rule in all these ensembles.
In particular, when race conditions are not allowed \cite{van2021profusion, teunisse2024transition, liu2024controlled}, constant-columns interactions restrict the avalanche size $A\le 2$, whereas 
flipping the most (or least) unstable elements leads to much larger $A$ (Fig. \ref{fig:large_N}a). Moreover, flipping all unstable elements simultaneously instead of only the most unstable one leads to a dominance of $L=2$ self-loops for {symmetric}, constant-columns and constant-rows interactions (Supplemental Material).

\begin{figure}[t!]
\centering
\begin{tikzpicture}

\draw[rounded corners=5pt] (0.0,0.0) rectangle (5.8,2.3);
\draw[dashed] (0.0,1.15) -- (5.8,1.15);

\node[rotate=90] at (-0.3,1.15) {\footnotesize Multistable media};

\node[anchor=west] at (0.05,2.03) {\footnotesize \textit{PASSIVE}};
\node[anchor=west] at (0.05,0.27) {\footnotesize \textit{ACTIVE}};

\node[rotate=0] at (1.1,1.45) {\footnotesize \textbf{PM}};
\node[rotate=0] at (1.1,0.85) {\footnotesize \textbf{AM}};

\node[rotate=0] at (3.9,1.45) {\footnotesize \textbf{PMH}};
\node[rotate=0] at (3.9,0.85) {\footnotesize \textbf{AMH}};

\draw[rounded corners=5pt] (2.2,-0.15) rectangle (7.2,2.15);

\node[rotate=90] at (7.5,1.0) {\footnotesize Hysteron models};

\node[rotate=0] at (6.9,1.0) {\footnotesize \textbf{H}};

\fill[blue!30, opacity=0.7] (5.8,1.4) ellipse (0.6cm and 0.4cm);
\draw[] (5.8,1.4) ellipse (0.6cm and 0.4cm);

\node[rotate=0] at (5.8,1.4) {\tiny $c_{ij} = c_{ji}$};

\fill[ColorSL, opacity=0.7] (5.8,0.5) ellipse (0.6cm and 0.4cm);
\draw[] (5.8,0.5) ellipse (0.6cm and 0.4cm);

\node[rotate=0] at (5.8,0.5) {\footnotesize \textbf{SL}};

\fill[ColorSL, opacity=0.7] (6.6,0.2) ellipse (0.25cm and 0.15cm);
\draw[] (6.6,0.2) ellipse (0.25cm and 0.15cm);

\fill[ColorSL, opacity=0.7] (6.6,0.48) ellipse (0.13cm and 0.08cm);
\draw[] (6.6,0.48) ellipse (0.13cm and 0.08cm);

\fill[ColorSL, opacity=0.7] (5.0,0.27) ellipse (0.2cm and 0.15cm);
\draw[] (5.0,0.27) ellipse (0.2cm and 0.15cm);

\fill[ColorSL, opacity=0.7] (4.90,0.62) ellipse (0.25cm and 0.15cm);
\draw[] (4.90,0.62) ellipse (0.25cm and 0.15cm);

\fill[ColorSL, opacity=0.7] (4.57,0.37) ellipse (0.18cm and 0.14cm);
\draw[] (4.57,0.37) ellipse (0.18cm and 0.13cm);

\end{tikzpicture}
~\vspace{-0.25cm}
\caption{Relations between active and passive
multistable physical systems, hysteron models, self-loops and symmetry of $c_{ij}$. Both active and passive systems can map to hysteron model with symmetric couplings ($c_{ij} = c_{ji}$, blue region); in this case, self-loops are forbidden.
The region where self-loops occur forms a complex cloud of polytopes in the space of hysteron models 
(SL, green regions).}
\label{fig:fig5}
\end{figure}

The strictly self-loop-free ensembles allow us to study the statistics of unprecedentedly large systems of interacting hysterons, including the distributions of avalanche sizes \(A\), transient times \(\tau\), and multiperiodicities \(T\) of orbits under cyclic drive (Fig.~\ref{fig:large_N}, Supplemental Material).
We find that these significantly depend on the ensemble, e.g.,  avalanches and transients are shorter in the constant-columns than in the symmetric
ensemble, and their dependence on $NJ_0$ is qualitatively different.

{\em Conlusion and Outlook.---}
The picture that emerges is that hysteron models connect with both passive and active multistable physical systems. We
summarize the relations between
networks of bistable elements, the symmetry of hysteron interactions and self-loops in Fig.~\ref{fig:fig5}. Strikingly, symmetric hysteron interactions can arise from networks of active or passive bistable elements, but cannot feature self-loops, and 
asymmetric hysteron interactions may lead to self-loops, but there is no simple criterion to delineate the parameter regions where self-loops can emerge.
Future challenges first concern expanding the range of mappings between physical networks and hysterons, to define additional self-loop-free hysteron ensembles, and to extend the hysteron model with a measure of energy dissipation.
Second, recent works have studied {the emergence of oscillatory dynamics through} non-reciprocal interactions
\cite{fruchart2021non,lorenzana2024non, avni2025dynamical, avni2025nonreciprocal}, and
our explorations suggest that  
non-reciprocally coupled passive bistable elements 
may be amenable to a hysteron-like description. Third, since dynamical effects \cite{jin2025dynamic} or geometric nonlinearities \cite{shohat2025geometric} 
lead to transitions not described by current hysteron models even for passive physical systems
(PM), it will be exciting to explore the interplay between dynamics, geometry, and bistability for
active systems (AM).
Fourth, we suggest that self-loops can be linked to multiperiodic responses under cyclic driving
using the concept of transition scaffold ~\cite{keim2021multiperiodic, teunisse2024transition}. 
Finally, it is an open question whether all sectors of the interacting hysteron model are realizable with physical systems (H).

\begin{acknowledgments}

PB, MHT and MvH acknowledge funding from European Research Council Grant ERC-$101019474$. We thank Dor Shohat, Yoav Lahini, Corentin Coulais, Jonas Veenstra, and Menachem Stern for valuable discussions, and an anonymous referee for astute criticism.

\end{acknowledgments}

\appendix

\section{Systematic convergence for symmetric interactions.}
\label{app:demo_reciprocal}
Here we show that symmetric interactions yield transitions that lower {a Lyapunov-like function}, first focusing on spins (for which $h_{i}^{+} = h_{i}^{-} = h_{i}^{c}$) {following standard approaches \cite{hopfield1982neural, mackay2003itprnn}}, and then extending the result to hysterons. Consider an initial state $S^{0}$ and a value of the drive $H$ such that element $p$ is unstable, which implies:
\begin{itemize}
 \item $H > h_{p}^{c} - \sum_{j \neq p}  c_{pj} s_{j}^{0}$, if $s_{p}^{0} = -1$~,
 \item $H < h_{p}^{c} - \sum_{j \neq p}  c_{pj}s_{j}^{0}$, if $s_{p}^{0} = 1$~,
\end{itemize}
which can be rewritten as:
\begin{equation}
 s_p^{0} \left( H - h_p^{c} + \sum_{j \neq p} c_{pj} s_j^{0}  \right) < 0~.\label{inst_eq}
\end{equation}
Let us now introduce the function $V$, reminiscent of a Sherrington-Kirkpatrick model with random thresholds:
\begin{equation}
\begin{aligned}
\label{eq:function_V}
 V(\boldsymbol{s}) &= - \sum_i \left[ s_i ( H - h_i^c ) + \frac{1}{2} \sum_{j \neq i} c_{ij} s_i s_j \right], \\
 &= - \sum_i s_i ( H - h_i^c)  - \frac{1}{2} \sum_i \sum_{j \neq i} c_{ij} s_i s_j,
\end{aligned}
\end{equation}
where the first and second term on the r.h.s. of Eq. (\ref{eq:function_V}) can be seen as a field and interaction term. We aim to compute $\Delta V = V(S^{1}) - V(S^{0})$, where $S^{1}$ and $S^{0}$
are the state before and after snapping element $p$, i.e. $s_{i \neq p}^{1} = s_{i \neq p}^{0}$, and $s_{p}^{1} = - s_{p}^{0}$. Clearly, the terms with elements different than $p$ will not contribute to $\Delta V$, and by splitting the sums accordingly, we obtain:
\begin{widetext}
\begin{equation}
\Delta V = -s_p^1(H-h_p^c) +s_p^0(H-h_p^c) -\frac{1}{2}\left(
\sum_{i\ne p}c_{ip}s_i s_p^1 
+\sum_{j\ne p}c_{pj}s_p^1 s_j
-\sum_{i\ne p}c_{ip}s_i s_p^0 
-\sum_{j\ne p}c_{pj}s_p^0 s_j\right)~,
\end{equation}
\end{widetext}
which can be simplified to:
\begin{equation}
 \Delta V = 2 s_p^0 \left[ (H - h_p^c) + \frac{1}{2} \sum_{j \neq p} \left( c_{jp} + c_{pj} \right) s_j \right],
\end{equation}
where we repeatedly use that for $j\ne p$, $s_j^0=s_j^1=:s_j$
For the case of symmetric interactions, i.e. $c_{jp} = c_{pj}$, we find:
\begin{equation}
 \Delta V = 2 s_p^0 \left[ (H - h_p^c) + \sum_{j \neq p} c_{pj} s_j  \right].
\end{equation}
Inserting the instability condition for element $p$, Eq.~(\ref{inst_eq}) yields $\Delta V < 0$. Therefore, the function $V$ is strictly decreasing for each single flip, which 
implies that the system cannot be trapped in a self-loop and must always evolve toward a stable state.
Note the importance of the factor $1/2$ in Eq. (\ref{eq:function_V}) in order to obtain the final result. This demonstration can be extended to finite span hysterons, by explicitly making the distinction between hysterons with positive and negative phase, and their respective thresholds, producing Eq.~(\ref{eq:energy_like_quantity}) of the main text.
\section{Positive interactions.}
\label{app:demo_positiveCouplings}
In this section, we show that for positive (ferromagnetic) interactions, i.e., $c_{ij} \ge  0$, the system cannot get trapped into a self-loop.
A self-loop is a cyclic avalanche: the system must come back to a previously visited unstable state. However, for positive interactions, each step in an avalanche 'goes in the same direction', i.e., the magnetization $m:=\sum_i s_i$ evolves monotonically
\cite{sethna1993hysteresis, van2021profusion}.
This prevents avalanches from revisiting earlier states, thus prohibiting self-loops, irrespective of the specific rule used to resolve race conditions (Supplemental Material).

\section{Constant-columns interactions.}
\label{app:demo_serial}
We now clarify why constant-columns ($c_{ik} = -d_k$, $d_k \geq 0$) define a self-loop-free interaction ensemble.
First, for negative (antiferromagnetic) interactions, avalanches (including self-loops) must be composed of alternating up/down transitions \cite{van2021profusion}. For $L \geq 6$, such self-loops exist
(Fig.~\ref{fig:self_loop_size_4}c, and Supplemental Material). 
However, all such self-loops violate loop-RPM, which requires scrambling: the ordering of the switching thresholds must be state-dependent \cite{van2021profusion}.
Constant-columns interactions do not allow for scrambling \cite{liu2024controlled}, therefore this ensemble strictly prevents all self-loops.

\section{Constant-rows interactions}
\label{app:demo_transposedserial}

We now show that in the constant-rows ensemble ($c_{ki} = -d_k$, $d_k \geq 0$), avalanches consist of at most 
two hysteron flips, which is too short to  allow self-loops. Without loss of generality, we consider an up avalanche initiated from state $S^0$ by an increase of $H$ up to $H_p^{+}(S^0)$, triggering the flipping of hysteron $p$ from $s_p = -1$ to $s_p = 1$, and leading to state $S^1$. Since we have negative interactions, avalanches must be composed of alternating up/down transitions, so that the next step would be the flipping of hysteron $q$ from $s_q = 1$ to $s_q = -1$ leading to state $S^2$.
To show that $S^2$ is stable to additional up flipping events, we first show that $H_i^+(S^0) = H_i^+(S^2)$ for $i \neq p,q$.
Using that in this ensemble $c_{ip} = c_{iq}$, and that
$s_p^0=-1$ and $s_q^0=1$, we find that 
\begin{eqnarray}
H_{i \neq p,q}^+(S^0)&=& h_i^+-\sum_j c_{ij}s_j^0 \\
&=& h_i^+ -\sum_{j \ne p,q} c_{ij}s_j^0 - c_{ip}s_p^0 -c_{iq}s_q^0\\
&=& h_i^+ -\sum_{j \ne p,q} c_{ij}s_j^0~.
\end{eqnarray}
Similarly, using that $s_p^2=1$ and $s_q^2=-1$,
we find that 
\begin{equation}
H_{i \neq p,q}^{+}(S^2) = h_i^+-\sum_{j\ne p,q} c_{ij}s_j^2~,\\
\end{equation}
and as $s_{j\ne p,q}^2=s_{j\ne p,q}^0$, we conclude that
$H_i^+(S^0)=H_i^+(S^2)$ for $i \neq p,q$. Since in state $S^0$ all hysterons $i \neq p,q$ are stable, they are also stable in state $S^2$. Moreover, both hysterons $p$ and $q$ are stable in state $S^2$ at $H = H_p^{+}(S^0)$: indeed $H_p^{-}(S^2) = H_p^{+}(S^0) - \sigma_p - 2 d_p < H_p^{+}(S^0)$, and hysteron $q$ just flipped. Therefore, the longest possible avalanche in this interaction ensemble consists of two steps.  
By the same argument, this result also holds for other race conditions which involve flipping hysterons one by one (Supplemental Material). \\
\section{Proliferation of self-loops}
\label{app:fundamental_loop}

In this section, we focus on {\em fundamental} loops, which are defined as the unique loops that involve all elements ($n_e=N$) up to permutations of the element indices (i.e., the two $L=4$ self-loops in Fig.~\ref{fig:self_loop_size_4}e are equivalent).
We determine the potential number of fundamental loop structures, \( M(n_{e}, L) \), from the combinatorics of flip sequences, and calculate the number of realizable loops with pairwise interactions, \( M_R(n_{e}, L) \) (Supplemental Material).
Both grow rapidly with $n_e$ and $L$ (Table~\ref{table:fundamental_loop}). In particular, for the shortest fundamental loops, 
$M(n_e,L=2n_e)$ grows as $1,6,56,796,\dots$ for $n_e=2,3,4,5,\dots$ (Figs. \ref{fig:self_loop_size_4}b and c for $n_e = 2,3$), and our data suggests that each of these is realizable.
The number of actual self-loops and polytopes grows even faster with \( N \). Introducing \( n_e \) elements into a larger group of \( N \) elements, and including permutations, maps each fundamental loop to a significantly larger number of actual loops and polytopes, fueling a further combinatorial explosion. So while in principle we can identify the self-loops and for each determine their polytope in parameter space \cite{teunisse2024transition}, in practice this is not feasible.

\begin{table}[h!]
\small
\vspace*{0.35cm}
\centering
\renewcommand{\arraystretch}{1.5}
\begin{tabular}{|p{.110\linewidth}|p{.13\linewidth}|p{.13\linewidth}|p{.200\linewidth}|p{.23\linewidth}|}
\hline
$L$/$n_e$ & $2$ & $3$ & $4$ & $5$ \\
\hline
$4$ & $1$/$1$/$0$ & -- & -- & -- \\
\hline
$6$ & -- & $6$/$6$/$2$ & -- & -- \\
\hline
$8$ & -- & $2$/$0$/$0$ & $56$/$56$/$24$ & -- \\
\hline
$10$ & -- & -- & $176$/$114$/$4$ & $796$/$796$/$376$ \\
\hline
$12$ & -- & -- & $420$/$145$/$1$ & $9028$/x/x \\
\hline
$14$ & -- & -- & $448$/$48$/$0$ & $76640$/x/x \\
\hline
$16$ & -- & -- & $112$/$4$/$0$ & $535584$/x/x \\
\hline
\end{tabular}
\vspace{-0.1cm}
\caption[table caption]{Numbers of fundamental self-loops of size $L$ involving $n_{e}$ elements. 
Note that \( 4 \leq L \leq 2^N \) and $ \log_2 L \leq n_e \leq L/2 $, as each element undergoes an even number of flips; loops with $ n_e $ elements can visit at most $ 2^{n_e} $ states; and self-loops of size 2 are excluded by \( h_i^+ \geq h_i^- \) (Supplemental Material). The numbers in each box represents $M(n_{e},L)$, $M_R(n_{e},L)$, and $M_W(n_{e},L)$, respectively. Note that the number of longest fundamental loops, $M(n_e, L=2^{n_e})$, are given by the number of directed Hamiltonian cycles in the binary $n_e$-cube ($1,2,112,15109096,\dots$ for $n_e=2,3,4,5,\dots$) \cite{knuth1975art, deza2010enumeration}.}
\label{table:fundamental_loop}
\end{table}

\bibliographystyle{apsrev4-2} 

\bibliography{bibli.bib}

\newpage
\widetext
\begin{center}
\textbf{\large Supplemental Material: Dynamic self-loops in networks of passive and active binary elements} \\
\vspace{0.4cm}
Paul Baconnier$^{1}$, Margot H. Teunisse$^{1, 2}$ and Martin van Hecke$^{1,2}$ \\
\vspace{0.2cm}
\textit{$^{1}$AMOLF, 1098 XG Amsterdam, The Netherlands.} \\
\textit{$^{2}$Huygens-Kamerlingh Onnes Laboratory, Leiden University, 2300 RA Leiden, The Netherlands.}
\end{center}
\vspace{0.5cm}
\setcounter{equation}{0}
\setcounter{figure}{0}
\setcounter{table}{0}
\setcounter{page}{1}
\makeatletter
\renewcommand{\theequation}{S\arabic{equation}}
\renewcommand{\thefigure}{S\arabic{figure}}
\renewcommand{\thetable}{S\Roman{table}}
\renewcommand{\theHtable}{S\Roman{table}}   

\setcounter{section}{0} 
\renewcommand\thesection{\arabic{section}} 
\renewcommand{\theHsection}{\arabic{section}} 
\makeatletter
\renewcommand{\@seccntformat}[1]{\csname the#1\endcsname.\space}
\renewcommand{\appendixname}{} 
\def\p@section{} 
\def\p@subsection{} 
\def\thesection{\arabic{section}}
\renewcommand\thesubsection{\thesection.\arabic{subsection}} 
\renewcommand{\@seccntformat}[1]{\csname the#1\endcsname.\space} 
\makeatother


\onecolumngrid

In the main manuscript, we investigate the root causes and statistics of self-loops in ensembles of interacting binary elements. In this document, we first elaborate on the conditions under which a self-loop emerges in the transition-graph shown in Fig. 1 of the main text. In Secs. \ref{sec:race_conditions} and \ref{sec:stable_states}, we discuss the probability of race conditions and the probability that a given state has a finite stability range as a function of coupling strength, for randomly-coupled spins and hysterons. In Secs. \ref{sec:spins} and \ref{sec:hysterons_spans}, we provide numerical evidence the problem of overwhelming self-loops is similar for spins instead of hysterons, and for ensembles of hysterons with equal spans.
In Sec. \ref{sec:gap_vs_selfLoops}, we further analyze the relationship between gaps and self-loops.
In Sec. \ref{sec:weak_nonreciprocity}, we provide numerical evidence that self-loops also overwhelm the response of large systems in the case of weakly asymmetric interactions.
In Sec. \ref{sec:size_4}, we derive the conditions to prevent $L=4$ self-loops for $N=2$ spins and $N=2$ hysterons with equal spans, and sample $L=4$ self-loops for large $N$.
In Sec. \ref{sec:active_serial_coupling}, we analyze the emergence of self-loops in system of serially-coupled two-state elements.
In Sec. \ref{sec:well_behaved_demo}, we provide physical illustrations for the different classes of well-behaved models.
In Sec. \ref{sec:rac_condition_rules}, we analyze the role of the different race conditions rules in the different families of models. In Sec. \ref{sec:selfLoop_structure}, we give details on the algorithm to generate all possible self-loops, and illustrate the different possible self-loop structures for different sizes $L$. Finally, in Sec. \ref{sec:large_avalanches}, we discuss numerical simulations of large systems of self-loop-free models.


\newpage
\section{Conditions for a self-loop of size $4$ in the general model}

In this section, we elaborate on the conditions under which a self-loop emerges in the graph represented in Fig. 1 of the main text. A sufficient condition for self-loops is that $H$ is in a range where all the states $S^{0}$, $S^{1}$, $S^{2}$ and $S^{3}$ are unstable. This situation is realized when $H$ is larger than the two up switching fields $H^{+}(S^{0})$ and $H^{+}(S^{1})$, smaller than the two down switching fields $H^{-}(S^{2})$ and $H^{-}(S^{3})$, and when both $H^{+}(S^{0})$ and $H^{+}(S^{1})$ are smaller than $H^{-}(S^{2})$ and $H^{-}(S^{3})$.

Starting from a stable state and driving the system up to instability, what are the conditions under which the transition triggers a self-loop? In this case, we evaluate the possibility for a self-loop for $H$ immediately above (below) the up (down) switching field of a given state. For each of the four possible starting states, we can write down the inequalities needed so that the self-loop shown on Fig. 1 of the main text is realized:

\begin{itemize}
\item \textbf{State} $S^{0}$: this state becomes unstable as soon as $H > H^{+}(S^{0})$, and a self-loop emerges at instability when:
\begin{equation}
\begin{aligned}
H^{+}(S^{1}) &< H^{+}(S^{0}), \\
H^{-}(S^{2}) &> H^{+}(S^{0}), \\
H^{-}(S^{3}) &> H^{+}(S^{0}),
\end{aligned}
\end{equation}
where each relationship enforces that a given state of the loop is unstable. Hence, a self-loop is triggered when: $H^{+}(S^{1}) < H^{+}(S^{0}) < \left(H^{-}(S^{2}),H^{-}(S^{3})\right)$ (see caption of Fig. 1 of the main text).
\item \textbf{State} $S^{1}$: this state becomes unstable as soon as $H > H^{+}(S^{1})$, and a self-loop emerges at instability when:
\begin{equation}
\begin{aligned}
H^{+}(S^{0}) &< H^{+}(S^{1}), \\
H^{-}(S^{2}) &> H^{+}(S^{1}), \\
H^{-}(S^{3}) &> H^{+}(S^{1}).
\end{aligned}
\end{equation}
This produces the condition: $H^{+}(S^{0}) < H^{+}(S^{1}) < \left(H^{-}(S^{2}),H^{-}(S^{3})\right)$.
\item \textbf{State} $S^{2}$: this state becomes unstable as soon as $H < H^{-}(S^{2})$, and a self-loop emerges at instability when:
\begin{equation}
\begin{aligned}
H^{-}(S^{3}) &> H^{-}(S^{2}), \\
H^{+}(S^{0}) &< H^{-}(S^{2}), \\
H^{+}(S^{1}) &< H^{-}(S^{2}).
\end{aligned}
\end{equation}
Yielding: $\left(H^{+}(S^{0}),H^{+}(S^{1})\right) < H^{-}(S^{2}) < H^{-}(S^{3})$.
\item \textbf{State} $S^{3}$: this state becomes unstable as soon as $H < H^{-}(S^{3})$, and a self-loop emerges at instability when:
\begin{equation}
\begin{aligned}
H^{-}(S^{2}) &> H^{-}(S^{3}), \\
H^{+}(S^{0}) &< H^{-}(S^{3}), \\
H^{+}(S^{1}) &< H^{-}(S^{3}).
\end{aligned}
\end{equation}
Yielding: $\left(H^{+}(S^{0}),H^{+}(S^{1})\right) < H^{-}(S^{3}) < H^{-}(S^{2})$.
\end{itemize}

\section{Race conditions} \label{sec:race_conditions}

Here, we investigate the probability of race conditions for collections of spins ($\sigma_i = 0$), hysterons ($\sigma_i$ flatly sampled from $\left[ 0, 0.5 \right]$), and hysterons with equal spans ($\sigma_i = 0.5$).

\begin{figure}[h!]
\hspace*{-0.15cm}
\begin{tikzpicture}

\node[rotate=0] at (-0.02,-8.6) {\includegraphics[height=4.3cm]{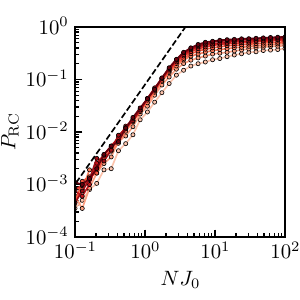}};

\node[rotate=0] at (-0.75,-7.15) {\small (a)};

\node[rotate=0] at (4.38,-8.6) {\includegraphics[height=4.3cm]{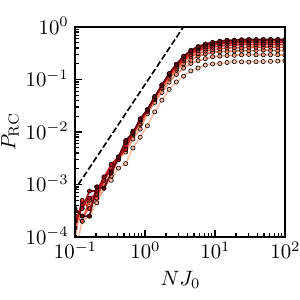}};

\node[rotate=0] at (3.65,-7.15) {\small (b)};

\node[rotate=0] at (8.78,-8.6) {\includegraphics[height=4.3cm]{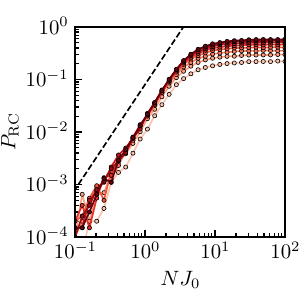}};

\node[rotate=0] at (8.05,-7.15) {\small (c)};


\end{tikzpicture}
\vspace*{-0.45cm}
\caption{\small{\textbf{Statistics of race conditions for randomly-coupled two-states elements.} Probability $P_{\rm{RC}}$ of race conditions (ensemble of $2 \times 10^4$ transitions $S^0 \rightarrow S^1$), for an initial state $S^0$ stable at $H = 0$; the black dashed line represents the slope $2$. Markers are color coded from light to dark red as $N$ increases, with $N \in \left[ 2, 3, 4, 5, 6, 7, 8, 9, 10\right]$. (a) Spins ($\sigma_i = 0$). (b) Hysterons ($\sigma_i$ flatly sampled from $\left[ 0, 0.5 \right]$). (c) Hysterons with equal spans ($\sigma_i = 0.5$).}}
\label{fig:race_conditions}
\end{figure}

When a state $S^0$ becomes unstable and one of the hysterons flips to produce state $S^1$, the number of unstable hysterons in $S^1$ can be either zero ($S^1$ is stable), one, or more than one. If more than one hysteron is unstable, this causes a race condition. We calculate the race-condition probability $P_{\rm{RC}}$ -- the probability of more than one hysteron being unstable in state $S^{1}$ -- by selecting a state $S^0$ that is stable at $H = 0$, increasing $H$ past $H^{+}(S^0)$, and investigating the number of unstable hysterons of $S^1$ at $H = H^{+}(S^0)$. We find that $P_{\rm{RC}}$ increases as $(N J_0)^2$ for $N J_0 \ll 1$, and saturates at a significant value that increases towards $1$ for large $N$ for $N J_0 \gg 1$ (Figs. \ref{fig:race_conditions}). We observe only minor differences between spins, hysterons, and hysterons with equal spans.

\section{Number of stable states} \label{sec:stable_states}

In this section, we measure the probability that a given state has a finite stability range for collections of spins ($\sigma_i = 0$), hysterons ($\sigma_i$ flatly sampled from $\left[ 0, 0.5 \right]$), and hysterons with equal spans ($\sigma_i = 0.5$).

\begin{figure}[h!]
\hspace*{-0.15cm}
\begin{tikzpicture}

\node[rotate=0] at (0.0,-4.3) {\includegraphics[height=4.3cm]{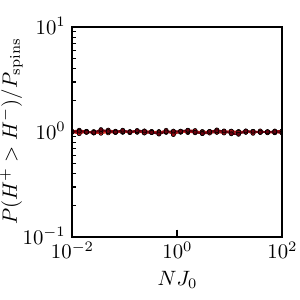}};

\node[rotate=0] at (-0.75,-2.85) {\small (a)};

\node[rotate=0] at (4.4,-4.3) {\includegraphics[height=4.3cm]{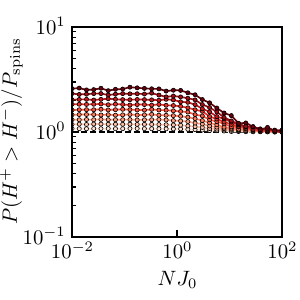}};

\node[rotate=0] at (3.65,-2.85) {\small (b)};

\draw[->] (4.5,-4.3) -- (5.0,-3.2);
\node[rotate=0] at (5.13,-2.97) {\small $N$};

\node[rotate=0] at (8.8,-4.3) {\includegraphics[height=4.3cm]{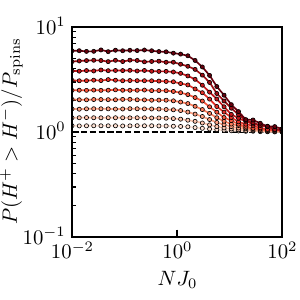}};

\draw[->] (9.5,-4.3) -- (10.0,-3.2);
\node[rotate=0] at (10.13,-2.97) {\small $N$};

\node[rotate=0] at (8.05,-2.85) {\small (c)};

\end{tikzpicture}
\vspace*{-0.45cm}
\caption{\small{\textbf{Fraction of potentially stable states.} Probability that a randomly chosen state $S$ has a finite stability range; scaled by $P_{\textrm{spins}} = (N+1)/2^N$. Markers are color coded from light to dark red as $N$ increases, with $N \in \left[ 2, 3, 4, 5, 6, 7, 8, 9, 10\right]$. (a) Spins ($\sigma_i = 0$). (b) Hysterons ($\sigma_i$ flatly sampled from $\left[ 0, 0.5\right]$). (c) Hysterons with equal spans ($\sigma_i = 0.5$).}}
\label{fig:stable_states}
\end{figure}

Independently of the microscopic hysteresis and coupling strength, the probability $P_s$ that a random state $S$ among the $2^N$ possible states is stable for some value of the driving $H$, i.e. $H^{+}(S) > H^{-}(S)$, asymptotes to $0$ for large $N$. Indeed, for spins, it is easy to show that there always exists only $N + 1$ potentially stable states, so that $P_s = P_{\rm{spins}} = (N + 1)/2^N$ (Fig. \ref{fig:stable_states}-a).

For hysterons and hysterons with equal spans, we find that $P_s$ asymptotes toward $P_{\rm{spins}}$ in the large coupling limit, i.e., $NJ_0 \gg 1$ (Figs. \ref{fig:stable_states}-b and c). However, in the small coupling limit, $P_s$ is larger than $P_{\rm{spins}}$. In this case, $P_s$ is dictated by the statistics of the Preisach graphs that are sampled, which is itself dictated by the statistics of the different orderings of the switching fields $h_i^{\pm}$. Interestingly, in the case of hysterons with equal spans, the Preisach graphs which are sampled generally contain more stable states. Note that in all case, $P_s$ still asymptotes to $0$ even in the small coupling limit, but slower than in the case of spins.

\section{Case of binary spins} \label{sec:spins}

\begin{figure}[t!]
\hspace*{-0.15cm}
\begin{tikzpicture}

\node[rotate=0] at (4.4,-8.6) {\includegraphics[height=4.3cm]{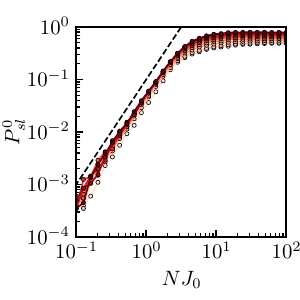}};
\node[rotate=0] at (4.4,-4.3) {\includegraphics[height=4.3cm]{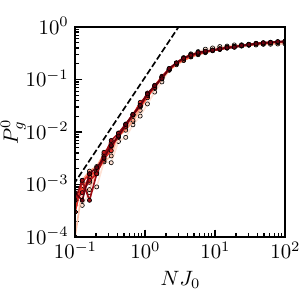}};

\node[rotate=0] at (8.8,-8.6) {\includegraphics[height=4.3cm]{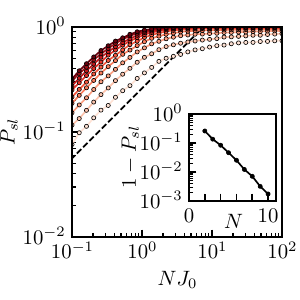}};
\node[rotate=0] at (8.8,-4.3) {\includegraphics[height=4.3cm]{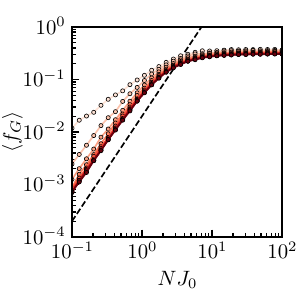}};

\node[rotate=0] at (3.65,-2.85) {\small (a)};
\node[rotate=0] at (3.65,-7.15) {\small (c)};

\node[rotate=0] at (8.05,-2.85) {\small (b)};
\node[rotate=0] at (8.05,-7.15) {\small (d)};

\end{tikzpicture}
\vspace*{-0.45cm}
\caption{\small{\textbf{Overwhelming self-loops for randomly-coupled spins.} Statistical measures for gaps and self-loops scale when plotted as function of $N J_0$ and dominate for $N J_0 \gg 1$ ($10^5$ samples; color from light to dark as $N$ increases from $2$ to $10$).
(a) Probability $P_g^0$ of finding a gap at $H = 0$  (dashed line indicates slope $2$).
(b) Fraction of gaps $f_g$, defined
as the mean of the ratio of the size of intervals where no stable states exist divided by $\left[ H^{+}(-\dots-), H^{-}(+\dots+) \right]$  (dashed line indicates slope $2$).
(c) Probability $P_{sl}^0$ of  self-loops at $H = 0$ (dashed line indicates slope $2$).
(d) Probability $P_{sl}$ of finding at least one self-loop for any value of $H$ (dashed line indicates slope $2/3$).
Inset: The probability to be self-loop free,
$1-P_{sl}$, decays to zero exponentially with $N$ for large couplings ($NJ_0 = 10^2$).}}
\label{fig:hysteron_problem}
\end{figure}

In this section, we reproduce the simulations of Fig. 2 of the main text, focusing on the case of randomly-coupled spins, i.e. $\sigma_i = 0$ for all the elements (Figs. \ref{fig:hysteron_problem}).

Let us focus on the similarities between spins and hysterons with distributed spans. In both cases, the statistical weight of gaps increases as a power law for $NJ_0 \ll 1$, and saturates toward a significant value for $NJ_0 \gg 1$ (Figs. \ref{fig:hysteron_problem}-a and b). Moreover, the probability of finding at least one self-loop for any $H$ increases as a power law for $N J_0 \ll 1$, and saturates toward a constant value for $N J_0 \gg 1$ (Fig. \ref{fig:hysteron_problem}-d), which asymptotes to $1$ as $N$ increases (Fig. \ref{fig:hysteron_problem}-d, inset). Therefore, the response of large, strongly coupled systems of spins is dominated by self-loops.

There are however subtle differences with the case of hysterons. In particular, for spins, all observables grow slower with $NJ_0$ but are bigger than for hysterons for $NJ_0 \ll 1$ (there are more gaps and self-loops for spins than for hysterons in the small coupling limit). Also, for $NJ_0 \gg 1$, self-loops are more likely with spins than with hysterons: for $N = 10$ and in the large coupling limit ($NJ_0 = 10^2$), $99.8\%$ of instances exhibit at least one self-loop for spins, while the probability is $98.5\%$ for hysterons. This is expected given that microscopic hysteresis contributes to preventing self-loops.

\section{Case of hysterons with equal spans} \label{sec:hysterons_spans}

\begin{figure}[t!]
\hspace*{-0.15cm}
\begin{tikzpicture}

\node[rotate=0] at (4.4,-8.6) {\includegraphics[height=4.3cm]{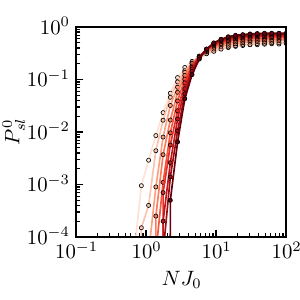}};
\node[rotate=0] at (4.38,-4.3) {\includegraphics[height=4.3cm]{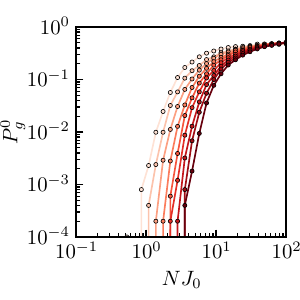}};

\node[rotate=0] at (8.8,-8.6) {\includegraphics[height=4.3cm]{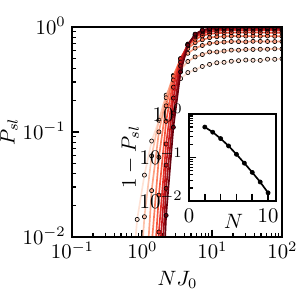}};
\node[rotate=0] at (8.8,-4.3) {\includegraphics[height=4.3cm]{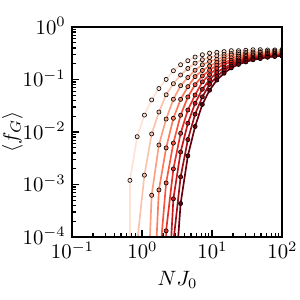}};

\node[rotate=0] at (3.65,-2.85) {\small (a)};
\node[rotate=0] at (3.65,-7.15) {\small (c)};

\node[rotate=0] at (8.05,-2.85) {\small (b)};
\node[rotate=0] at (8.05,-7.15) {\small (d)};

\end{tikzpicture}
\vspace*{-0.45cm}
\caption{\small{\textbf{Overwhelming self-loops for randomly-coupled hysterons with equal spans.} Statistical measures for gaps and self-loops scale when plotted as function of $N J_0$ and dominate for $N J_0 \gg 1$ ($10^5$ samples; color from light to dark as $N$ increases from $2$ to $10$).
(a) Probability $P_g^0$ of finding a gap at $H = 0$.
(b) Fraction of gaps $f_g$, defined
as the mean of the ratio of the size of intervals where no stable states exist divided by $\left[ H^{+}(-\dots-), H^{-}(+\dots+) \right]$.
(c) Probability $P_{sl}^0$ of self-loops at $H = 0$.
(d) Probability $P_{sl}$ of finding at least one self-loop for any value of $H$.
Inset: The probability to be self-loop free,
$1-P_{sl}$, decays to zero exponentially with $N$ for large couplings ($NJ_0 = 10^2$).}}
\label{fig:hysteron_problem_equal_spans}
\end{figure}

In this section, we reproduce the simulations of Fig. 2 of the main text, focusing on the case of randomly-coupled hysterons with equal spans, i.e. $\sigma_i = 0.5$ for all the elements (Figs. \ref{fig:hysteron_problem_equal_spans}).

We find that the probability of finding at least one self-loops for any $H$ saturates toward a constant value for $N J_0 \gg 1$ (Fig. \ref{fig:hysteron_problem_equal_spans}-d), which asymptotes to $1$ as $N$ increases (Fig. \ref{fig:hysteron_problem_equal_spans}-d, inset). However, in the case of hysterons with equal spans, we find that all observables seem to have a lower cutoff in $NJ_0$ below which no gaps or self-loops exist. This is expected given the physics of self-loops for $N=2$ hysterons (see Fig. 3-d of the main text and next section). Moreover, also as expected, we find an even lower probability of $98.3\%$ of finding at least one self-loop for any $H$ for $N=10$ and $NJ_0 = 10^{2}$ (Fig. \ref{fig:hysteron_problem_equal_spans}-d).

\section{Relationship between gap and self-loops} \label{sec:gap_vs_selfLoops}

\begin{figure}[t!]
\hspace*{-0.15cm}
\begin{tikzpicture}

\node[rotate=0] at (0.0,0.0) {\includegraphics[height=4.3cm]{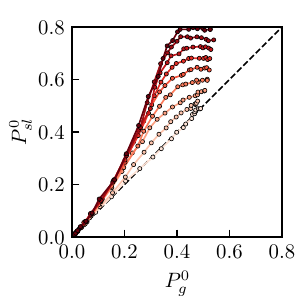}};

\node[rotate=0] at (4.4,0.0) {\includegraphics[height=4.3cm]{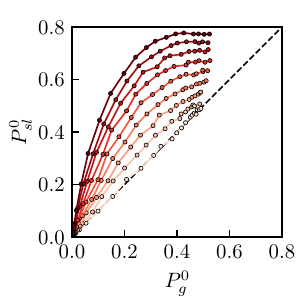}};

\node[rotate=0] at (8.8,0.0) {\includegraphics[height=4.3cm]{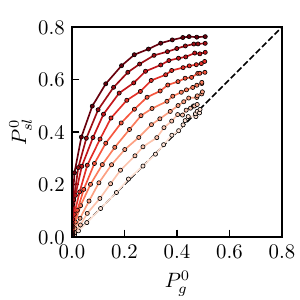}};

\node[rotate=0] at (-0.8,1.45) {\small (a)};
\node[rotate=0] at (3.6,1.45) {\small (b)};
\node[rotate=0] at (8.0,1.45) {\small (c)};

\draw[<-] (-0.1,1.25) -- (0.2,-0.15);
\node[rotate=0] at (-0.12,1.47) {\small $N$};


\end{tikzpicture}
\vspace*{-0.45cm}
\caption{\small{\textbf{Probability of self-loops as a function of the probability of gaps} at $H = 0$, for $NJ_0 \in \left[ 10^{-1}, 10^{2} \right]$. (a) Spins ($\sigma_i = 0$). (b) Hysterons with distributed spans ($\sigma_i$ flatly sampled from $\left[ 0, 0.5 \right]$). (c) Hysterons with equal spans ($\sigma_i = 0.5$). The black dashed lines represent $y = x$; markers are color coded from light to dark red as $N$ increases, with $N \in \left[ 2, 3, 4, 5, 6, 7, 8, 9, 10\right]$.}}
\label{fig:gap_vs_self_loops}
\end{figure}

In this section, we provide further evidence that self-loops can also occur outside of gaps. We restrict to $H = 0$, and we measure the probability of finding a gap $P_g^0$, and of the emergence of at least one self-loop starting from any of the $2^N$ states $P_{sl}^0$, while varying $J_0$.

For $N=2$ spins (Fig. \ref{fig:gap_vs_self_loops}-a), hysterons (Fig. \ref{fig:gap_vs_self_loops}-b), and hysterons with equal spans (Fig. \ref{fig:gap_vs_self_loops}-c), there is a one-to-one correspondence between the probability of gaps $P_{g}^{0}$ and the probability of self-loops $P_{sl}^0$: indeed, in this case, all $4$ possible states must be unstable to lead to a self-loop. However, for larger $N$, we systematically find more self-loops than gaps, and the difference grows with $N$ and $J_0$. Interestingly, in the limit of small coupling, self-loops and gaps seem to remain equally likely for spins. In contrast, for weakly-coupled hysterons, the excess of self-loops grows with $N$.

\section{Weak asymmetry} \label{sec:weak_nonreciprocity}

\begin{figure}[t!]
\hspace*{-0.15cm}
\begin{tikzpicture}

\node[rotate=0] at (0.0,4.2) {\includegraphics[height=4.3cm]{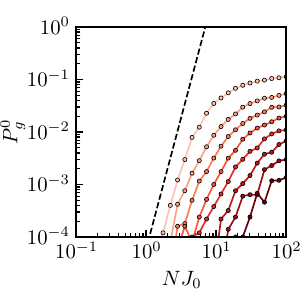}};

\draw[->] (1.5,5.4) -- (1.77,3.37);
\node[rotate=0] at (1.75,3.2) {\small $N$};

\node[rotate=0] at (4.5,4.2) {\includegraphics[height=4.3cm]{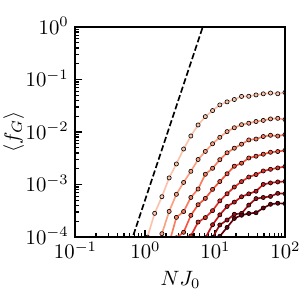}};

\draw[->] (6.0,5.2) -- (6.27,3.37);
\node[rotate=0] at (6.25,3.2) {\small $N$};

\node[rotate=0] at (0.0,0.0) {\includegraphics[height=4.3cm]{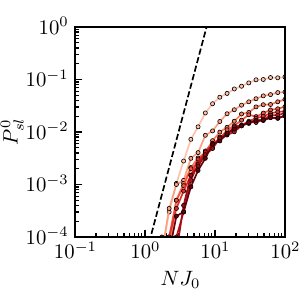}};

\draw[->] (1.45,1.2) -- (1.72,0.08);
\node[rotate=0] at (1.7,-0.15) {\small $N$};

\node[rotate=0] at (4.5,0.0) {\includegraphics[height=4.3cm]{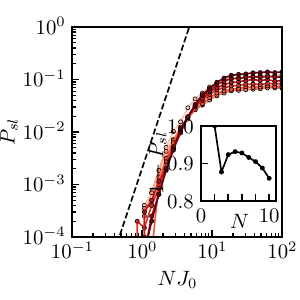}};

\node[rotate=0] at (-0.7,5.62) {\small (a)};
\node[rotate=0] at (3.79,5.62) {\small (b)};

\node[rotate=0] at (-0.7,1.42) {\small (c)};
\node[rotate=0] at (3.79,1.42) {\small (d)};

\end{tikzpicture}
\vspace*{-0.45cm}
\caption{\small{\textbf{Self-loops statistics for weak asymmetric interactions between hysterons.} Statistical measures for self-loops scale when plotted as function of $N J_0$ ($10^5$ samples; color from light to dark as $N$ increases from $2$ to $10$).
(a) Probability $P_g^0$ of finding a gap at $H = 0$ (dashed line indicates slope $5$).
(b) Fraction of gaps $f_G$, defined as the mean of the ratio of
the size of intervals where no stable states exist divided by $H^{+}(-\dots-), H^{-}(+\dots+)$ (dashed line indicates slope $4$).
(c) Probability $P_{sl}^0$ of self-loops at $H = 0$ (dashed line indicates slope $5$).
(d) Probability $P_{sl}$ of finding at least one self-loop for any value of $H$; the black dashed line represents the slope $4$.
Inset: The probability to be self-loop free,
$1 - P_{sl}$, decays monotonously with $N$ for large enough $N$, and for large couplings ($N J_0 = 10^2$).}}
\label{fig:hysteron_weak_NR}
\end{figure}

In this section, we reproduce the simulations of Fig. 2 of the main text, focusing on the case of hysterons with distributed spans ($\sigma_i$ flatly sampled from $\left[ 0, 0.5 \right]$), and restricted to weakly asymmetric interactions, i.e. $c_{ij}c_{ji} > 0$ for all pairs $(i,j)$ (Figs. \ref{fig:hysteron_weak_NR}).

There are two main differences with purely random interactions. First, the probability $P_{sl}^0$ of at least one self-loop occuring at $H = 0$ saturates toward a constant value for $N J_0 \gg 1$ (Fig. \ref{fig:hysteron_weak_NR}-a), which {\em decreases} with $N$: the larger the system, the smaller the probability for self-loops to occur at $H=0$. Moreover, the probability $P_{sl}$ of finding at least one self-loops for any $H$ saturates toward a constant value for $N J_0 \gg 1$ (Fig. \ref{fig:hysteron_weak_NR}-b), which does not asymptote to $1$ as fast as for purely random interactions, but which nevertheless increases for large $N$. Therefore, weakly asymmetric interactions lead to fewer self-loops than random asymmetric interactions, but we still expect that self-loops dominate the response of large, strongly coupled systems.

\section{Self-loops of size $L=4$}
\label{sec:size_4}

In this section, we focus on the emergence of $L = 4$ self-loops.

\subsection{Gap formation mechanism for $N=2$ coupled spins}

We consider two coupled spins indexed $1$ and $2$, such that $h_{1}^+ = h_{1}^- = h_{1}^c$ and $h_{2}^+ = h_{2}^- = h_{2}^c$ (i.e. $\sigma_1 = \sigma_2 = 0$), where we take $h_1^c < h_2^c$. Importantly, the only possible self-loops have size $L=4$, such that the system visits all four possible states in a cycle, and thus the self-loop must occur within a gap. We can compute the upper and lower switching fields of individual spins for each state:

\begin{equation}
\begin{aligned}
 H_1^+(--) &= h_1^c + c_{12}, \\
 H_2^+(--) &= h_2^c + c_{21},
 \end{aligned}
\end{equation}

\begin{equation}
\begin{aligned}
 H_1^-(+-) &= h_1^c + c_{12}, \\
 H_2^+(+-) &= h_2^c - c_{21},
 \end{aligned}
\end{equation}

\begin{equation}
\begin{aligned}
 H_1^+(-+) &= h_1^c - c_{12}, \\
 H_2^-(-+) &= h_2^c + c_{21},
 \end{aligned}
\end{equation}

\begin{equation}
\begin{aligned}
 H_1^-(++) &= h_1^c - c_{12}, \\
 H_2^-(++) &= h_2^c - c_{21},
 \end{aligned}
\end{equation}

and, in the limit of small couplings ($|c_{12}|$,$|c_{21}| \ll \Delta h^c$, with $\Delta h^c = h_2^c - h_1^c > 0$), the upper and lower switching fields of each state are:

\begin{equation}
 H^{+} (--) = H_1^+(--) = h_1^c + c_{12},
\end{equation}

\begin{equation}
\begin{aligned}
 H^{-} (+-) &= H_1^-(+-) = h_1^c + c_{12}, \\
 H^{+} (+-) &= H_2^+(+-) = h_2^c - c_{21},
\end{aligned}
\end{equation}

\begin{equation}
\begin{aligned}
 H^{-} (-+) &= H_2^-(-+) = h_2^c + c_{21}, \\
 H^{+} (-+) &= H_1^+(-+) = h_1^c - c_{12},
\end{aligned}
\end{equation}

\begin{equation}
 H^{-} (++) = H_2^-(++) = h_2^c - c_{21}.
\end{equation}

\begin{figure}[t!]
\centering
\begin{tikzpicture}


\draw[->, thick] (0.0,0.0) -- (6.0,0.0);
\node[rotate=0] at (6.35,0.0) {\small $H$};

\draw[thick] (2.0,-0.1) -- (2.0,0.1);
\node[rotate=0] at (2.0,0.35) {\small $h_1^c$};
\draw[thick] (4.0,-0.1) -- (4.0,0.1);
\node[rotate=0] at (4.0,0.35) {\small $h_2^c$};

\draw[dashed] (2.0,-0.2) -- (2.0,-3.4);
\draw[dashed] (4.0,-0.2) -- (4.0,-3.4);

\draw[color=lightgray, thin] (0.5,-1.2) -- (5.5,-1.2);
\draw[color=lightgray, thin] (0.5,-1.7) -- (5.5,-1.7);
\draw[color=lightgray, thin] (0.5,-2.2) -- (5.5,-2.2);
\draw[color=lightgray, thin] (0.5,-2.7) -- (5.5,-2.7);

\draw[very thick] (0.5,-2.7) -- (2.0,-2.7);
\draw[very thick] (2.0,-2.8) -- (2.0,-2.6);
\draw[very thick] (1.9,-2.6) -- (2.0,-2.6);
\draw[very thick] (1.9,-2.8) -- (2.0,-2.8);

\draw[very thick] (4.0,-1.2) -- (5.5,-1.2);
\draw[very thick] (4.0,-1.3) -- (4.0,-1.1);
\draw[very thick] (4.1,-1.1) -- (4.0,-1.1);
\draw[very thick] (4.1,-1.3) -- (4.0,-1.3);

\draw[very thick] (2.0,-2.2) -- (4.0,-2.2);

\draw[very thick] (4.0,-2.3) -- (4.0,-2.1);
\draw[very thick] (3.9,-2.1) -- (4.0,-2.1);
\draw[very thick] (3.9,-2.3) -- (4.0,-2.3);

\draw[very thick] (2.0,-2.3) -- (2.0,-2.1);
\draw[very thick] (2.1,-2.1) -- (2.0,-2.1);
\draw[very thick] (2.1,-2.3) -- (2.0,-2.3);

\draw[very thick, dotted] (2.0,-1.7) -- (4.0,-1.7);

\draw[very thick] (4.0,-1.8) -- (4.0,-1.6);
\draw[very thick] (4.1,-1.6) -- (4.0,-1.6);
\draw[very thick] (4.1,-1.8) -- (4.0,-1.8);

\draw[very thick] (2.0,-1.8) -- (2.0,-1.6);
\draw[very thick] (1.9,-1.6) -- (2.0,-1.6);
\draw[very thick] (1.9,-1.8) -- (2.0,-1.8);

\node[rotate=0] at (-0.4,-1.2) {\small $\boldsymbol{+1+1}$};
\node[rotate=0] at (-0.4,-1.7) {\small $\boldsymbol{-1+1}$};
\node[rotate=0] at (-0.4,-2.2) {\small $\boldsymbol{+1-1}$};
\node[rotate=0] at (-0.4,-2.7) {\small $\boldsymbol{-1-1}$};

\node[rotate=0] at (0.0,0.6) {\small (a)};

\draw[->, very thick, color=ColorSP] (2.0,-2.7) to[out=0, in=-90] (2.3,-2.2);
\draw[->, very thick, color=ColorSP] (4.0,-2.2) to[out=0, in=-90] (4.3,-1.2);

\draw[->, very thick, color=ColorSM] (4.0,-1.2) to[out=180, in=90] (3.7,-2.2);
\draw[->, very thick, color=ColorSM] (2.0,-2.2) to[out=180, in=90] (1.7,-2.7);


\fill[color=lightgray!30] (10.8,-3.4) rectangle (14.2,0.0);

\draw[->, thick] (8.8,0.0) -- (16.2,0.0);
\node[rotate=0] at (16.55,0.0) {\small $H$};

\draw[color=lightgray, thin] (9.3,-1.2) -- (15.7,-1.2);
\draw[color=lightgray, thin] (9.3,-1.7) -- (15.7,-1.7);
\draw[color=lightgray, thin] (9.3,-2.2) -- (15.7,-2.2);
\draw[color=lightgray, thin] (9.3,-2.7) -- (15.7,-2.7);

\draw[thick] (12.8,-0.1) -- (12.8,0.1);
\node[rotate=0] at (12.8,0.35) {\small $h_2^c$};
\draw[thick] (12.2,-0.1) -- (12.2,0.1);
\node[rotate=0] at (12.2,0.35) {\small $h_1^c$};

\draw[dashed] (12.2,-0.2) -- (12.2,-3.4);
\draw[dashed] (12.8,-0.2) -- (12.8,-3.4);

\draw[thick] (14.8,-0.1) -- (14.8,0.1);
\draw[thick] (14.2,-0.1) -- (14.2,0.1);
\node[rotate=0] at (14.5,0.35) {\small $+c$};

\draw[dashed] (14.2,-0.2) -- (14.2,-3.4);
\draw[dashed] (14.8,-0.2) -- (14.8,-3.4);

\draw[thick] (10.8,-0.1) -- (10.8,0.1);
\draw[thick] (10.2,-0.1) -- (10.2,0.1);
\node[rotate=0] at (10.5,0.35) {\small $-c$};

\draw[dashed] (10.2,-0.2) -- (10.2,-3.4);
\draw[dashed] (10.8,-0.2) -- (10.8,-3.4);

\draw[very thick] (9.3,-2.7) -- (10.8,-2.7);
\draw[very thick] (10.8,-2.8) -- (10.8,-2.6);
\draw[very thick] (10.7,-2.6) -- (10.8,-2.6);
\draw[very thick] (10.7,-2.8) -- (10.8,-2.8);

\draw[very thick] (14.8,-1.2) -- (15.7,-1.2);
\draw[very thick] (14.8,-1.3) -- (14.8,-1.1);
\draw[very thick] (14.9,-1.1) -- (14.8,-1.1);
\draw[very thick] (14.9,-1.3) -- (14.8,-1.3);

\draw[very thick, dotted] (10.2,-1.7) -- (10.8,-1.7);

\draw[very thick] (10.8,-1.8) -- (10.8,-1.6);
\draw[very thick] (10.9,-1.6) -- (10.8,-1.6);
\draw[very thick] (10.9,-1.8) -- (10.8,-1.8);

\draw[very thick] (10.2,-1.8) -- (10.2,-1.6);
\draw[very thick] (10.1,-1.6) -- (10.2,-1.6);
\draw[very thick] (10.1,-1.8) -- (10.2,-1.8);

\draw[very thick] (14.2,-2.2) -- (14.8,-2.2);

\draw[very thick] (14.8,-2.3) -- (14.8,-2.1);
\draw[very thick] (14.7,-2.1) -- (14.8,-2.1);
\draw[very thick] (14.7,-2.3) -- (14.8,-2.3);

\draw[very thick] (14.2,-2.3) -- (14.2,-2.1);
\draw[very thick] (14.3,-2.1) -- (14.2,-2.1);
\draw[very thick] (14.3,-2.3) -- (14.2,-2.3);

\draw[->, very thick, color=ColorSP] (11.8,-2.7) to[out=60, in=-60] (11.8,-1.7);
\draw[->, very thick, color=ColorSP] (11.8,-1.7) to[out=60, in=-60] (11.8,-1.2);

\draw[->, very thick, color=ColorSM] (11.8,-1.2) to[out=240, in=120] (11.8,-2.2);
\draw[->, very thick, color=ColorSM] (11.8,-2.2) to[out=240, in=120] (11.8,-2.7);

\draw[->, very thick, color=ColorSP] (14.8,-2.2) to[out=0, in=-90] (15.1,-1.2);
\draw[->, very thick, color=ColorSM] (14.8,-1.2) to[out=180, in=90] (14.5,-2.2);

\draw[->, very thick, color=ColorSP] (10.8,-2.7) to[out=0, in=-90] (11.1,-1.7);

\draw[->, very thick, color=ColorSM] (14.2,-2.2) to[out=180, in=90] (13.9,-2.7);

\draw[<->, thick] (10.8,-3.7) -- (14.2,-3.7);
\node[rotate=0] at (12.5,-4.0) {\small $2c - \Delta h^c$};

\node[rotate=0] at (8.4,-1.2) {\small $\boldsymbol{+1+1}$};
\node[rotate=0] at (8.4,-1.7) {\small $\boldsymbol{-1+1}$};
\node[rotate=0] at (8.4,-2.2) {\small $\boldsymbol{+1-1}$};
\node[rotate=0] at (8.4,-2.7) {\small $\boldsymbol{-1-1}$};

\node[rotate=0] at (8.8,0.6) {\small (b)};


\fill[color=lightgray!30] (10.8,-8.4) rectangle (14.2,-5.0);

\draw[->, thick] (8.8,-5.0) -- (16.2,-5.0);
\node[rotate=0] at (16.55,-5.0) {\small $H$};

\draw[color=lightgray, thin] (9.3,-6.2) -- (15.7,-6.2);
\draw[color=lightgray, thin] (9.3,-6.7) -- (15.7,-6.7);
\draw[color=lightgray, thin] (9.3,-7.2) -- (15.7,-7.2);
\draw[color=lightgray, thin] (9.3,-7.7) -- (15.7,-7.7);

\draw[thick] (12.8,-5.1) -- (12.8,-4.9);
\node[rotate=0] at (12.8,-4.65) {\small $h_2^c$};
\draw[thick] (12.2,-5.1) -- (12.2,-4.9);
\node[rotate=0] at (12.2,-4.65) {\small $h_1^c$};

\draw[dashed] (12.2,-5.2) -- (12.2,-8.4);
\draw[dashed] (12.8,-5.2) -- (12.8,-8.4);

\draw[thick] (14.8,-5.1) -- (14.8,-4.9);
\draw[thick] (14.2,-5.1) -- (14.2,-4.9);
\node[rotate=0] at (14.5,-4.65) {\small $-c$};

\draw[dashed] (14.2,-5.2) -- (14.2,-8.4);
\draw[dashed] (14.8,-5.2) -- (14.8,-8.4);

\draw[thick] (10.8,-5.1) -- (10.8,-4.9);
\draw[thick] (10.2,-5.1) -- (10.2,-4.9);
\node[rotate=0] at (10.5,-4.65) {\small $+c$};

\draw[dashed] (10.2,-5.2) -- (10.2,-8.4);
\draw[dashed] (10.8,-5.2) -- (10.8,-8.4);

\draw[very thick] (9.3,-7.7) -- (10.2,-7.7);
\draw[very thick] (10.2,-7.8) -- (10.2,-7.6);
\draw[very thick] (10.1,-7.6) -- (10.2,-7.6);
\draw[very thick] (10.1,-7.8) -- (10.2,-7.8);

\draw[very thick] (14.2,-6.2) -- (15.7,-6.2);
\draw[very thick] (14.2,-6.3) -- (14.2,-6.1);
\draw[very thick] (14.3,-6.1) -- (14.2,-6.1);
\draw[very thick] (14.3,-6.3) -- (14.2,-6.3);

\draw[very thick, dotted] (14.2,-6.7) -- (14.8,-6.7);

\draw[very thick] (14.8,-6.8) -- (14.8,-6.6);
\draw[very thick] (14.9,-6.6) -- (14.8,-6.6);
\draw[very thick] (14.9,-6.8) -- (14.8,-6.8);

\draw[very thick] (14.2,-6.8) -- (14.2,-6.6);
\draw[very thick] (14.1,-6.6) -- (14.2,-6.6);
\draw[very thick] (14.1,-6.8) -- (14.2,-6.8);

\draw[very thick] (10.2,-7.2) -- (10.8,-7.2);

\draw[very thick] (10.8,-7.3) -- (10.8,-7.1);
\draw[very thick] (10.7,-7.1) -- (10.8,-7.1);
\draw[very thick] (10.7,-7.3) -- (10.8,-7.3);

\draw[very thick] (10.2,-7.3) -- (10.2,-7.1);
\draw[very thick] (10.3,-7.1) -- (10.2,-7.1);
\draw[very thick] (10.3,-7.3) -- (10.2,-7.3);

\draw[->, very thick, color=ColorSP] (11.8,-7.7) to[out=60, in=-60] (11.8,-7.2);
\draw[->, very thick, color=ColorSP] (11.8,-7.2) to[out=60, in=-60] (11.8,-6.2);

\draw[->, very thick, color=ColorSM] (11.8,-6.2) to[out=240, in=120] (11.8,-6.7);
\draw[->, very thick, color=ColorSM] (11.8,-6.7) to[out=240, in=120] (11.8,-7.7);

\draw[->, very thick, color=ColorSP] (10.2,-7.7) to[out=0, in=-90] (10.5,-7.2);
\draw[->, very thick, color=ColorSM] (10.2,-7.2) to[out=180, in=90] (9.9,-7.7);

\draw[->, very thick, color=ColorSP] (10.8,-7.2) to[out=0, in=-90] (11.1,-6.2);

\draw[->, very thick, color=ColorSM] (14.2,-6.2) to[out=180, in=90] (13.9,-6.7);

\draw[<->, thick] (10.8,-8.7) -- (14.2,-8.7);
\node[rotate=0] at (12.5,-9.0) {\small $- 2c - \Delta h^c$};

\node[rotate=0] at (8.4,-6.2) {\small $\boldsymbol{+1+1}$};
\node[rotate=0] at (8.4,-6.7) {\small $\boldsymbol{-1+1}$};
\node[rotate=0] at (8.4,-7.2) {\small $\boldsymbol{+1-1}$};
\node[rotate=0] at (8.4,-7.7) {\small $\boldsymbol{-1-1}$};

\node[rotate=0] at (8.8,-4.4) {\small (c)};

\node[rotate=0] at (3.0,-6.2) {\includegraphics[height=4.5cm]{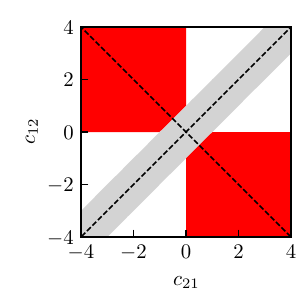}};

\node[rotate=0] at (1.1,-4.3) {\small (d)};

\node[rotate=45] at (4.36,-4.9) {\small $c_{12} = c_{21}$};
\node[rotate=-45] at (4.5,-6.65) {\small $c_{12} = -c_{21}$};


\draw[<->] (2.6,-6.51) -- (2.6,-7.24);
\node[] at (3.1,-6.95) {\footnotesize $2|\Delta h^c|$};

\end{tikzpicture}
\vspace*{-0.7cm}
\caption{\small{\textbf{Stability ranges for $N=2$ binary spins.} (a) Zero coupling ($c_{12} = c_{21} = 0$).  (b-c) Completely asymmetric couplings $c_{12} = - c_{21} = c > 0$, i.e. $\Delta c > 0$ (b); $c_{12} = - c_{21} = c < 0$, i.e. $\Delta c < 0$ (c). The solid black, dotted black, and solid gray lines represent stable configuration, unstable configuration with $2$ unstable spins, and with $1$ unstable spin, respectively. Red (resp. blue) arrows represent up (resp. down) transitions. The gray area in (b-c) indicates the range of $H$ with a gap, and the colored transitions represent the self-loop occuring within this range. (d) Portions of the $(c_{12},c_{21})$-plane leading to $L=4$ self-loops (red areas).}}
\label{fig:holes_2spins}
\end{figure}

Note that $H^+(--) = H^-(+-)$ and $H^+(+-) = H^-(++)$: the upper switching field of the state "below" coincides with the lower switching field of the state "above" (Fig. \ref{fig:holes_2spins}-a). Therefore, in the limit of small couplings, the convention $h_1^c < h_2^c$ enforces the structure of the Preisach graph represented in Fig. \ref{fig:holes_2spins}-a. Moreover, as the zero-coupling ordering of the states $(--) \leftrightarrow (+-) \leftrightarrow (++)$ is preserved, no gap can open, even with finite interactions.

One hole may open in two mutually-excluding cases: when the ordering of the switching fields in state $(--)$ (resp. $(++)$) is reversed. We focus on the first case without loss of generality. The condition $H_2^+(--) < H_1^+(--)$ translates into:

\begin{equation} \label{eq:reverse_ordering}
 \Delta c > \Delta h^c,
\end{equation}

\noindent where $\Delta c = c_{12} - c_{21}$. Let us assume the condition given by Eq. (\ref{eq:reverse_ordering}) is satisfied, and analyze the other states' switching fields. First, the condition to keep the same switching fields ordering in state $(++)$ can be written as:

\begin{equation}
H_2^-(++) > H_1^-(++) \Leftrightarrow \Delta c > - \Delta h^c,
\end{equation}

\noindent which is necessarily true when Eq. (\ref{eq:reverse_ordering}) is satisfied. Then, the condition to open up a range of $H$ in between the saturating states' stability ranges can be written as:

\begin{equation}
 H_2^-(++) > H_2^+(--) \Leftrightarrow c_{21} < 0,
\end{equation}

\noindent imposing the sign of $c_{21}$. Finally, we consider the states $(+-)$ and $(-+)$. Let us start with the first one. Given $H^{+}(+-) = H^-(++)$, the opening of a gap inside $\left[ H^+(--), H^-(++) \right]$ requires $H^-(+-) > H^+(--)$. This condition translates into:

\begin{equation}
 H_2^+(--) < H_1^-(+-) \Leftrightarrow \Delta c > \Delta h^c,
\end{equation}

\noindent which is equivalent to Eq. (\ref{eq:reverse_ordering}) characterizing the reversal of the critical hysteron in state $(--)$. The last condition to open a gap is that there is a finite range of $H$ in between $H^{+}(-+)$ and $H^{-}(+-)$:

\begin{equation}
  H^{-}(+-) > H^{+}(-+) \Leftrightarrow c_{12} > 0,
\end{equation}

\noindent imposing the sign of $c_{12}$. Altogether, the necessary and sufficient conditions for the emergence of a self-loop of size $L=4$ in a system of $N=2$ coupled spins can be written as:

\begin{equation}
\begin{aligned}
 \Delta c &> \Delta h^c, \\
 c_{12}c_{21} &< 0.
\end{aligned}
\end{equation}

These two conditions lead to the self-loop of size $L=4$ represented in Fig. \ref{fig:holes_2spins}-b. The second scenario mentioned above leads to the following conditions:

\begin{equation}
\begin{aligned}
 \Delta c &< - \Delta h^c, \\
 c_{12}c_{21} &< 0,
\end{aligned}
\end{equation}

\noindent which lead to the self-loop of size $L=4$ with the opposite "chirality", as shown in Fig. \ref{fig:holes_2spins}-c. In conclusion, couplings of opposite signs (strong asymmetry) and large enough asymmetry $|\Delta c| = |c_{12} - c_{21}|$ lead to a gap in between the saturating states (Fig. \ref{fig:holes_2spins}-d). Inside this gap, no stable state exists, which guarantees the existence of a self-loop of size $L=4$. This is best illustrated by representing the states' stability ranges in the case of completely asymmetric interactions, i.e. $c_{12} = - c_{21} = c$, where $\Delta c = 2 c$ (Figs. \ref{fig:holes_2spins}-b and c). We find that a hole of size $|\Delta c| - |\Delta h^c|$ opens in  between the saturating states as soon as $|\Delta c| > |\Delta h^c|$.

\subsection{$N=2$ coupled hysterons}

\begin{figure}[t!]
\hspace*{-0.15cm}
\begin{tikzpicture}

\node[rotate=0] at (0.0,0.0) {\includegraphics[height=4.5cm]{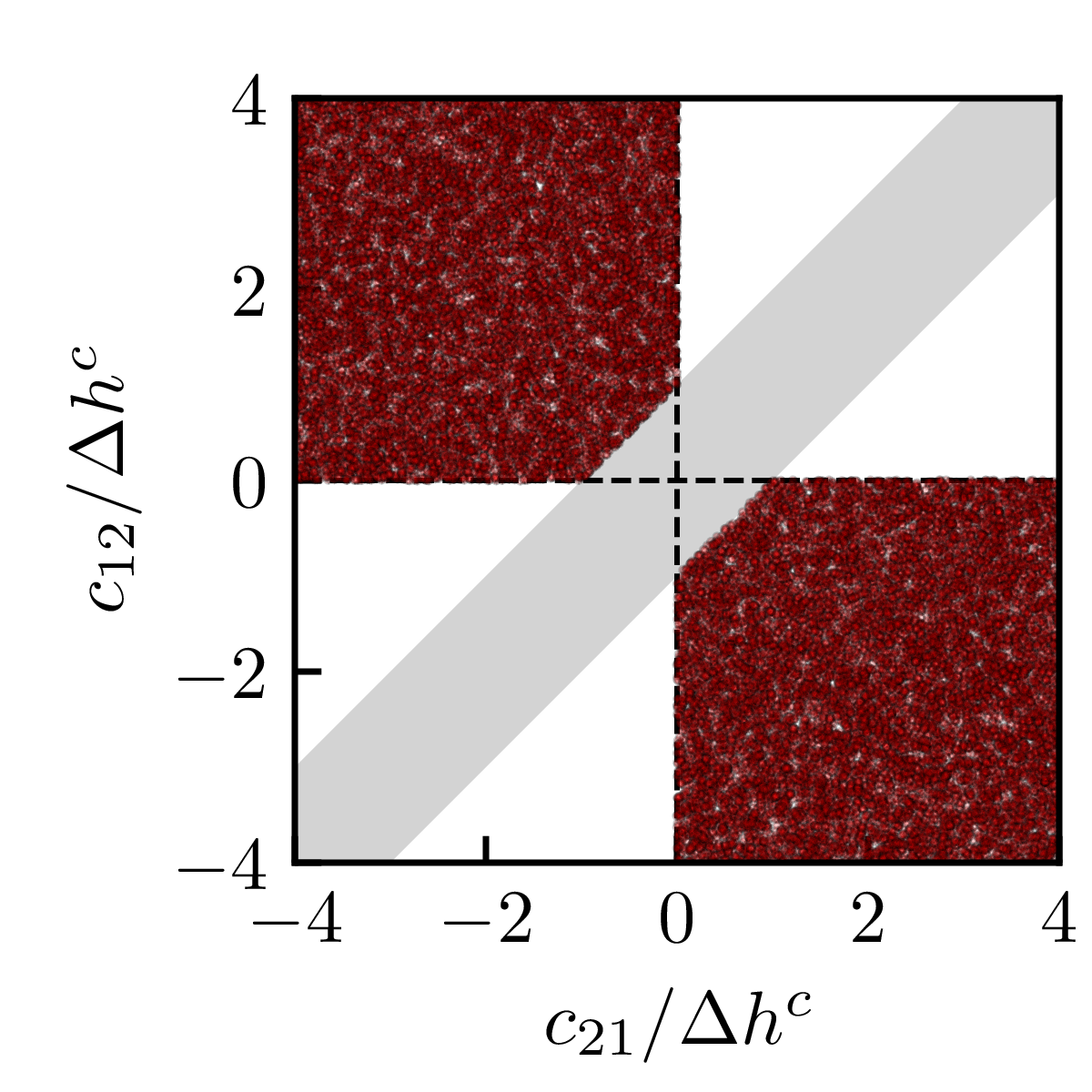}};

\draw[<->] (-0.6,-1.22) -- (-0.6,-0.53);
\node[rotate=0] at (-0.44,-0.875) {\footnotesize $2$};

\node[rotate=0] at (1.76,1.5) {\small (a)};

\node[rotate=0] at (4.5,0.0) {\includegraphics[height=4.5cm]{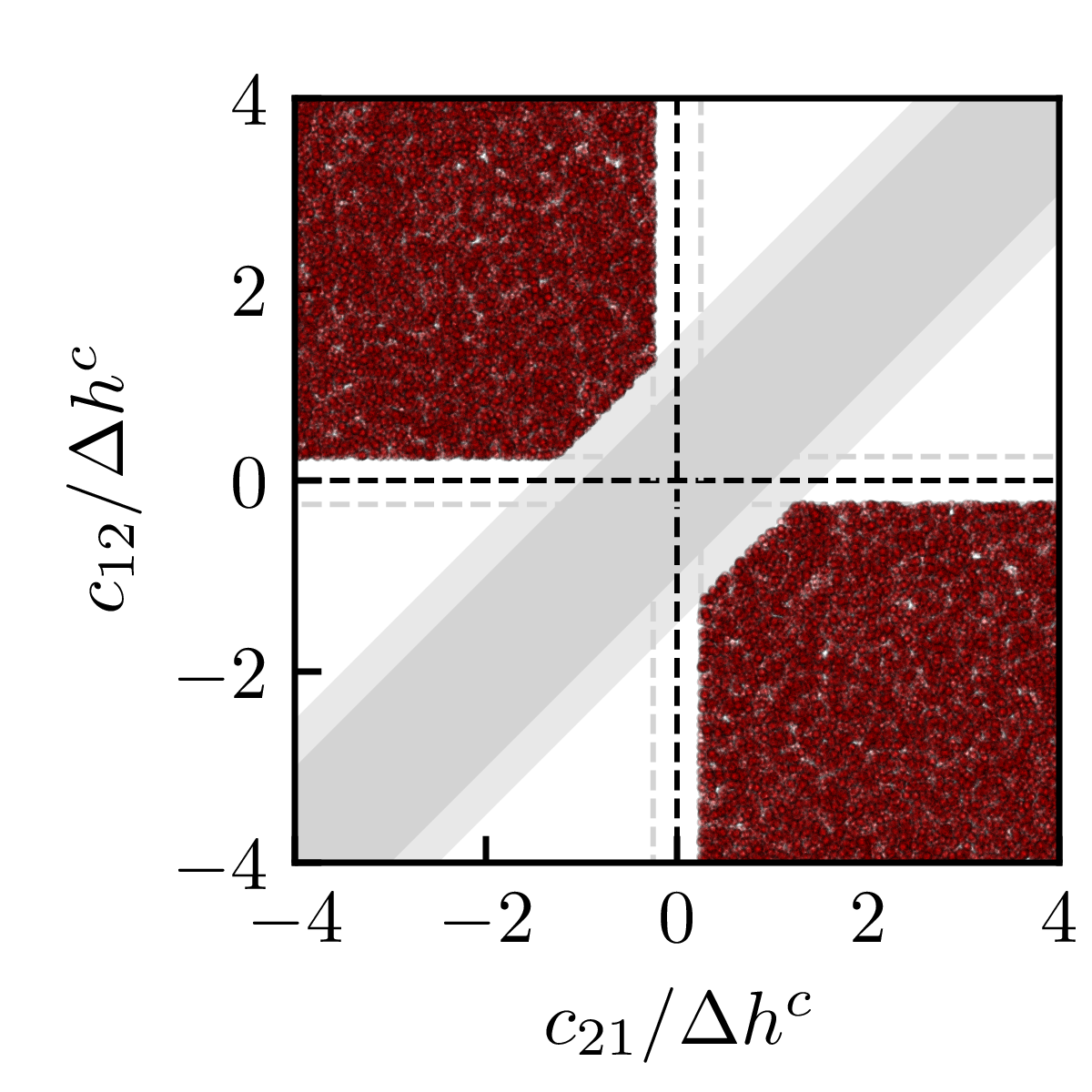}};

\draw[<->] (4.1,-1.23) -- (4.1,-0.12);
\node[rotate=0] at (4.6,-0.675) {\footnotesize $2 + 2\sigma$};

\node[rotate=0] at (6.26,1.5) {\small (b)};

\node[rotate=0] at (9.0,0.0) {\includegraphics[height=4.5cm]{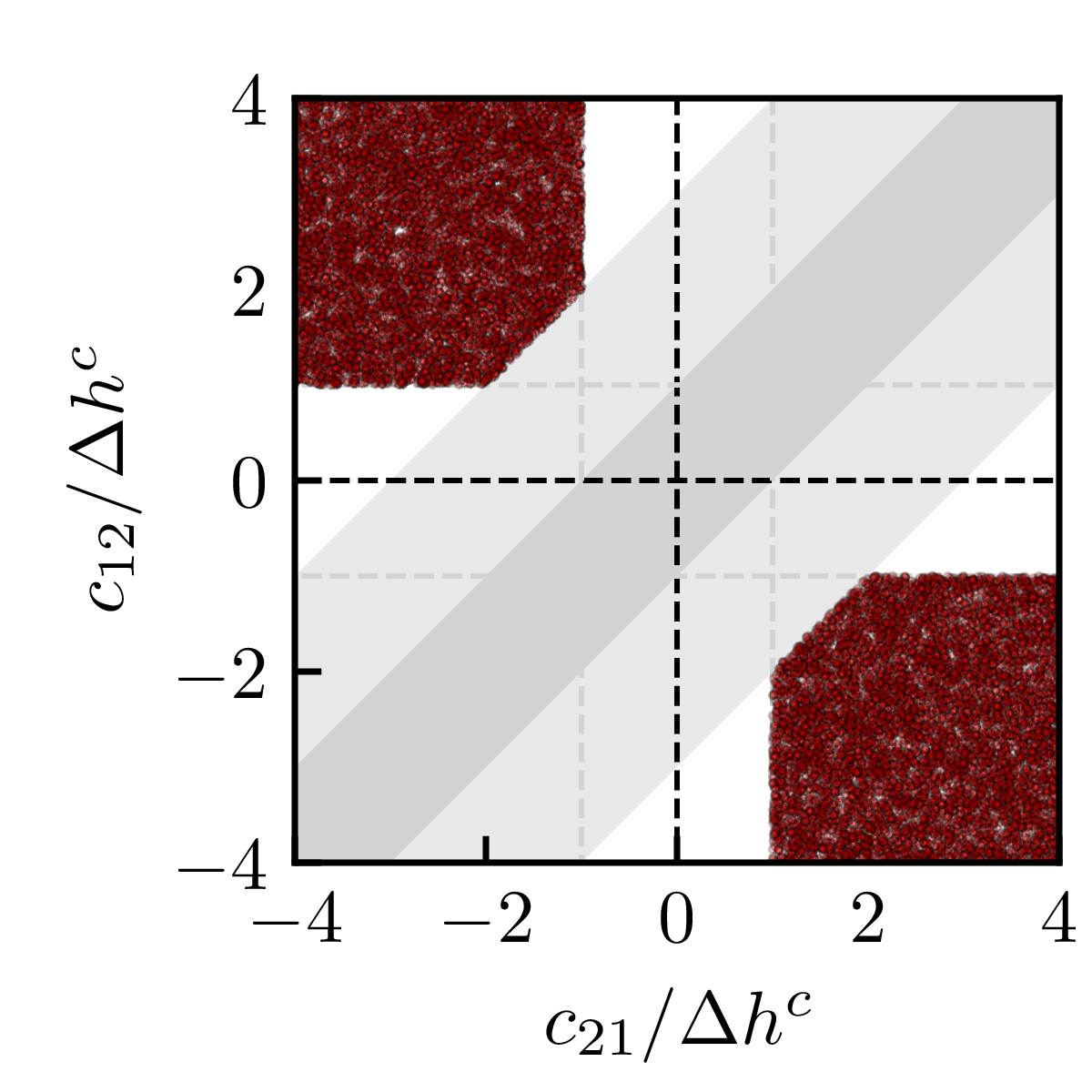}};

\draw[<->] (8.33,-0.1) -- (8.33,0.63);
\node[rotate=0] at (8.5,0.14) {\footnotesize $\sigma$};

\draw[<->] (9.17,-0.9) -- (9.9,-0.9);
\node[rotate=0] at (9.72,-0.74) {\footnotesize $\sigma$};

\node[rotate=0] at (10.76,1.5) {\small (c)};

\node[rotate=0] at (0.53,2.15) {\small \textbf{Spins} ($\sigma = 0$)};

\draw[->] (3.7,2.15) -- (10.5,2.15);
\node[rotate=0] at (10.8,2.15) {\small $\sigma$};

\node[rotate=0] at (7.1,2.45) {\small \textbf{Hysterons} ($\sigma > 0$)};

\end{tikzpicture}
\vspace*{-0.45cm}
\caption{\small{\textbf{Sampling self-loops for $N = 2$ interacting elements.} Self-loops of size $4$ for arbitrary interactions in the ($c_{12}$,$c_{21}$)-plane (red transparent markers); (a) spins, i.e. $\sigma = 0$; (b-c) hysterons, for $\sigma = 0.5$ (b) and $\sigma = 2$ (c).}}
\label{fig:spins_vs_hysterons}
\end{figure}

The conditions above can be extended to the case of hysterons with finite microscopic hysteresis:

\begin{equation} \label{eq:condition_sl_equal_spans}
\begin{aligned}
 |\Delta c| &> |\Delta h^c + \sigma|, \\
 c_{12}c_{21} &< 0, \\
 |c_{12}| &> \sigma/2, \\
 |c_{21}| &> \sigma/2,
\end{aligned}
\end{equation}

\noindent where we have considered that the two hysterons have the same microscopic hysteresis, i.e. $\sigma_1 = \sigma_2 = \sigma$. Expectedly, for larger microscopic hysteresis $\sigma > 0$, some self-loops that would have been possible for binary spins are forbidden. More precisely, self-loops emerge for larger asymmetry, i.e. $|\Delta c| > |\Delta h^c + \sigma|$, and large enough couplings, i.e. $|c_{12}| > \sigma/2$ and $|c_{21}| > \sigma/2$. Note that, similarly as for spins, self-loops emerge only when the coupling coefficients have opposite signs, i.e. $c_{12}c_{21} < 0$ (strong asymmetry).

Sampling $10^5$ different instances for different $\sigma$ and fixed $\Delta h^c = 1$, we represent the parameters leading to self-loops in the ($c_{12}$,$c_{21}$)-plane (Figs. \ref{fig:spins_vs_hysterons}). In the limit of binary spins ($\sigma = 0$), we recover the necessary and sufficient conditions for self-loops to emerge, i.e. $c_{12}c_{21} < 0$ and $|\Delta c| > |\Delta h^c|$ (Fig. \ref{fig:spins_vs_hysterons}-a). For $\sigma > 0$, numerical results are consistent with the conditions given by Eqs. (\ref{eq:condition_sl_equal_spans}) (Figs. \ref{fig:spins_vs_hysterons}-b and c). We find similar results when the two hysterons' spans are different (not shown here). Altogether, at the level of $N = 2$ interacting elements, the microscopic hysteresis tends to prevent $L=4$ self-loops that would have been possible otherwise, but in all cases strong asymmetry is a necessary condition for self-loops to emerge.

\subsection{Large systems}

The sufficient and necessary conditions for $L=4$ self-loops to emerge in systems of $N=2$ spins can be extended into necessary conditions for arbitrary $N$. Starting with spins and noting that each $L\!=\!4$ self-loop only involves two spins ($k$ and $l$), we find that such $L\!=\! 4$ loop can only occur when
$c_{kl}c_{lk} < 0$ and $|\Delta c_{kl}/\Delta \tilde{h}_{kl}^c|>1$, where $\Delta c_{kl} = c_{kl} - c_{lk}$ and where

\begin{equation} \label{eq:self_loop_4_largeN}
\tilde{h}_i^c = h_i^c - \sum_{j \neq k,l} c_{ij} s_j.
\end{equation}

Hence, for arbitrary $N$, the effect of the $N-2$ spins that do not flip is to effectively rescale the difference between the switching fields of the two flipping spins. Therefore, the same as for $N=2$, self-loops of size $L=4$ only emerge for strong asymmetry ($c_{kl}c_{lk} < 0$) and large enough asymmetry $|\Delta c_{kl}|$. However, in contrast with $N=2$, when a pair of spins satisfies the conditions above, it does not imply that the system necessarily reaches this cycle, as it might be disconnected from the stable states.

We confirm the results above by sampling $10^5$ systems with $N = 1024$ elements, and looking for self-loops at any $H$. We plot a red marker in the rescaled ($c_{12}$,$c_{21}$)-plane when we find a $L=4$ self-loop, where $1$ and $2$ now represent the two flipping elements. We focus on the large coupling limit ($NJ_0 = 10$), for collections of spins (Fig. \ref{fig:self_loop_size_4}-a) and hysterons (Fig. \ref{fig:self_loop_size_4}-b).

We first focus on the case of spins. When self-loops occur, we find they are necessary included in the region of the rescaled ($c_{12}$,$c_{21}$)-plane where $L=4$ self-loops occur for $N=2$ spins. However, for some systems there exist pairs of spins which are in the \textit{red} region (for a given configuration of the rest of the system), but which are never involved in a $L=4$ self-loop. The situation is very similar for hysterons with distributed spans: we find self-loops in the same region as for spins, though there are fewer self-loops, especially close to the boundaries.

\begin{figure}[h!]
\hspace*{-0.3cm}
\begin{tikzpicture}

\node[rotate=0] at (0.0,0.0) {\includegraphics[height=4.3cm]{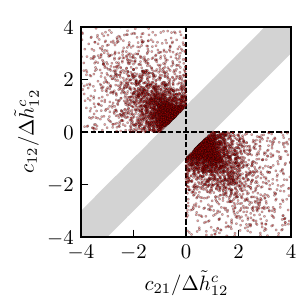}};
\node[rotate=0] at (4.4,0.0) {\includegraphics[height=4.3cm]{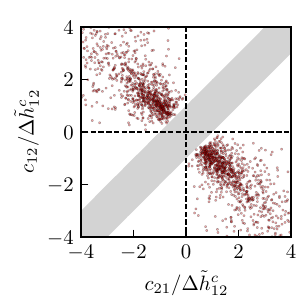}};

\node[rotate=0] at (1.65,1.4) {\small (a)};
\node[rotate=0] at (6.05,1.4) {\small (b)};

\end{tikzpicture}
\vspace*{-0.45cm}
\caption{\small{\textbf{Self-loops of size $4$ in large systems.} $L=4$ Self-loops for arbitrary interactions in the $(c_{12}-c_{21})$ plane (red transparent markers); in the limit of binary spins (a), i.e. $\sigma_i = 0$, and for hysterons (b), i.e. with $\sigma_i$ flatly sampled from $\left[ 0,0.5 \right]$; fixed $N = 1024$, and $NJ_0 = 10$.}}
\label{fig:self_loop_size_4}
\end{figure}

\subsection{Preventing $L=4$ self-loops}

We note that the condition to prevent $L=4$ self-loops consists of a simple and a complex part, and we refer to the simple part - $c_{ij} c_{ji} >0$ for all pairs $(i,j)$ - as the condition of weak asymmetry. This simple condition is sufficient to guarantee the absence of $L\!=\!4$ self-loops for any system size, although, as there are other parts in hysteron parameter space where such self-loops are prohibited - it is overly restrictive.

\section{Serially-coupled two-state elements} \label{sec:active_serial_coupling}

In this section, we summarize the main theoretical and numerical insights from a model of serially coupled bilinear force-displacement curves with arbitrary force jumps.

\subsection{Conventions} \label{sec:active_conventions}
\begin{figure}[h!]
\hspace*{-0.6cm}
\begin{tikzpicture}

\node[rotate=0] at (0.0,0.0) {\includegraphics[height=4.3cm]{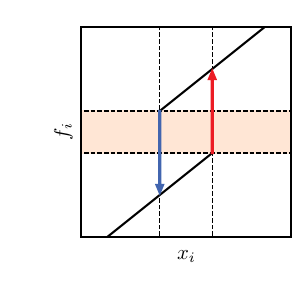}};
\node[rotate=0] at (4.4,0.0) {\includegraphics[height=4.3cm]{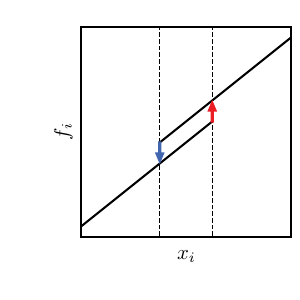}};
\node[rotate=0] at (8.8,0.0) {\includegraphics[height=4.3cm]{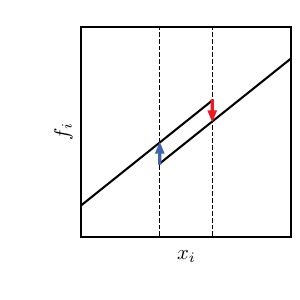}};
\node[rotate=0] at (13.2,0.0) {\includegraphics[height=4.3cm]{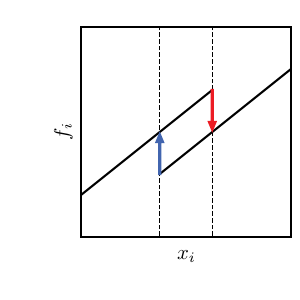}};

\draw[->] (-0.8,2.2) -- (14.8,2.2);

\node[] at (15.1,2.2) {\small $d_i$};

\end{tikzpicture}
\vspace*{-0.8cm}
\caption{\small{\textbf{Bilinear force-displacement curves with different force jumps.} From left to right, $d_i = (-2.0, -0.5, 0.5, 1)$; fixed $(x_i^-, x_i^+) = (-1, +1)$. In the left panel, $d_i < -\sigma_i^0 /2$, leading to a range of force with no stable states (highlighted in green).}}
\label{fig:active_and_passive_elements}
\end{figure}

Each element is associated with a bilinear force-displacement curve:
\begin{equation}
 f_i = x_i - d_i s_i,
\end{equation}
where $d_i$ is the force jump of element $i$,
$s_i = \pm 1$ denotes the phase of element $i$,
and transitions occur at $x_i = x_i^{\pm}$, with $\sigma_i^0 = x_i^{+} - x_i^{-} > 0$ (Fig. \ref{fig:active_and_passive_elements}). We assume that the curves have a constant slope $1$, so that the upper and lower force jumps are equal; more general cases can also be considered. Note that in earlier work, the phases where taken to have values $0$ and $1$ (instead of $\pm 1$), which leads to factors of two differences between our mapping here and those earlier mappings \cite{liu2024controlled}.
Such bilinear models are well-known for allowing simple analytics and direct mappings to the coupled hysteron model \cite{liu2024controlled, shohat2025geometric}.
However, in contrast to previous work, here we consider force jumps of arbitrary signs: $d_i > 0$ corresponds to a regular dissipative bistable mechanical element, and $d_i < 0$ corresponds to an active element that injects energy into the system. This distinction is captured by the energy $\Delta E_i$ dissipated by element $i$ during a quasistatic drive cycle (area enclosed by the hysteresis loop):
\begin{equation}
 \Delta E_i = d_i (x_i^+ - x_i^-) = d_i \sigma_i^0,
\end{equation}
which has the same sign as $d_i$ for $\sigma_i^0 > 0$. Note that when one single such element is controlled in position, i.e. through the $x_i$, no self-loop can occur because at least one of the two states ($s_1 = \pm 1$) is stable. However, when the element is controlled in force, i.e. through the $f_i$, self-loops can occur when $d_1 < -\sigma_1^0 / 2$ as this condition leads to a range of force where no states are stable (Fig. \ref{fig:active_and_passive_elements}-left). As we show below, coupling such an element with a spring or another bilinear element can lead to self-loops when the system is controlled in position.

\subsection{One element in series with a spring}
\begin{figure}[t!]
\centering
\vspace{0.2cm}
\begin{tikzpicture}

\foreach \y in {-0.6,-0.4,...,0.6} {
    \draw[] (-1.2,\y) -- (-1,\y+0.2);
}
\draw[thick] (-1.0,-0.8) -- (-1.0,0.8);

\draw[thick] (-1.0,0.0) -- (-0.15,0.0);

\draw[black, thick, domain=-1.08:0.27, samples=100]
    plot (\x + 1.5, {1.0*(\x) + 0.45});
\draw[black, thick, domain=-0.27:1.10, samples=100]
    plot (\x + 1.5, {1.0*(\x) - 0.25});
\draw[ColorSP, ->, >=stealth] (1.77,0.72) -- (1.77,0.02);
\draw[ColorSM, ->, >=stealth] (1.23,-0.52) -- (1.23,0.18);

\draw[thick] (-0.15,-1.2) rectangle (3.1,1.2);
\draw[->] (0.28,-0.75) -- (2.4,-0.75);
\node[] at (2.7,-0.75) {\footnotesize $x_1$};
\draw[->] (0.28,-0.75) -- (0.28,0.52);
\node[] at (0.28,0.79) {\footnotesize $f_1$};

\node[rotate=45] at (2.45,0.44) {\scriptsize $+1$};
\node[rotate=45] at (0.55,-0.28) {\scriptsize $-1$};

\draw[thick] (3.1,0) -- (3.9,0);

\draw[thick] (3.9,0.0) -- (4.0,0.4);
\draw[thick] (4.0,0.4) -- (4.2,-0.4);
\draw[thick] (4.2,-0.4) -- (4.4,0.4);
\draw[thick] (4.4,0.4) -- (4.6,-0.4);
\draw[thick] (4.6,-0.4) -- (4.8,0.4);
\draw[thick] (4.8,0.4) -- (5.0,-0.4);
\draw[thick] (5.0,-0.4) -- (5.2,0.4);
\draw[thick] (5.2,0.4) -- (5.4,-0.4);
\draw[thick] (5.4,-0.4) -- (5.6,0.4);
\draw[thick] (5.6,0.4) -- (5.8,-0.4);
\draw[thick] (5.8,-0.4) -- (5.9,0.0);

\draw[thick, ->] (5.9,0) -- (6.8,0);
\node[] at (7.1,0.0) {\footnotesize $U$};

\node[] at (4.9,0.7) {\footnotesize $k$};

\end{tikzpicture}
\vspace*{-0.1cm}
\caption{\textbf{Schematic of one bilinear force-displacement curve in series with a spring of stiffness $k$.} In this specific example, $d_1 > 0$.}
\label{fig:schematic_fordis_curves_single}
\end{figure}
We consider the mapping from a physical network composed one bilinear element (indexed $1$) in series with a spring of stiffness $k$ (Fig.~\ref{fig:schematic_fordis_curves_single}) to the abstract hysteron model. Combining the condition of fixed imposed displacement $U = x_1 + x_2$ and mechanical equilibrium allows to express the global switching thresholds $U_1^{\pm}$ as function of the force jump, spring stiffness, and local thresholds:
\begin{subequations} \label{eq:mapping_1element_1spring}
\begin{align}
 U_1^+ &= \frac{1+k}{k} x_1^{+} + \frac{d_1}{k}, \\
 U_1^- &= \frac{1+k}{k} x_1^{+} - \frac{d_1}{k}.
\end{align}
\end{subequations}
Following \cite{liu2024controlled}, we identify the physical critical thresholds $U_1^{\pm}$, as given by Eqs. (\ref{eq:mapping_1element_1spring}), to those of an abstract hysteron with a global driving $H$. Considering $H = U$, we simply find:
\begin{equation} \label{eq:condition_L2_1element}
 \sigma_1 = H_1^+ - H_1^- = \frac{1}{k} \left[ (k+1) \sigma_1^0 + 2 d_1 \right].
\end{equation}
Whereas for a dissipative element ($d_1 > 0$), the abstract hysteron always has a positive span, larger than the one of the corresponding mechanical elements, for an active element ($d_1 < 0$), the hysterons can have a negative span, and be intrinsically unstable in some range of $H$. Therefore, a self-loop of length $L=2$ oscillating between states $s_1 = \pm 1$ can emerge when $\sigma_1 < 0$. Using Eq. (\ref{eq:condition_L2_1element}), this condition can be rewritten:
\begin{equation} \label{eq:condition_L2_single}
 d_1 < - \frac{k+1}{2} \sigma_1^0.
\end{equation}
Note that Eq. (\ref{eq:condition_L2_single}) takes the form $d_1 < - \sigma_1^0$ for $k=1$. Moreover, using the asymptotic limits $k \rightarrow 0$ and $k \rightarrow +\infty$, we recover the cases discussed in section \ref{sec:active_conventions} for force and position control, respectively.
\subsection{Two elements in series}
\begin{figure}[b!]
\centering
\vspace{0.2cm}
\begin{tikzpicture}

\foreach \y in {-0.6,-0.4,...,0.6} {
    \draw[] (-1.2,\y) -- (-1,\y+0.2);
}
\draw[thick] (-1.0,-0.8) -- (-1.0,0.8);

\draw[thick] (-1.0,0.0) -- (-0.15,0.0);

\draw[black, thick, domain=-1.08:0.27, samples=100]
    plot (\x + 1.5, {1.0*(\x) + 0.45});
\draw[black, thick, domain=-0.27:1.10, samples=100]
    plot (\x + 1.5, {1.0*(\x) - 0.25});
\draw[ColorSP, ->, >=stealth] (1.77,0.72) -- (1.77,0.02);
\draw[ColorSM, ->, >=stealth] (1.23,-0.52) -- (1.23,0.18);

\draw[thick] (-0.15,-1.2) rectangle (3.1,1.2);
\draw[->] (0.28,-0.75) -- (2.4,-0.75);
\node[] at (2.7,-0.75) {\footnotesize $x_1$};
\draw[->] (0.28,-0.75) -- (0.28,0.52);
\node[] at (0.28,0.79) {\footnotesize $f_1$};

\node[rotate=45] at (2.45,0.44) {\scriptsize $+1$};
\node[rotate=45] at (0.55,-0.28) {\scriptsize $-1$};

\draw[thick] (3.1,0) -- (3.9,0);

\draw[black, thick, domain=-0.6:0.27, samples=100]
    plot (\x + 5.55, {1.0*(\x) - 0.05});
\draw[black, thick, domain=-0.27:0.6, samples=100]
    plot (\x + 5.55, {1.0*(\x) + 0.25});
\draw[ColorSP, <-, >=stealth] (5.82,0.52) -- (5.82,0.22);
\draw[ColorSM, <-, >=stealth] (5.28,-0.32) -- (5.28,-0.02);

\draw[thick] (3.9,-1.2) rectangle (7.15,1.2);
\draw[->] (4.33,-0.75) -- (6.45,-0.75);
\node[] at (6.75,-0.75) {\footnotesize $x_2$};
\draw[->] (4.33,-0.75) -- (4.33,0.52);
\node[] at (4.33,0.79) {\footnotesize $f_2$};

\node[rotate=45] at (5.9,0.84) {\scriptsize $+1$};
\node[rotate=45] at (4.98,-0.39) {\scriptsize $-1$};

\draw[thick, ->] (7.15,0) -- (8.05,0);
\node[] at (8.35,0.0) {\footnotesize $U$};

\end{tikzpicture}
\vspace*{-0.1cm}
\caption{\textbf{Schematic of two bilinear force-displacement curves in series.} In this specific example, $d_1 > 0$  and $d_2 < 0$, corresponding to an active, energy injecting element.}
\label{fig:schematic_fordis_curves}
\end{figure}
We now consider the mapping from a physical network composed of two serially coupled bilinear elements (Fig.~\ref{fig:schematic_fordis_curves}) to the abstract hysteron model. Once again, combining the condition of fixed imposed displacement $U = x_1 + x_2$, mechanical equilibrium, and defining the abstract driving field $H = U$ allows to express the global switching thresholds $H_i^{\pm}$ as function of the state, force jumps, and local thresholds \cite{liu2024controlled}:
\begin{subequations} \label{eq:physical_critical_thresholds}
\begin{align}
 U_1^{\pm}(S) &= 2 x_1^{\pm} - d_1 s_1 + d_2 s_2, \\
 U_2^{\pm}(S) &= 2 x_2^{\pm} - d_2 s_2 + d_1 s_1,
\end{align}
\end{subequations}
And again, following \cite{liu2024controlled}, we compare the physical critical thresholds $H_i^\pm(S)$, as given by Eqs. (\ref{eq:physical_critical_thresholds}), to those of an abstract hysteron model,
\begin{equation}
 H_i^{\pm}(S) = h_i^{\pm} - \sum_{j \neq i} c_{ij} s_j,
\end{equation}
enabling to determine the bare switching fields $h_i^\pm$ and interaction coefficients $c_{ij}$. Identifying $H = U$, and using the convention that self-interactions are absent ($c_{ii} = 0$) this yields, up to an overall scale transformation and shift \cite{van2021profusion}:
\begin{subequations}
 \begin{align}
  h_1^{+} &= 2 x_{1}^{+} + d_{1}, \label{eq:mapping_v1+} \\
  h_1^{-} &= 2 x_{1}^{-} - d_{1}, \label{eq:mapping_v1-} \\
  h_2^{+} &= 2 x_{2}^{+} + d_{2}, \label{eq:mapping_v2+} \\
  h_2^{-} &= 2 x_{2}^{-} - d_{2}, \label{eq:mapping_v2-} \\
  c_{12} &= - d_2, \label{eq:mapping_c12} \\
  c_{21} &= - d_1. \label{eq:mapping_c21}
 \end{align}
\end{subequations}
The force jumps $d_i$ play a crucial role in the mapping. Not only do they directly set the interaction coefficients $c_{ij}$ (Eqs. (\ref{eq:mapping_c12}-\ref{eq:mapping_c21}), but, similarly as in the case discussed above, they also contribute to the bare switching thresholds of the abstract hysterons (Eqs. (\ref{eq:mapping_v1+}-\ref{eq:mapping_v2-})). In other words, the span of the hysterons can be written $\sigma_i = h_i^+ - h_i^- = 2 (\sigma_i^0 + d_i)$. We recover the result found in Eq. (\ref{eq:condition_L2_1element}) for $k=1$.
\subsection{Self-loops}
\begin{figure}[h!]
\centering
\hspace*{-1.1cm}
\begin{tikzpicture}

\fill[colorMixed] (-6.35,-4.6) rectangle (-4.0,-3.0);
\draw[] (-6.35,-4.6) rectangle (-4.0,-3.0);
\draw[] (-5.185,-4.6) -- (-5.185,-3.0);

\node[] at (-5.775,-2.7) {\small SL$_1$};
\node[] at (-4.575,-2.7) {\small SL$_2$};

\node[rotate=0] at (-5.75,-3.8) {\includegraphics[height=2.1cm]{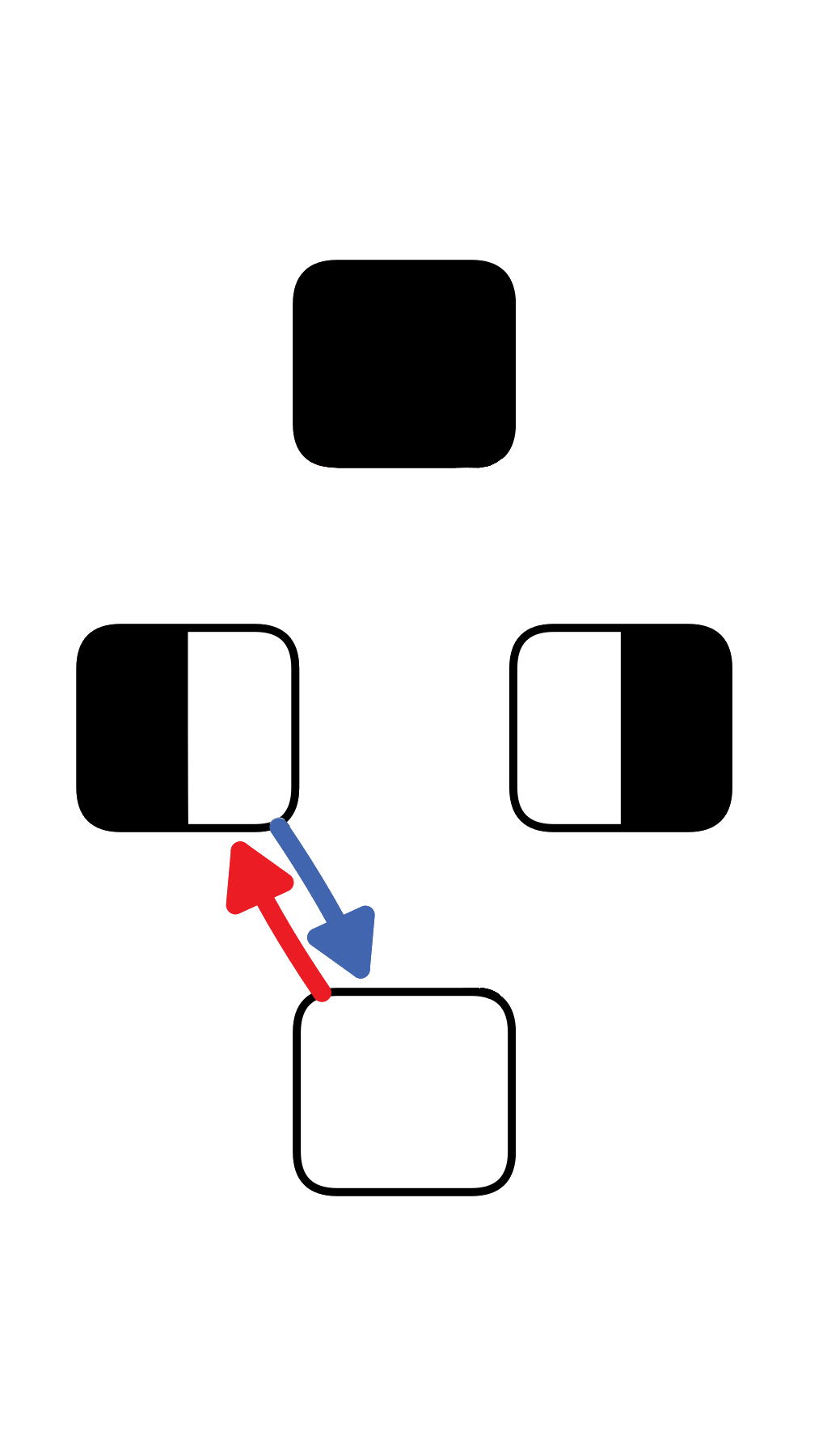}};

\node[rotate=0] at (-5.765,-4.325) {\small \scalebox{0.45}{$\boldsymbol{--}$}};
\node[rotate=0] at (-5.765,-3.275) {\small \scalebox{0.45}{$\color{white}\boldsymbol{++}$}};
\node[rotate=0] at (-6.08,-3.795) {\small \scalebox{0.45}{$\color{white}\boldsymbol{+}\color{black}\boldsymbol{-}$}};
\node[rotate=0] at (-5.445,-3.795) {\small \scalebox{0.45}{$\color{black}\boldsymbol{-}\color{white}\boldsymbol{+}$}};

\node[rotate=0] at (-4.6,-3.8) {\includegraphics[height=2.1cm]{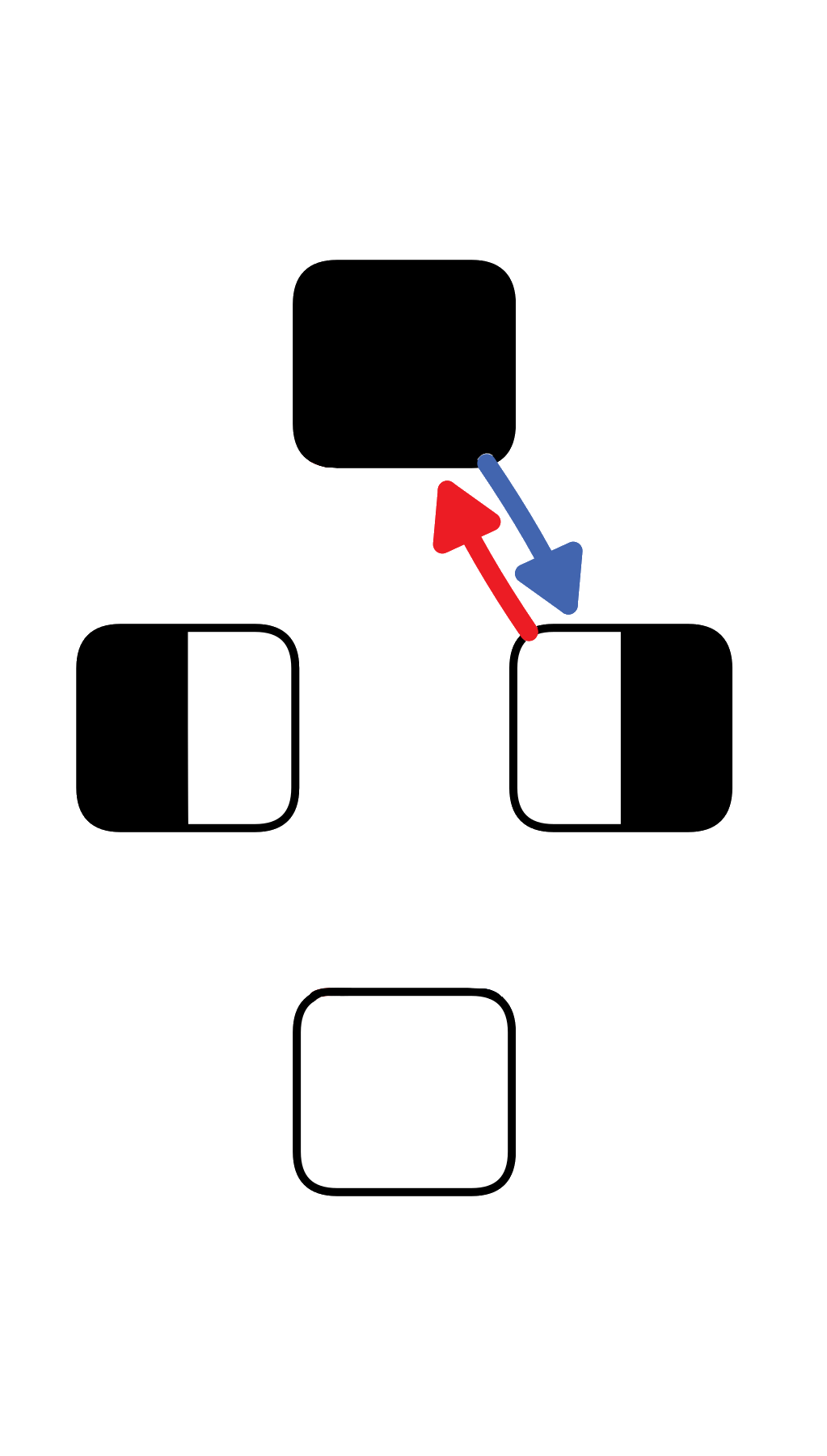}};

\node[rotate=0] at (-4.615,-4.325) {\small \scalebox{0.45}{$\boldsymbol{--}$}};
\node[rotate=0] at (-4.615,-3.275) {\small \scalebox{0.45}{$\color{white}\boldsymbol{++}$}};
\node[rotate=0] at (-4.93,-3.795) {\small \scalebox{0.45}{$\color{white}\boldsymbol{+}\color{black}\boldsymbol{-}$}};
\node[rotate=0] at (-4.295,-3.795) {\small \scalebox{0.45}{$\color{black}\boldsymbol{-}\color{white}\boldsymbol{+}$}};

\fill[colorFerro] (-3.5,-4.6) rectangle (-1.15,-3.0);
\draw[] (-3.5,-4.6) rectangle (-1.15,-3.0);
\draw[] (-2.335,-4.6) -- (-2.335,-3.0);

\node[] at (-2.925,-2.7) {\small SL$_3$};
\node[] at (-1.725,-2.7) {\small SL$_4$};

\node[rotate=0] at (-2.9,-3.8) {\includegraphics[height=2.1cm]{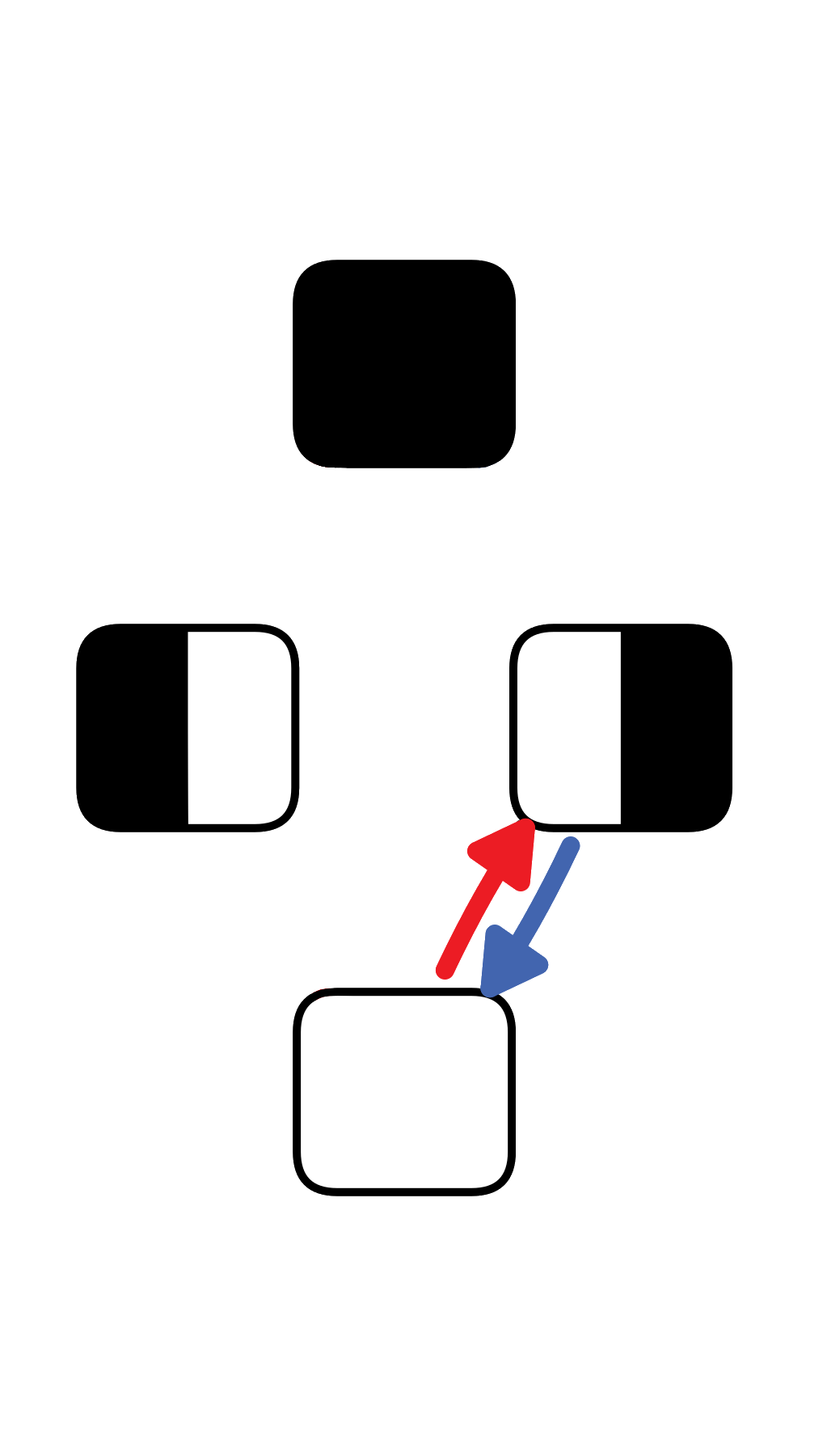}};

\node[rotate=0] at (-2.915,-4.325) {\small \scalebox{0.45}{$\boldsymbol{--}$}};
\node[rotate=0] at (-2.915,-3.275) {\small \scalebox{0.45}{$\color{white}\boldsymbol{++}$}};
\node[rotate=0] at (-3.23,-3.795) {\small \scalebox{0.45}{$\color{white}\boldsymbol{+}\color{black}\boldsymbol{-}$}};
\node[rotate=0] at (-2.595,-3.795) {\small \scalebox{0.45}{$\color{black}\boldsymbol{-}\color{white}\boldsymbol{+}$}};

\node[rotate=0] at (-1.75,-3.8) {\includegraphics[height=2.1cm]{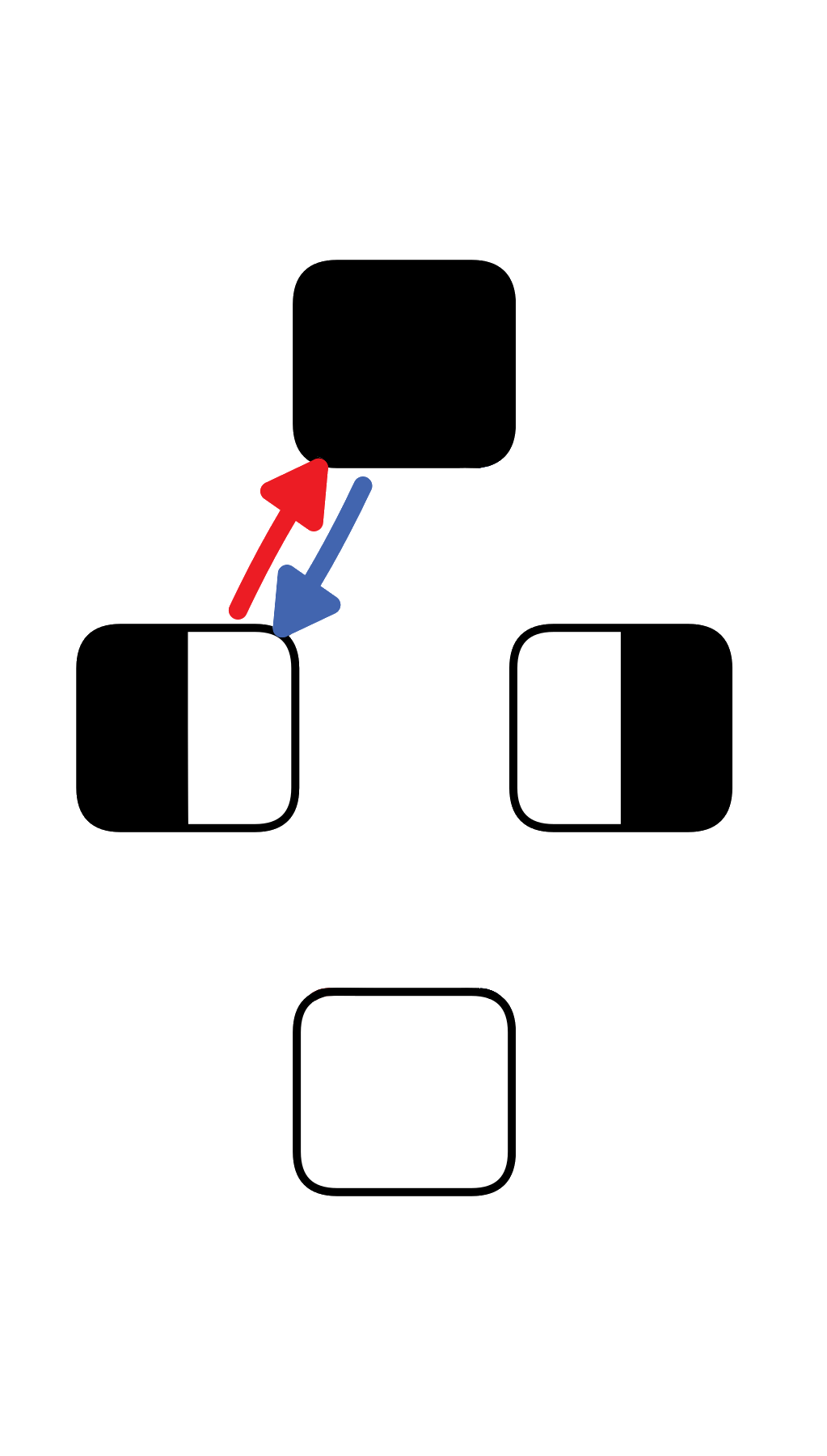}};

\node[rotate=0] at (-1.765,-4.325) {\small \scalebox{0.45}{$\boldsymbol{--}$}};
\node[rotate=0] at (-1.765,-3.275) {\small \scalebox{0.45}{$\color{white}\boldsymbol{++}$}};
\node[rotate=0] at (-2.08,-3.795) {\small \scalebox{0.45}{$\color{white}\boldsymbol{+}\color{black}\boldsymbol{-}$}};
\node[rotate=0] at (-1.445,-3.795) {\small \scalebox{0.45}{$\color{black}\boldsymbol{-}\color{white}\boldsymbol{+}$}};

\fill[ColorSL] (-0.65,-4.6) rectangle (1.7,-3.0);
\draw[] (-0.65,-4.6) rectangle (1.7,-3.0);
\draw[] (0.515,-4.6) -- (0.515,-3.0);

\node[] at (-0.075,-2.7) {\small SL$_5$};
\node[] at (1.125,-2.7) {\small SL$_6$};

\node[rotate=0] at (-0.05,-3.8) {\includegraphics[height=2.1cm]{self_loop_4_0_transparent.png}};

\node[rotate=0] at (-0.065,-4.325) {\small \scalebox{0.45}{$\boldsymbol{--}$}};
\node[rotate=0] at (-0.065,-3.275) {\small \scalebox{0.45}{$\color{white}\boldsymbol{++}$}};
\node[rotate=0] at (-0.38,-3.795) {\small \scalebox{0.45}{$\color{white}\boldsymbol{+}\color{black}\boldsymbol{-}$}};
\node[rotate=0] at (0.255,-3.795) {\small \scalebox{0.45}{$\color{black}\boldsymbol{-}\color{white}\boldsymbol{+}$}};

\node[rotate=0] at (1.075,-3.8) {\includegraphics[height=2.1cm]{self_loop_4_1_transparent.png}};

\node[rotate=0] at (1.085,-4.325) {\small \scalebox{0.45}{$\boldsymbol{--}$}};
\node[rotate=0] at (1.085,-3.275) {\small \scalebox{0.45}{$\color{white}\boldsymbol{++}$}};
\node[rotate=0] at (0.77,-3.795) {\small \scalebox{0.45}{$\color{white}\boldsymbol{+}\color{black}\boldsymbol{-}$}};
\node[rotate=0] at (1.405,-3.795) {\small \scalebox{0.45}{$\color{black}\boldsymbol{-}\color{white}\boldsymbol{+}$}};

\end{tikzpicture}
\vspace*{-0.5cm}
\caption{\small{\textbf{Possible self-loops for two coupled bilinear force-displacement curves.} (SL$_1$-SL$_4$) $L=2, n=1$; (SL$_5$-SL$_6$) $L=4, n=2$. (SL$_1$-SL$_2$) Only element $1$ is involved; (SL$_3$-SL$_4$) only element $2$ is involved; (SL$_5$-SL$_6$) both elements are involved.}}
\label{fig:bilinear_SL}
\end{figure}

We have determined the region of parameter space where self-loops are present numerically for two serially coupled elements. For each set of parameters, this is done by systematically analyzing the whole range of driving fields. We consider both elements that mimic
spins ($x_i^-=x_i^+$), and hysteretic elements ($x_i^-<x_i^+$), and scan $d_1$ and $d_2$ for different orderings of $x_i^{\pm}$. We resolve race conditions where multiple elements (here two) become simultaneously unstable the same way as in the abstract hysteron model: we flip first the most unstable element, i.e. the one furthest away from its stability threshold \cite{keim2021multiperiodic}. Multiple distinct self-loop structures are found with sizes $L = 2$ and $L = 4$ (Fig. \ref{fig:bilinear_SL}). Each self-loop is associated with a set of linear inequalities, so that they form polytopes in parameter space \cite{van2021profusion, teunisse2024transition}.
\subsubsection{Spin-like bilinear elements}
\begin{figure}[t!]
\centering
\hspace*{-0.7cm}
\begin{tikzpicture}

\node[] at (-5.0,0.0) {\includegraphics[width=5.0cm]{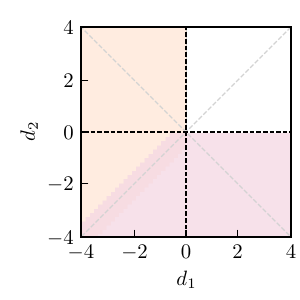}};

\node[] at (0.0,0.0) {\includegraphics[width=5.0cm]{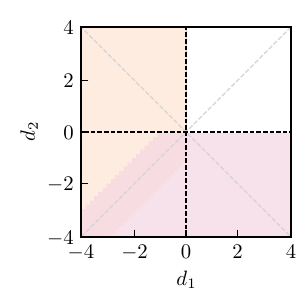}};

\node[] at (5.0,0.0) {\includegraphics[width=5.0cm]{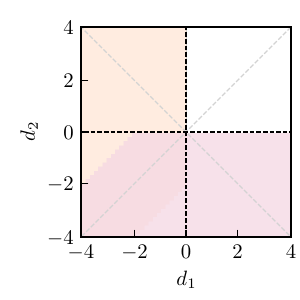}};

\fill[black] (-0.27,-0.57) circle (0.05);
\node[] at (-0.0,-0.3) {\small (ii)};

\draw[<->] (-0.98,-0.9) -- (-0.2,-0.9);
\node[] at (0.1,-1.15) {\small $2 \Delta x^c$};

\fill[black] (1.47,-0.57) circle (0.05);
\node[] at (1.79,-0.3) {\small (iii)};

\fill[black] (-0.27,1.17) circle (0.05);
\node[] at (-0.0,1.34) {\small (i)};

\draw[->] (-6.2,2.5) -- (7.2,2.5);

\node[] at (7.7,2.5) {\small $\Delta x^c$};

\fill[colorFerro] (7.9,-1.5) rectangle (9.3,0.1);
\draw[] (7.9,-1.5) rectangle (9.3,0.1);

\fill[colorMixed] (7.9,0.4) rectangle (9.3,2.0);
\draw[] (7.9,0.4) rectangle (9.3,2.0);

\node[] at (8.6,-0.7) {\includegraphics[width=1.3cm]{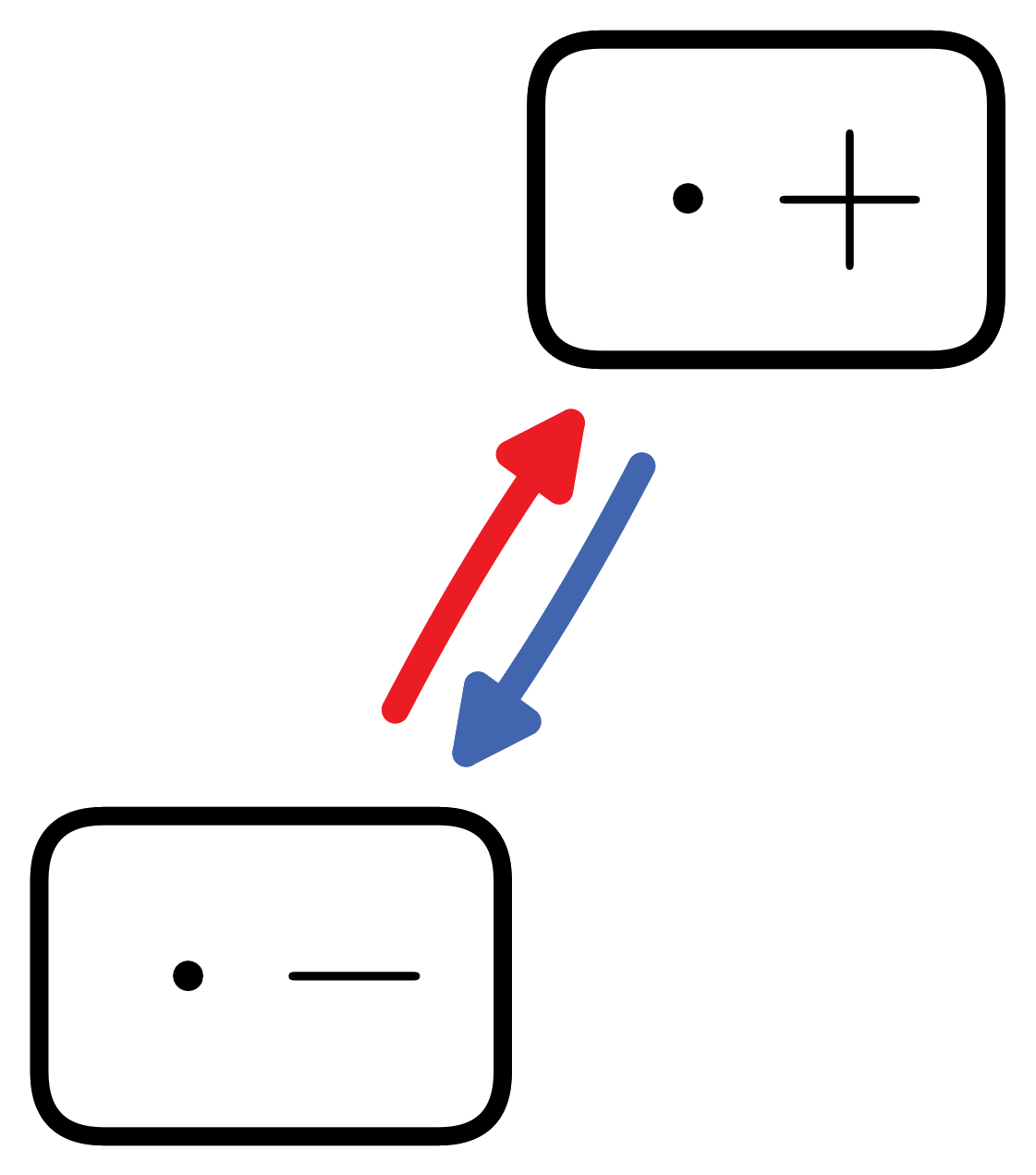}};
\node[] at (8.6,1.2) {\includegraphics[width=1.3cm]{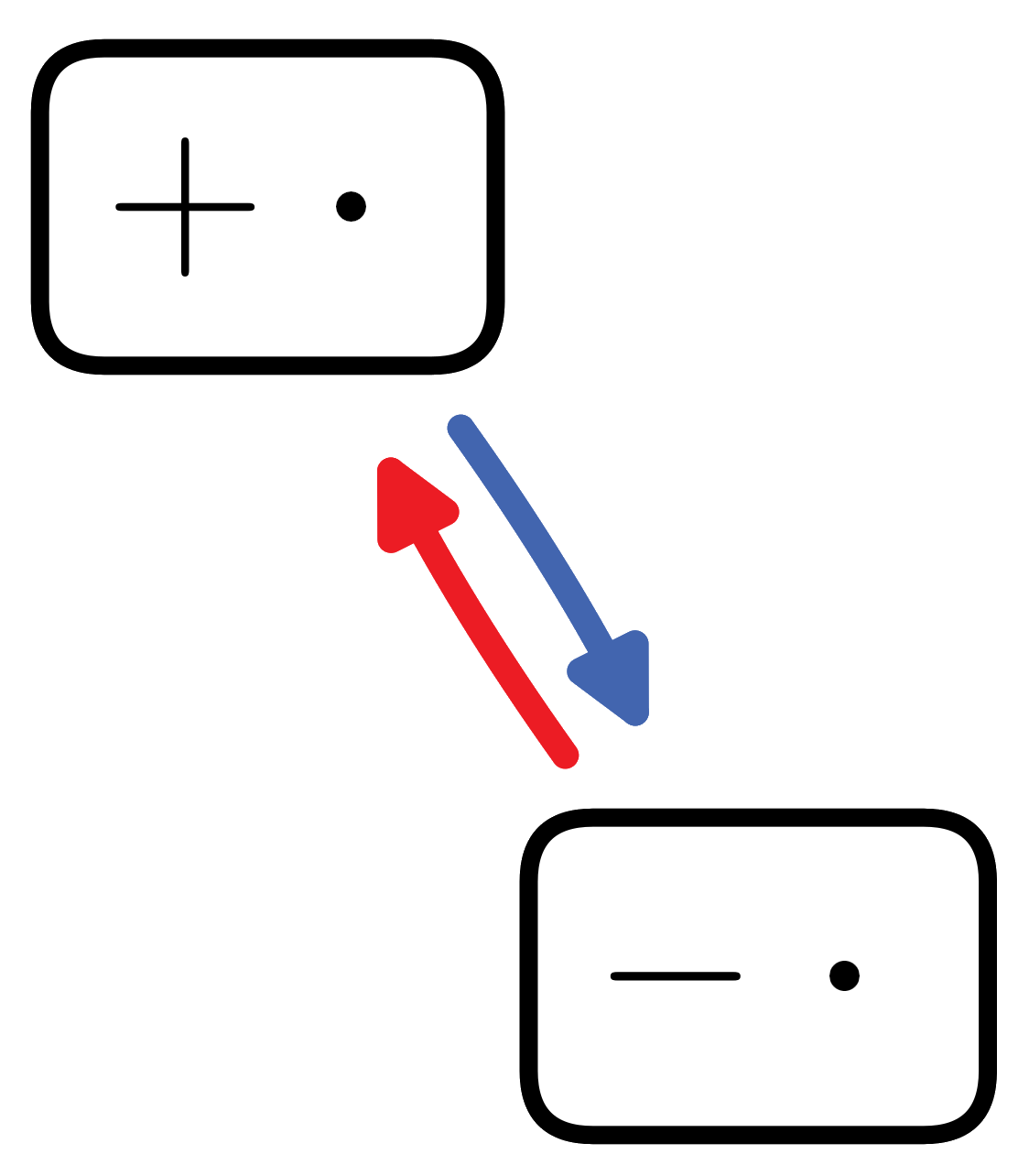}};

\draw[->] (7.9,-0.7) -- (6.9,-0.7);
\draw[->] (7.9,1.2) -- (4.9,1.2);

\node[] at (-4.5,-5.0) {\includegraphics[width=4.4cm]{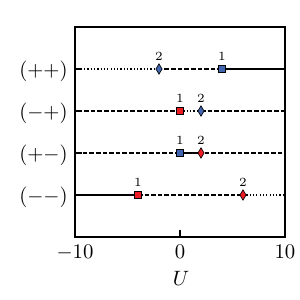}};

\node[] at (-4.7,-8.5) {\includegraphics[width=5.0cm]{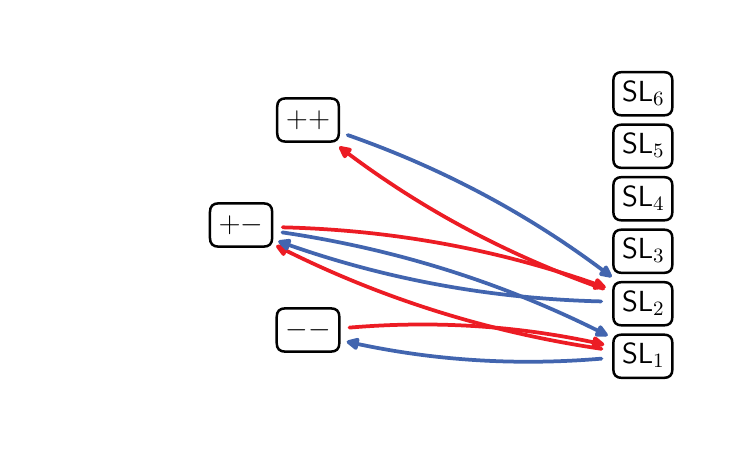}};

\draw[] (-6.6,-9.85) rectangle (-2.2,-2.9);

\node[] at (0.5,-5.0) {\includegraphics[width=4.4cm]{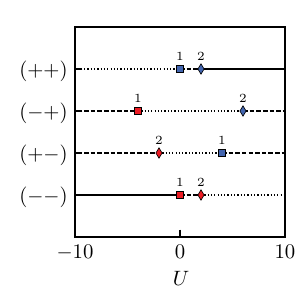}};

\node[] at (0.3,-8.5) {\includegraphics[width=5.0cm]{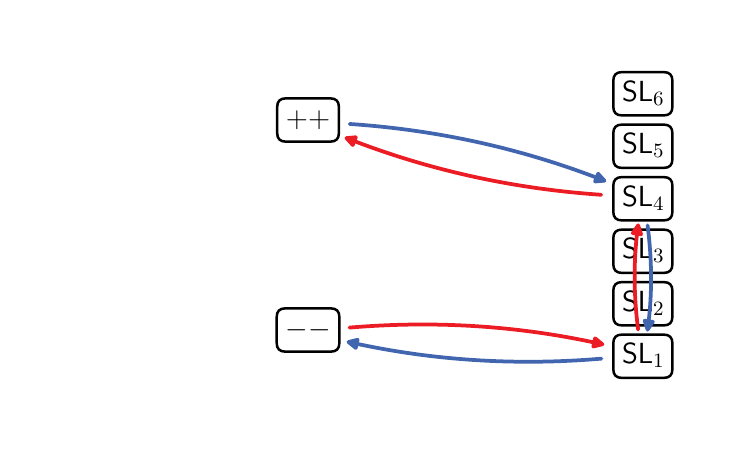}};

\draw[] (-1.6,-9.85) rectangle (2.8,-2.9);

\node[] at (5.5,-5.0) {\includegraphics[width=4.4cm]{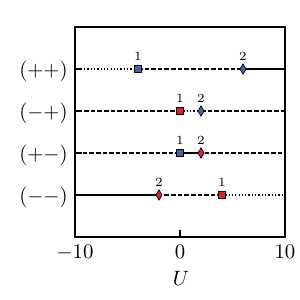}};

\node[] at (5.3,-8.5) {\includegraphics[width=5.0cm]{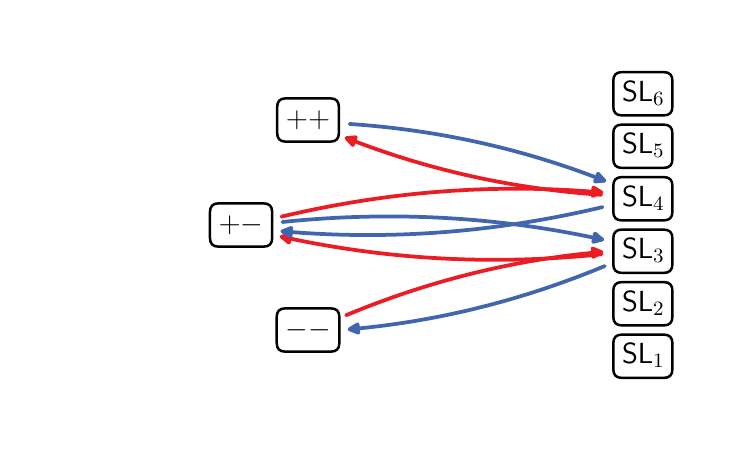}};

\draw[] (3.4,-9.85) rectangle (7.8,-2.9);

\node[] at (-6.85,2.55) {\small (a)};
\node[] at (-6.85,-2.5) {\small (b)};

\node[] at (-6.3,-9.52) {\small (i)};
\node[] at (-1.25,-9.52) {\small (ii)};
\node[] at (3.8,-9.52) {\small (iii)};

\fill[white] (-4.26,-7.09) rectangle (-3.86,-6.69);
\node[] at (-4.06,-6.89) {\small $H$};

\fill[white] (0.74,-7.09) rectangle (1.14,-6.69);
\node[] at (0.94,-6.89) {\small $H$};

\fill[white] (5.74,-7.09) rectangle (6.14,-6.69);
\node[] at (5.94,-6.89) {\small $H$};

\end{tikzpicture}
\vspace*{-0.25cm}
\caption{\small{\textbf{Presence of different self-loops in the case of spin-like bilinear elements.} (a) The colored regions indicate the presence of the different families of self-loops as a function of $d_1$ and $d_2$, color coded as in Fig. 3 of the main text. In all panels, $x_1^c = 0$; and from left to right: $x_2^c = 0.5$; $x_2^c = 1.0$; $x_2^c = 2.0$. (b) State switching fields for different values ($\pm2$) of $d_1$ and $d_2$, and associated transition graph.}}
\label{fig:bilinear_sim_spins}
\end{figure}
We first discuss spin-like elements, i.e. $x_i^+ = x_i^- = x_i^c$, varying $\Delta x^c = x_2^c - x_1^c > 0$. Strikingly,
only size $L=2$ self-loops can occur (Fig. \ref{fig:bilinear_sim_spins}).
For purely dissipative cases, i.e. $d_1>0$ and $d_2 > 0$, no self-loops emerge
\cite{liu2024controlled}.
For mixed cases, i.e. when $d_i < 0$ and $d_j > 0$, self-loops involving element $i$ can emerge (Figs.~\ref{fig:bilinear_sim_spins}-b-i,iii). Finally, when both elements are non-dissipative, i.e.
$d_1 < 0$ and $d_2 < 0$, large negative $d_i$ favors oscillations of element $i$, with a crossover region  around the line $d_1 = d_2$
where both hysterons become unstable. The
width of this region scales linearly with $\Delta x^c$.

While a gap of finite size emerges within this range of parameters, the associated transition-graphs include both $L=2$ self-loops involving elements $1$ and $2$, but no $L=4$ self-loops (Fig. \ref{fig:bilinear_sim_spins}-b-ii). It is instructive to discuss the absence of $L=4$ self-loops for this case. In the gap, all states are unstable, e.g. in states $(+-)$ and $(-+)$ both elements are unstable. Therefore, the transitions between unstable states depend on which is the most unstable element, thus on the value of $H$. At the beginning of the gap (for $H$ just above $H^+(--)$), the only periodic attractor is SL$_1$, where only element $1$ flips back and forth. Indeed, the two other unstable states not involved in SL$_1$, $(-+)$ and $(++)$, flip towards the stables involved in SL$_1$.
Remarkably, in the middle of the gap, the most unstable element of states $(+-)$ and $(-+)$ changes simultaneously. This leads to a different periodic attractor SL$_4$, where only element $2$ flips back and forth. The absence of $L=4$ self-loops is therefore related to the way we resolve race conditions and to the simultaneous change of the most unstable element in states $(+-)$ and $(-+)$.
\subsubsection{Hysteretic bilinear elements}
\begin{figure}[p!]
\centering
\hspace*{-0.5cm}
\begin{tikzpicture}

\node[] at (-5.0,0.0) {\includegraphics[width=5.0cm]{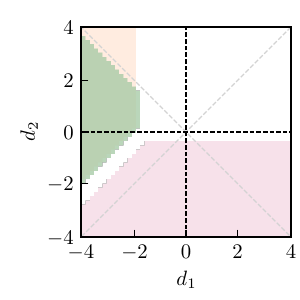}};

\node[] at (0.0,0.0) {\includegraphics[width=5.0cm]{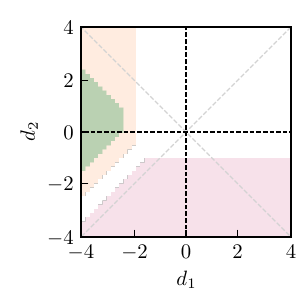}};

\node[] at (5.0,0.0) {\includegraphics[width=5.0cm]{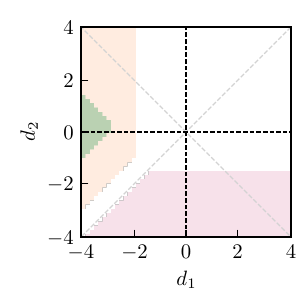}};

\draw[->] (-6.2,2.5) -- (7.2,2.5);
\node[] at (7.7,2.5) {\small $x_2^+$};

\fill[ColorSL] (7.9,-0.75) rectangle (9.6,1.35);
\draw[] (7.9,-0.75) rectangle (9.6,1.35);

\node[] at (8.75,0.3) {\includegraphics[width=1.55cm]{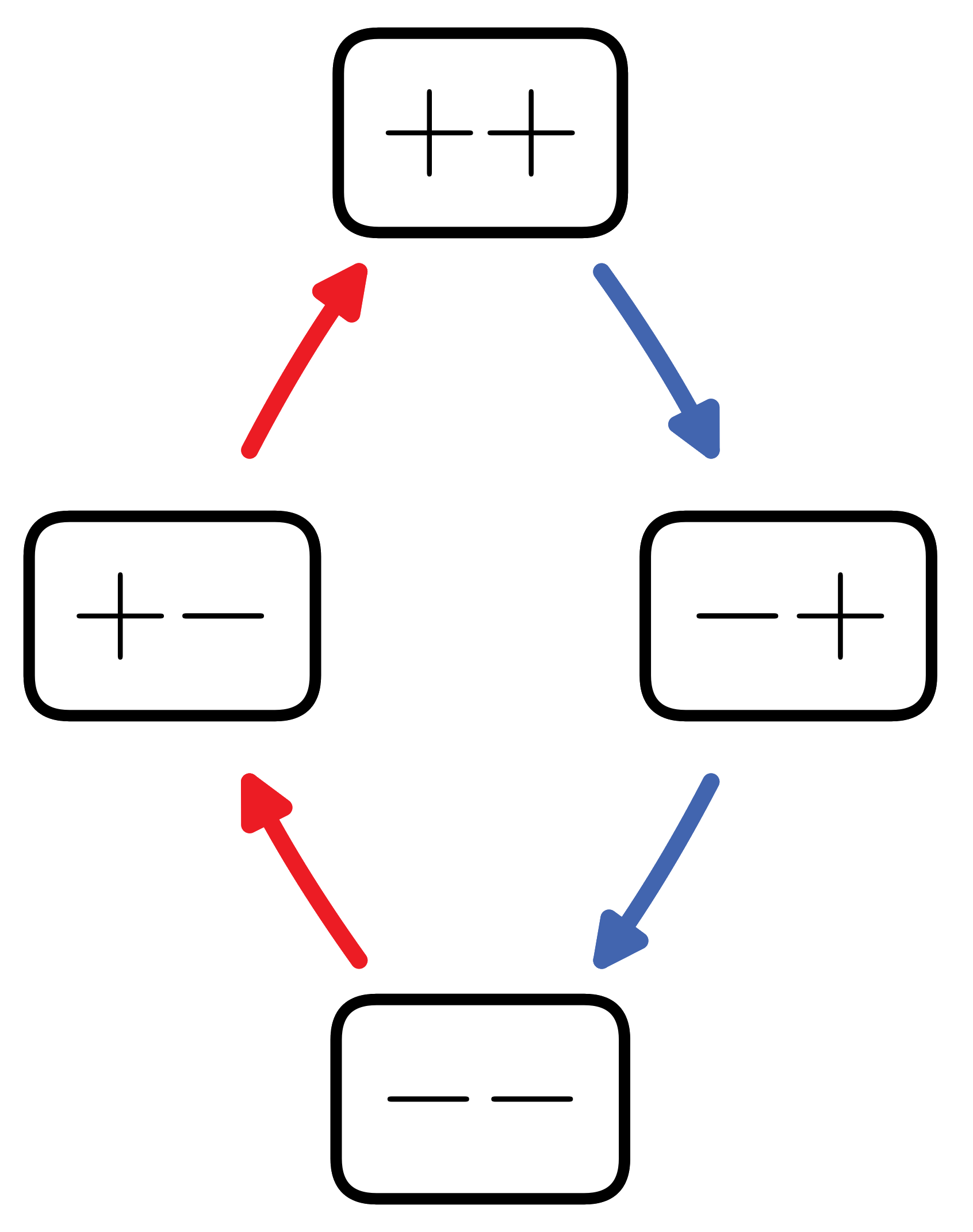}};

\draw[->] (7.9,-0.1) -- (4.0,0.1);

\node[] at (-5.0,-5.5) {\includegraphics[width=5.0cm]{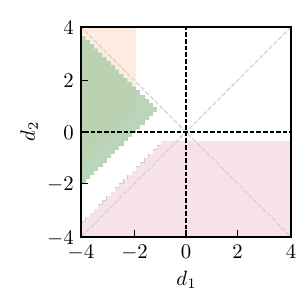}};

\node[] at (0.0,-5.5) {\includegraphics[width=5.0cm]{selfloop_plane_x1M_-1.0_x1P_1.0_x2M_0.5_x2P_0.75.pdf}};

\node[] at (5.0,-5.5) {\includegraphics[width=5.0cm]{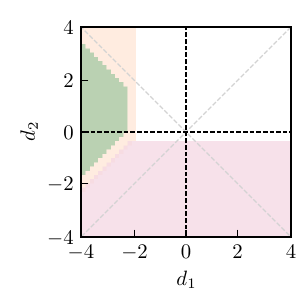}};

\draw[->] (-6.2,-3.0) -- (7.2,-3.0);
\node[] at (7.7,-3.0) {\small $x_2^c$};

\node[] at (-5.0,-11.0) {\includegraphics[width=5.0cm]{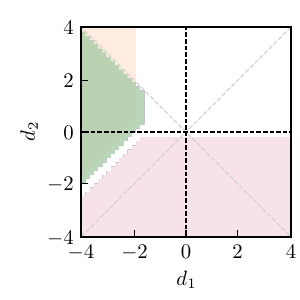}};

\node[] at (0.0,-11.0) {\includegraphics[width=5.0cm]{selfloop_plane_x1M_-1.0_x1P_1.0_x2M_0.5_x2P_0.75.pdf}};

\node[] at (5.0,-11.0) {\includegraphics[width=5.0cm]{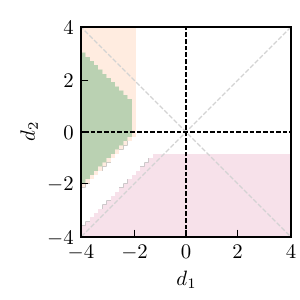}};

\draw[->] (-6.2,-8.5) -- (7.2,-8.5);
\node[] at (7.7,-8.5) {\small $\sigma_2^0$};

\node[] at (-5.0,-16.5) {\includegraphics[width=5.0cm]{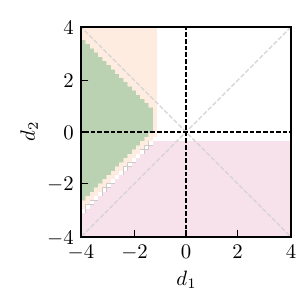}};

\node[] at (0.0,-16.5) {\includegraphics[width=5.0cm]{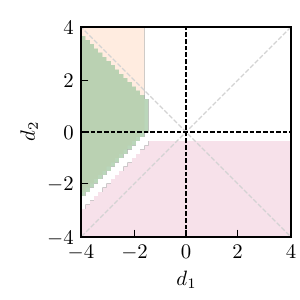}};

\node[] at (5.0,-16.5) {\includegraphics[width=5.0cm]{selfloop_plane_x1M_-1.0_x1P_1.0_x2M_0.5_x2P_0.75.pdf}};

\draw[->] (-6.2,-14.0) -- (7.2,-14.0);
\node[] at (7.7,-14.0) {\small $\sigma_1^0$};

\node[] at (-6.85,2.55) {\small (a)};
\node[] at (-6.85,-2.95) {\small (b)};
\node[] at (-6.85,-8.45) {\small (c)};
\node[] at (-6.85,-13.95) {\small (d)};

\end{tikzpicture}
\vspace*{-0.5cm}
\caption{\small{\textbf{Presence of different self-loops in the case of hysteretic bilinear elements.} The colored regions indicate the presence of the different families of self-loops as a function of $d_1$ and $d_2$, color coded as in Fig. 3 of the main text. (a) In all panels, $x_1^+ = 1$, $x_1^- = -1$, $x_2^- = 0.5$; from left to right: $x_2^+ = 0.75$, $1.5$; $2.0$. (b) In all panels, $\sigma_1^0 = 2$, $\sigma_2^0 = 0.25$, $x_1^c = 0$; from left to right: $x_2^c = 0$; $0.625$; $1.2$. (c) In all panels, $\sigma_1^0 = 2$, $x_1^c = 0.0$, $x_2^c = 0.625$; from left to right: $\sigma_2^0 = 0.1$,$0.25$,$0.85$. (d) In all panels, $x_1^c = 0$, $x_2^c = 0.625$, $\sigma_2^0 = 0.25$; from left to right: $\sigma_1^0 = 1.2$,$1.6$,$2.0$.}}
\label{fig:bilinear_sim_hysterons}
\end{figure}

We then explore the case of hysteretic elements, i.e. $\sigma_i^0 = x_i^+ - x_i^- > 0$. As compared to the case above, in addition to the distance between the hysteresis midpoints $\Delta x^c = x_2^c - x_1^c$, with $ x_1^c = (x_1^+ + x_1^-)/2$ and $x_2^c = (x_2^+ + x_2^-)/2$, we now have two additional parameters corresponding to the spans of the two elements, i.e. $\sigma_1^0 = x_1^+ - x_1^-$ and $\sigma_2^0 = x_2^+ - x_2^-$. We explored the self-loops that emerge as a function of $d_1$ and $d_2$, varying $x_2^+$ (Fig. \ref{fig:bilinear_sim_hysterons}-a), $x_2^c$ (Fig. \ref{fig:bilinear_sim_hysterons}-b), $\sigma_2^0$ (Fig. \ref{fig:bilinear_sim_hysterons}-c), and $\sigma_1^0$ (Fig. \ref{fig:bilinear_sim_hysterons}-d).
The presence of a finite span then allows
self-loops of size $L = 4$ (Fig. \ref{fig:bilinear_sim_hysterons}).

\subsection{Conditions for self-loops}
\label{sec:conditions_self_loops}
In this section, we derive the necessary and sufficient conditions for the different self-loops to emerge in systems of two serially coupled elements.
\begin{figure}[b!]
\centering
\hspace*{-0.5cm}
\begin{tikzpicture}

\node[] at (-1.6,4.9) {\small $1$};
\draw[dashed] (-1.2,5.15) -- (2.3,5.15);
\node[] at (2.6,5.15) {\small $+$};
\draw[dashed] (-1.2,4.65) -- (2.3,4.65);
\node[] at (2.6,4.65) {\small $-$};

\draw[thick] (-1.2,4.65) -- (1.55,4.65);
\draw[thick] (-0.45,5.15) -- (2.3,5.15);
\draw[->, thick, color=ColorSP] (1.55,4.65) -- (1.55,5.13);
\draw[->, thick, color=ColorSM] (-0.45,5.15) -- (-0.45,4.67);

\draw[thick] (-1.2,3.65) -- (1.35,3.65);
\draw[thick] (0.9,4.15) -- (2.3,4.15);
\draw[->, thick, color=ColorSP] (1.35,3.65) -- (1.35,4.13);
\draw[->, thick, color=ColorSM] (0.9,4.15) -- (0.9,3.67);

\node[] at (-1.6,3.9) {\small $2$};
\draw[dashed] (-1.2,4.15) -- (2.3,4.15);
\node[] at (2.6,4.15) {\small $+$};
\draw[dashed] (-1.2,3.65) -- (2.3,3.65);
\node[] at (2.6,3.65) {\small $-$};

\draw[->] (-1.4,3.05) -- (2.45,3.05);
\node[] at (2.75,3.05) {\small $H$};

\node[] at (0.0,0.0) {\includegraphics[width=5.0cm]{selfloop_plane_x1M_-1.0_x1P_1.0_x2M_0.5_x2P_0.75.pdf}};

\draw[<->] (-0.13,0.8) -- (0.56,0.8);
\node[] at (0.215,1.07) {\small $W$};

\draw[dashed] (1.5,2.06) -- (-1.15,-0.59);
\draw[dashed] (-1.15,1.94) -- (2.25,-1.46);

\draw[<->] (-0.54,-0.78) -- (-0.92,-0.4);

\draw[<->] (-0.21,2.2) -- (0.57,2.2);
\node[] at (0.18,2.49) {\small $\sigma_1^0$};

\draw[<-] (2.2,0.13) -- (2.2,-0.11);
\draw[->] (2.2,0.57) -- (2.2,0.33);
\node[] at (2.65,0.23) {\small $\sigma_2^0$};

\fill[black] (-0.52,1.57) circle (0.05);
\node[] at (-0.25,1.74) {\small (i)};

\fill[black] (-0.52,0.87) circle (0.05);
\node[] at (-0.7,0.56) {\small (ii)};

\fill[black] (-0.50,-0.13) circle (0.05);
\node[] at (-0.19,-0.38) {\small (iii)};

\node[] at (-4.5,-5.0) {\includegraphics[width=4.4cm]{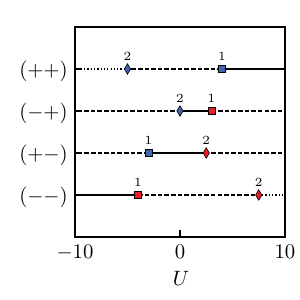}};

\node[] at (-4.7,-8.5) {\includegraphics[width=5.0cm]{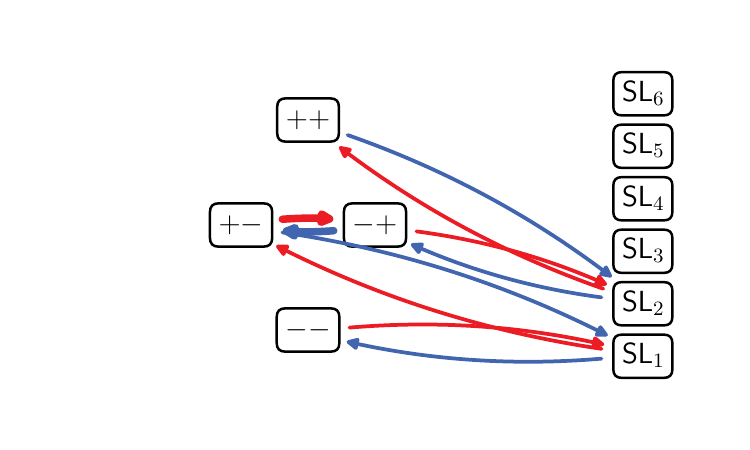}};

\draw[] (-6.6,-9.85) rectangle (-2.2,-2.9);

\node[] at (0.5,-5.0) {\includegraphics[width=4.4cm]{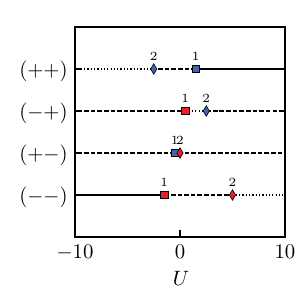}};

\node[] at (0.3,-8.5) {\includegraphics[width=5.0cm]{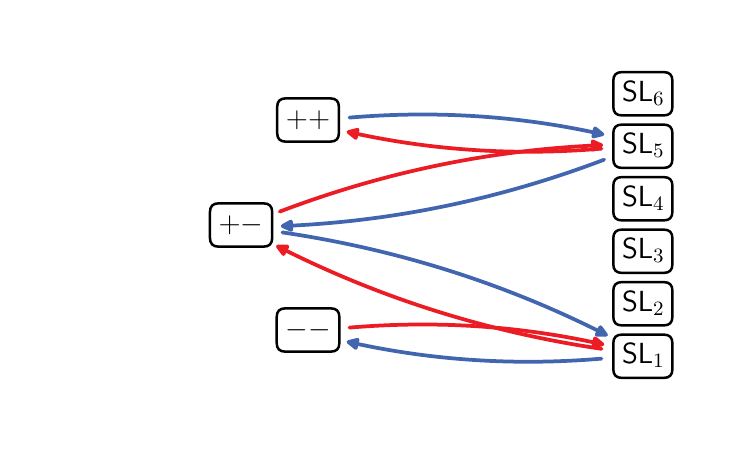}};

\draw[] (-1.6,-9.85) rectangle (2.8,-2.9);

\node[] at (5.5,-5.0) {\includegraphics[width=4.4cm]{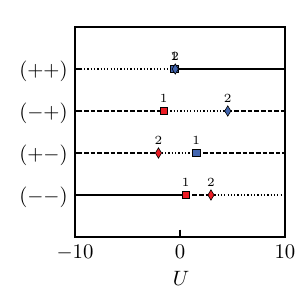}};

\node[] at (5.3,-8.5) {\includegraphics[width=5.0cm]{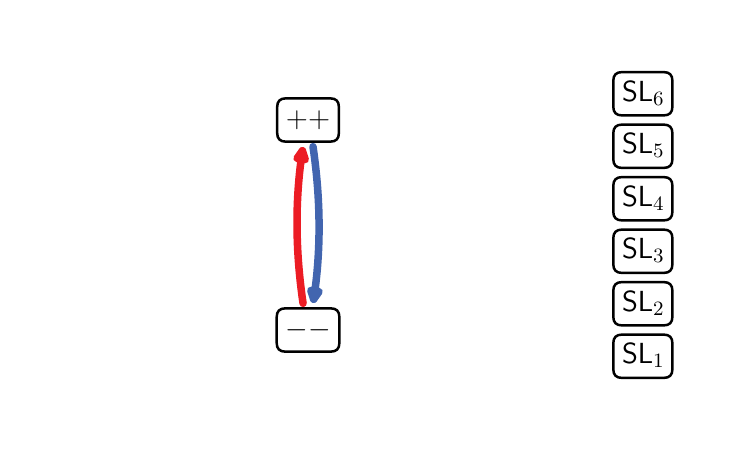}};

\draw[] (3.4,-9.85) rectangle (7.8,-2.9);

\node[] at (-6.85,-2.5) {\small (b)};

\node[] at (-6.3,-9.52) {\small (i)};
\node[] at (-1.25,-9.52) {\small (ii)};
\node[] at (3.8,-9.52) {\small (iii)};

\fill[white] (-4.26,-7.09) rectangle (-3.86,-6.69);
\node[] at (-4.06,-6.89) {\small $H$};

\fill[white] (0.74,-7.09) rectangle (1.14,-6.69);
\node[] at (0.94,-6.89) {\small $H$};

\fill[white] (5.74,-7.09) rectangle (6.14,-6.69);
\node[] at (5.94,-6.89) {\small $H$};

\end{tikzpicture}
\vspace*{-0.4cm}
\caption{\small{\textbf{Summary of the conditions for the emergence of the different self-loops in the case of hysteretic bilinear elements.} (a) Self-loops in the $d_1 - d_2$ plane. The colors indicate the presence of the different families of self-loops as a function of $d_1$ and $d_2$, color coded as in Fig. 3 of the main text. In the example, $x_1^+ = 1$, $x_1^- = -1$, $x_2^+ = 0.75$, $x_2^- = 0.5$. In the schematic, $W = \min \left( (x_2^- - x_1^+), (x_1^- - x_2^+) \right)$, as given in Eq. (\ref{eq:final_eq_loop6:3}).}}
\label{fig:bilinear_sim_hysterons_SL}
\end{figure}
\subsubsection{$L=2$}
Self-loops of length $L=2$ are associated to one given element $i$ flipping back and forth at a given value of the driving $H$. This is only possible if element $i$ is intrinsically unstable for this $H$, i.e. $h_i^+ < h_i^-$, which can be rewritten $d_i < - \sigma_i^0 < 0$. As expected, we recover the result found for a single element in series with a spring of stiffness $k=1$. These conditions explain the vertical and horizontal separations between $L=2$ self-loops and the $x$- an $y$-axis in the $d_1 - d_2$ plane (Fig. \ref{fig:bilinear_sim_spins}a, Fig. \ref{fig:bilinear_sim_hysterons}, and Fig. \ref{fig:bilinear_sim_hysterons_SL}). As we see below, the conditions above are necessary but not sufficient. \\

\paragraph{Self-loop $(--) \leftrightarrow (+-)$}
~\\

In order that the self-loop $(--) \leftrightarrow (+-)$ (SL$_1$, first of Fig. \ref{fig:bilinear_SL}) emerges, the first condition to be met is that element $1$ is unstable, thus we must have $d_1 < - \sigma_1^0$. This is equivalent to say $H_1^+ (--) < H_1^- (+-)$.

However, when state $(+-)$ is unconditionaly unstable, the self-loop SL$_1$ involves resolving a race condition. State $(+-)$ is unconditionaly unstable when:
\begin{equation}
 H_2^+ (+-) < H_1^- (+-),
\end{equation}
which can be rewritten as:
\begin{equation} \label{eq:cond:SL1}
 d_2 + d_1 < x_1^- - x_2^+.
\end{equation}
In this range of parameter space ($d_1 < - \sigma_1^0$ and $d_2 + d_1 < x_1^- - x_2^+$), we must impose that there exists a range of $H$ where element $1$ is more unstable than element $2$ in state $(+-)$:
\begin{equation}
 H_1^+ (--) < \frac{H_2^+ (+-) + H_1^- (+-)}{2},
\end{equation}
which can be rewritten as:
\begin{equation}
 d_2 - d_1 > - \Delta x^c + \frac{3}{2} \sigma_1^0 - \frac{1}{2} \sigma_2^0.
\end{equation}
\paragraph{Self-loop $(-+) \leftrightarrow (++)$}
~\\

We now apply the same reasoning as above for the self-loop $(-+) \leftrightarrow (++)$ (SL$_2$, second of Fig. \ref{fig:bilinear_SL}). The first condition to be met is that element $1$ is unstable, thus we must have $d_1 < - \sigma_1^0$. This is equivalent to say $H_1^+ (-+) < H_1^- (++)$.
However, when state $(-+)$ is unconditionaly unstable, the self-loop involves resolving a race condition. State $(-+)$ is unconditionaly unstable when:
\begin{equation}
 H_1^+ (-+) < H_2^- (-+),
\end{equation}
which can be rewritten as:
\begin{equation}
 d_2 + d_1 < x_2^- - x_1^+.
\end{equation}
In this range of parameter space ($d_1 < - \sigma_1^0$ and $d_2 + d_1 < x_2^- - x_1^+$), we must impose that there exists a range of $H$ where element $1$ is more unstable than element $2$ in state $(-+)$:
\begin{equation}
 H_1^- (++) > \frac{H_1^+ (-+) + H_2^- (-+)}{2},
\end{equation}
which can be rewritten as:
\begin{equation}
 d_2 - d_1 > \Delta x^c + \frac{3}{2} \sigma_1^0 - \frac{1}{2} \sigma_2^0.
\end{equation}
\paragraph{Summary for self-loops involving element $1$}
~\\

$L=2$ self-loops involving element $1$ (SL$_1$ and SL$_2$) can emerge only when $d_1 < -\sigma_1^0$. Moreover, within the range $d_2 + d_1 < x_1^- - x_2^+$ (resp. $d_2 + d_1 < x_2^- - x_1^+$), the self-loop $(--) \leftrightarrow (+-)$ (resp. $(-+) \leftrightarrow (++)$) only emerges for $d_2 - d_1 > - \Delta x^c + \frac{3}{2} \sigma_1^0 - \frac{1}{2} \sigma_2^0$ (resp. $d_2 - d_1 > \Delta x^c + \frac{3}{2} \sigma_1^0 - \frac{1}{2} \sigma_2^0$). There is therefore a diagonal band above which both SL$_1$ and SL$_2$ are present (Fig. \ref{fig:bilinear_sim_spins}b-i), inside which only SL$_1$ is present (Fig. \ref{fig:bilinear_sim_spins}b-ii), and below which none are present (Fig. \ref{fig:bilinear_sim_spins}b-iii).
\subsubsection{$L=4$}
\paragraph{Self-loop $(--) \rightarrow (+-) \rightarrow (++) \rightarrow (-+) \rightarrow \dots$}
~\\

Here, we derive the conditions under which the
$L=4$ self-loop SL$_5$  emerges (Fig. \ref{fig:bilinear_SL}).

First, there are four conditions to impose that the four states are unstable:
\begin{subequations} \label{eq:states_instability}
\begin{align}
 H &> H_1^+(--) = h_1^+ + c_{12}, \label{eq:states_instability:1} \\
 H &< H_1^-(++) = h_1^- - c_{12}, \label{eq:states_instability:2} \\
 H &> H_2^+(+-) = h_2^+ - c_{21}, \label{eq:states_instability:3} \\
 H &< H_2^-(-+) = h_2^- + c_{21}. \label{eq:states_instability:4}
\end{align}
\end{subequations}
Moreover, there are $2$ conditions to impose the scaffold, i.e. that element $1$ flips when states $(++)$ and $(--)$ are unstable:
\begin{subequations}
\begin{align}
 H_2^+(--) = h_2^+ + c_{21} &> H_1^+(--) = h_1^+ + c_{12}, \\
 H_2^-(++) = h_2^- - c_{21} &< H_1^-(++) = h_1^- - c_{12}.
\end{align}
\end{subequations}
These two last equations translate directly into:
\begin{subequations}
\begin{align}
 d_2 - d_1 &> x_1^+ - x_2^+, \\
 d_2 - d_1 &> x_2^- - x_1^-.
\end{align}
\end{subequations}
Thus, because the two conditions must be satisfied:
\begin{equation}
 d_2 - d_1 > \max \left( (x_1^+ - x_2^+), (x_2^- - x_1^-) \right).
\end{equation}
Coming back to Eqs. (\ref{eq:states_instability}), we can construct a set of four linear inequalities to make sure there exist a finite range of $H$ inside which there is a gap:
\begin{subequations} \label{eq:states_instability_gap}
\begin{align}
 d_2 - d_1 &> \sigma_1^0, \\
 d_1 + d_2 &< - \sigma_2^0, \\
 d_1 &< x_2^- - x_1^+, \\
 d_1 &< x_1^- - x_2^+,
\end{align}
\end{subequations}
where the first, second, third and fourth equations come from combining Eq. (\ref{eq:states_instability:1}) and Eq. (\ref{eq:states_instability:2}), Eq. (\ref{eq:states_instability:3}) and Eq. (\ref{eq:states_instability:4}), Eq. (\ref{eq:states_instability:1}) and Eq. (\ref{eq:states_instability:4}), Eq. (\ref{eq:states_instability:2}) and Eq. (\ref{eq:states_instability:3}), respectively. Altogether, we find a set of three linear inequalities:
\begin{subequations} \label{eq:final_eq_loop5}
\begin{align}
 d_2 - d_1 &> \max \left( (x_1^+ - x_2^+), (x_2^- - x_1^-), \sigma_1^0 \right), \\
 d_1 + d_2 &< - \sigma_2^0 < 0, \\
 d_1 &< \min \left( (x_2^- - x_1^+), (x_1^- - x_2^+) \right).
\end{align}
\end{subequations}
Importantly, these equations do not take into account race conditions: if the two elements are unstable in state $(+-)$ or $(-+)$, this might lead to a race condition. Taking into account the race conditions lead to an additional condition $\sigma_1^0 > \sigma_2^0$ (consistent with simulations of Figs. \ref{fig:bilinear_sim_hysterons} and \ref{fig:bilinear_sim_hysterons_SL}). The latter corresponds to impose that the \textit{midpoint} of state $(-+)$ is larger than the \textit{midpoint} of state $(+-)$; where the \textit{midpoint} of state $S$ writes $\textrm{mp}(S) = (H^+(S) + H^-(S))/2$. Another way to phrase it is that, within the gap, the range of $H$ where $L=4$ self-loops emerge scale like $\sigma_1^0 - \sigma_2^0$. \\

\paragraph{Self-loop $(--) \rightarrow (-+) \rightarrow (++) \rightarrow (+-) \rightarrow \dots$}
~\\

Here, we derive the conditions under which the self-loop SL$_6$ emerges (Fig. \ref{fig:bilinear_SL}).
By exchanging the roles of elements $1$ and $2$ in Eqs. (\ref{eq:final_eq_loop5}), we find:
\begin{subequations} \label{eq:final_eq_loop6}
\begin{align}
 d_2 - d_1 &< - \max \left( (x_2^+ - x_1^+), (x_1^- - x_2^-), \sigma_2^0 \right), \label{eq:final_eq_loop6:1} \\
 d_1 + d_2 &< - \sigma_1^0 < 0, \label{eq:final_eq_loop6:2} \\
 d_2 &< \min \left( (x_2^- - x_1^+), (x_1^- - x_2^+) \right). \label{eq:final_eq_loop6:3}
\end{align}
\end{subequations}
Similarly to above, taking into account race conditions leads to the additional condition $\sigma_2^0 > \sigma_1^0$.
\subsection{The case of symmetric elements}
Finally, we focus on the case of symmetrically-coupled bistable elements, i.e., on equal force jumps $d_1 = d_2 =d$. The case with positive force jumps $d > 0$ corresponds to the purely dissipative case \cite{liu2024controlled}, where self-loops are forbidden. However, it is interesting to ask if self-loops can emerge in the active case, i.e. $d < 0$, but focusing on the case where the mapping still holds, i.e. $d > - \min_i \sigma_i^0$ (otherwise, at least one element maps with a hysteron with a negative span). The question of what happens when the mapping breaks down is left for future research.

As expected from the mapping ($c_{ij} = -d_j = -d$), when the force jumps are the same, the physical system maps to a hysteron model with a symmetric interaction matrix $c_{ij}$. Therefore, self-loops are forbidden when $d_1 = d_2$, even though the system is active, highlighting that energy injection is a necessary but not sufficient condition for the emergence of self-loops.

\begin{figure}[t!]
\centering
\hspace*{-0.5cm}
\begin{tikzpicture}

\node[] at (-4.5,-5.0) {\includegraphics[width=4.4cm]{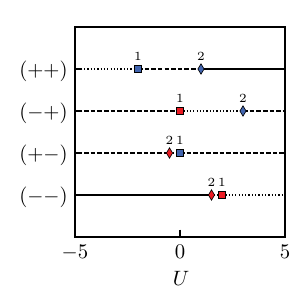}};

\node[] at (-4.7,-8.5) {\includegraphics[width=5.0cm]{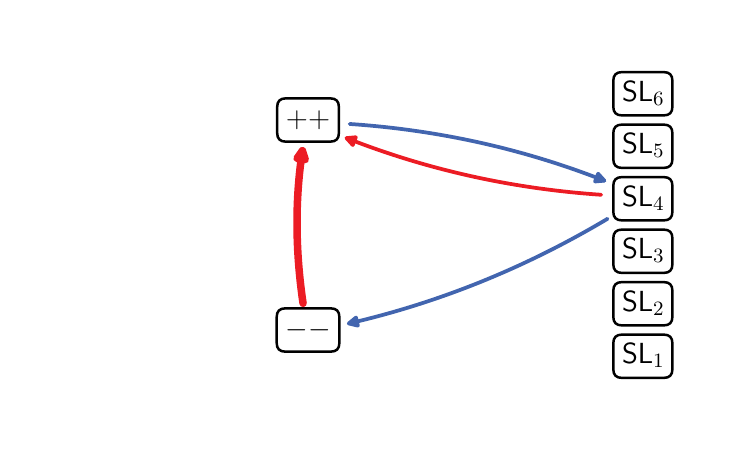}};

\draw[] (-6.6,-9.85) rectangle (-2.2,-2.9);

\node[] at (0.5,-5.0) {\includegraphics[width=4.4cm]{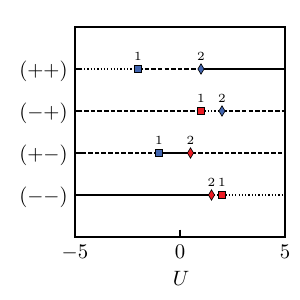}};

\node[] at (0.3,-8.5) {\includegraphics[width=5.0cm]{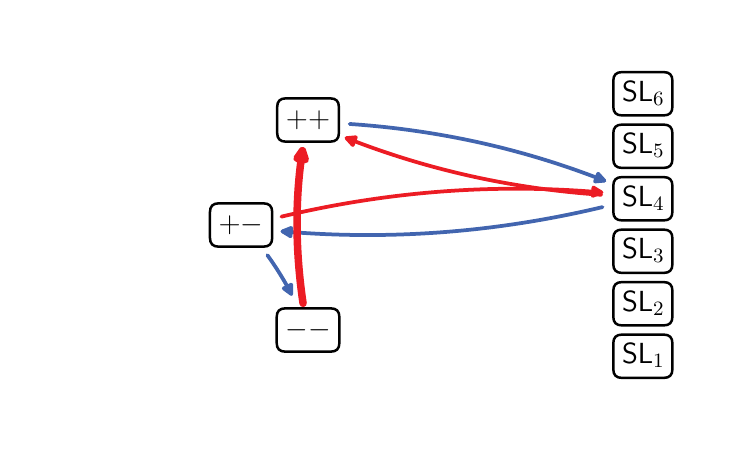}};

\draw[] (-1.6,-9.85) rectangle (2.8,-2.9);

\node[] at (5.5,-5.0) {\includegraphics[width=4.4cm]{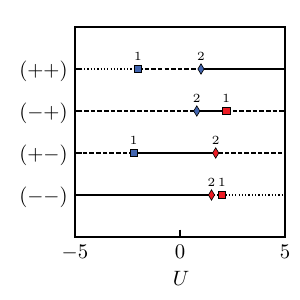}};

\node[] at (5.3,-8.5) {\includegraphics[width=5.0cm]{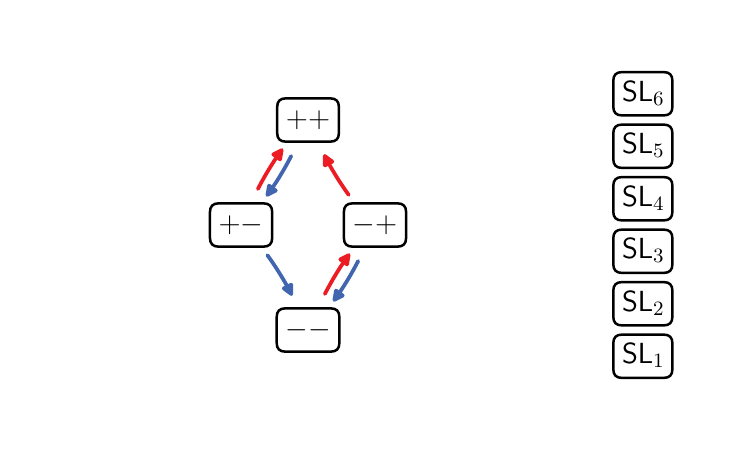}};

\draw[] (3.4,-9.85) rectangle (7.8,-2.9);

\fill[white] (-4.26,-7.09) rectangle (-3.86,-6.69);
\node[] at (-4.06,-6.89) {\small $H$};

\fill[white] (0.74,-7.09) rectangle (1.14,-6.69);
\node[] at (0.94,-6.89) {\small $H$};

\fill[white] (5.74,-7.09) rectangle (6.14,-6.69);
\node[] at (5.94,-6.89) {\small $H$};

\draw[->] (-6.8,-2.5) -- (8.0,-2.5);
\node[] at (8.25,-2.48) {\small $d$};

\draw[] (3.1,-2.4) -- (3.1,-2.6);
\node[] at (3.1,-2.2) {\small $-\min_i \sigma_i^0$};

\draw[] (5.6,-2.5) -- (5.6,-2.9);
\draw[] (0.6,-2.5) -- (0.6,-2.9);
\draw[] (-4.4,-2.5) -- (-4.4,-2.9);

\node[] at (-6.3,-9.52) {\small (i)};
\node[] at (-1.25,-9.52) {\small (ii)};
\node[] at (3.8,-9.52) {\small (iii)};

\end{tikzpicture}
\vspace*{-0.4cm}
\caption{\small{\textbf{Presence of different self-loops in the case of symmetrically-coupled hysteretic bilinear elements.} From left to right:$d = -1.0, -0.5, 0.1$, with $d_1 = d_2 = d$. In the example, $x_1^+ = 1$, $x_1^- = -1$, $x_2^+ = 0.75$, $x_2^- = 0.5$, and the transition between case (ii) and (iii) happens precisely at $d = - \min_i \sigma_i^0 = -0.25$.}}
\label{fig:bilinear_sim_hysterons_SL_symmetric}
\end{figure}

This is confirmed by exploring the transition graphs that emerge, restricting to $d_1 = d_2$ (Fig. \ref{fig:bilinear_sim_hysterons_SL_symmetric}). Varying the driving parameter $U$ back-and-forth, we never find self-loops when $d > - \min_i \sigma_i^0$. However, as soon as the mapping breaks down, self-loops can emerge in the transition-graph.

\section{Strictly self-loop free ensembles} \label{sec:well_behaved_demo}

In this section, we discuss the different classes of strictly well-behaved hysteron model ensembles.

\subsection{Symmetric interactions}

\begin{figure}[t!]
\centering
\begin{tikzpicture}

\tikzstyle{block} = [draw, fill=orange!80, rectangle, minimum height=2em, minimum width=3em]


\draw[very thick] (0.5,-0.3) -- (0.5,0.4); 
\foreach \i in {0.25, 0.4, 0.55, 0.7, 0.85} {
    \draw[thick] (0.2, \i-0.8) -- (0.5, \i-0.5); 
    }

\node[block] (S1) at (1.7, 0) {$s_1$};
\node[block] (S2) at (3.4, 0) {$s_2$};
\node[block] (S3) at (6.6, 0) {$s_N$};

\node[rotate=0] at (5.0,0.0) {\small $\dots$};

\draw[-, thick] (0.5, 0.0) -- (S1);
\draw[-, thick] (S1.east) -- (S2.west);
\draw[-, thick] (S2.east) -- (4.5,0.0);
\draw[-, thick] (5.5,0.0) -- (S3.west);
\draw[->, thick] (S3.east) -- (7.9,0.0);


\draw[->, thick] (6.6,-3.4) -- (7.4,-3.4);
\node[rotate=0] at (7.7,-3.4) {\small $F$};

\draw[-, thick] (6.5,-3.3) -- (6.6,-3.4);
\draw[-, thick] (6.3,-3.5) -- (6.5,-3.3);
\draw[-, thick] (6.1,-3.3) -- (6.3,-3.5);
\draw[-, thick] (5.9,-3.5) -- (6.1,-3.3);
\draw[-, thick] (5.7,-3.3) -- (5.9,-3.5);
\draw[-, thick] (5.5,-3.5) -- (5.7,-3.3);
\draw[-, thick] (5.4,-3.4) -- (5.5,-3.5);

\node[rotate=0] at (6.0,-2.95) {\small $k$};

\draw[-, thick] (5.0,-3.4) -- (5.4,-3.4);

\draw[-, thick] (5.0,-2.1) -- (5.0,-4.7);
\draw[-, dashed] (5.0,-1.9) -- (5.0,-1.2);

\node[block] (S1b) at (3.9,-2.1) {$s_1$};
\node[block] (S2b) at (3.9,-3.1) {$s_2$};
\node[rotate=90] at (3.9,-3.9) {\small $\dots$};
\node[block] (S3b) at (3.9,-4.7) {$s_N$};

\draw[-, thick] (S1b.east) -- (5.0,-2.1);
\draw[-, thick] (S1b.west) -- (2.8,-2.1);

\draw[-, thick] (S2b.east) -- (5.0,-3.1);
\draw[-, thick] (S2b.west) -- (2.8,-3.1);

\draw[-, thick] (S3b.east) -- (5.0,-4.7);
\draw[-, thick] (S3b.west) -- (2.8,-4.7);

\draw[-, thick] (2.8,-2.1) -- (2.8,-4.7);
\draw[-, dashed] (2.8,-1.9) -- (2.8,-1.2);

\draw[-, dashed] (2.8,-4.9) -- (2.8,-5.6);
\draw[-, dashed] (7.0,-3.6) -- (7.0,-5.6);

\draw[<->, thick] (6.9,-5.4) -- (2.9,-5.4);
\node[rotate=0] at (4.9,-5.7) {\small $U$};

\draw[<->, thick] (2.9,-1.4) -- (4.9,-1.4);
\node[rotate=0] at (3.9,-1.1) {\small $U_{\parallel}$};

\draw[-, thick] (2.2,-3.4) -- (2.8,-3.4);

\draw[very thick] (2.2,-3.75) -- (2.2,-3.05); 
\foreach \i in {0.25, 0.4, 0.55, 0.7, 0.85} {
    \draw[thick] (1.9, \i-4.25) -- (2.2, \i-3.95); 
    }

\node[rotate=0] at (3.5,0.9) {\small (b)};
\node[rotate=0] at (1.9,-1.8) {\small (c)};

\node[rotate=0] at (10.0,-3.5) {\includegraphics[height=4.5cm]{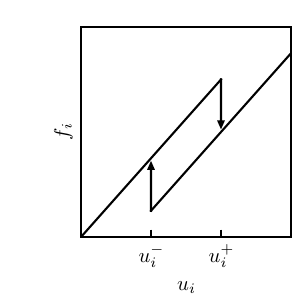}};

\node[rotate=0] at (10.5,-1.3) {\small (d)};

\node[rotate=0] at (11.3,-2.6) {\small $d_i$};
\node[rotate=45] at (9.5,-3.87) {\small $\propto u_i$};

\node[rotate=0] at (-1.7,-3.5) {\includegraphics[height=4.5cm]{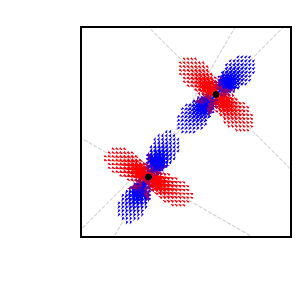}};

\node[rotate=0] at (-1.2,-1.3) {\small (a)};

\draw[<->, thick] (-1.25,-3.8) to[out=35, in=-100] (-0.7,-3.1);
\node[rotate=0] at (-0.27,-3.7) {\small $c_{ij} = c_{ji}$};

\end{tikzpicture}
\vspace*{-0.0cm}
\caption{\small{\textbf{Realizations of the different interaction classes.} (a) Symmetric interactions: two interacting soft spots, associated with quadrupolar displacement fields, interact symmetrically: blue arrows: inward displacements; red arrows: outward displacements. In this example, the orientations of the quadrupoles are such that $c_{ij} = c_{ji} < 0$ due to the incompatibility of the quadrupolar fields. (b) Constant-columns interactions: linear chain of mechanical hysterons. (c) Ferromagnetic interactions: parallel arrangement of mechanical hysterons in series with a spring of stiffness $k$ and zero-rest-length. (d) Bilinear force-displacement curve.}}
\label{fig:geometries}
\end{figure}

Symmetric interactions are reminiscent of dilute interacting soft spots, as discussed in
\cite{keim2021multiperiodic}. Soft spots correspond to local rearrangements associated with quadrupolar Eshelby-like displacement fields (Fig. \ref{fig:geometries}-a). In this context, the binary elements are modeled as having the same sizes, which implies $c_{ij} = c_{ji}$, and the sign of the interaction coefficient is given by their relative orientations. When compatible lobes of the fields face each other, the interaction is rather ferromagnetic ($c_{ij} > 0$), while it is rather antiferromagnetic ($c_{ij} < 0$) when incompatible lobes face each other.

\subsection{Constant-columns interactions}

Constant-columns interactions occur in a linear chain of bistable springs (Fig. \ref{fig:geometries}-b), if we assume that the force-displacement curve of such two-state elements is given by a bilinear relation:

\begin{equation}
 f_i = x_i - d_i s_i,
\end{equation}

\noindent with $f_i$ the force, $x_i$ the displacement, $d_i$ the force jump and $s_i$ the state (Fig. \ref{fig:geometries}-d) \cite{liu2024controlled, shohat2025geometric}.

\subsection{Ferromagnetic interactions}

Ferromagnetic interactions occur, for example, in a parallel arrangement of bistable springs in series with a harmonic spring of stiffness $k$ \cite{shohat2025geometric} (Fig. \ref{fig:geometries}-d). The mapping then is $c_{ij} = d_j/k$  with $d_j > 0$, where $H = U$ is the total displacement \cite{shohat2025geometric}. Note that a collection of bistable springs arranged in parallel as represented in Fig. \ref{fig:geometries}-c interact ferromagnetically, and the interaction matrix has a constant-columns structure.

\subsection{Constant-rows interactions}

The case of constant-rows interactions is relatively artificial as it corresponds to a case where hysterons are affected by the flipping of any other hysteron in exactly the same way. We do not know if this interaction matrix can be derived from mechanical equilibrium of mechanical two-state elements.

\section{Role of the race condition rules} \label{sec:rac_condition_rules}

In this section, we analyze the impact of different race condition rules on the probability of finding self-loops and on the self-loop size distribution. We explore four different dynamical rules. Rule $0$ considers the model ill-defined whenever race conditions occur \cite{van2021profusion, liu2024controlled} (discarding the associated instance), and prevents sampling large systems (see section \ref{sec:race_conditions}). Rule $1$ (focus of the main text) and $1'$ specify to flip only the most and the least unstable element, respectively. Finally, under rule $2$, all unstable elements are flipped simultaneously. Table \ref{table:all_models} summarizes the ill-defined conditions encountered in each interaction ensemble for the different race condition rules.

\begin{table}[h!]
\small
\vspace*{0.35cm}
\begin{center}
\renewcommand{\arraystretch}{1.5}
\begin{tabular}{|p{.21\textwidth}|p{.085\textwidth}|p{.07\textwidth}|p{.07\textwidth}|p{.07\textwidth}|}
\hline
\textbf{Interaction ensemble} & \textbf{rule} $\boldsymbol{0}$ & \textbf{rule} $\boldsymbol{1}$ & \textbf{rule} $\boldsymbol{1'}$ & \textbf{rule} $\boldsymbol{2}$ \\
\hline
Arbitrary & RC/G/SL & G/SL & G/SL & G/SL$^{\dagger}$ \\
\hline
Symmetric & RC & -- & -- & SL$^*$ \\
\hline
Constant-columns & RC & -- & -- & SL$^*$ \\
\hline
Constant-rows & RC & -- & -- & SL$^*$ \\
\hline
Ferromagnetic & RC & -- & -- & -- \\
\hline
\end{tabular}
\end{center}
\begin{flushleft}
$^*$ $L = 2$ self-loops only. $^{\dagger}$ All $L \geq 2$ self-loops allowed.
\end{flushleft}
\vspace{-0.4cm}
\caption[table caption]{\small{Summary of the occurrence of ill-defined conditions for various interaction ensembles and race condition rules: RC: race conditions; G: gaps; SL: self-loops.}}
\label{table:all_models}
\end{table}

\subsection{Arbitrarily-coupled hysterons}

\begin{figure}[b!]
\hspace*{-0.3cm}
\begin{tikzpicture}

\node[rotate=0] at (4.6,-2.5) {\includegraphics[height=4.3cm]{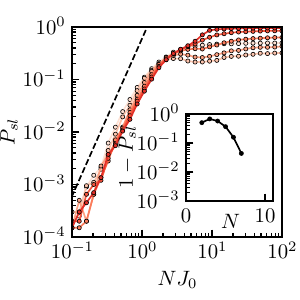}};
\node[rotate=0] at (9.2,-2.5) {\includegraphics[height=4.3cm]{selfLoopGlobal_scaled_hysterons.pdf}};
\node[rotate=0] at (13.8,-2.5) {\includegraphics[height=4.3cm]{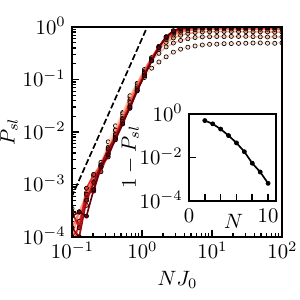}};
\node[rotate=0] at (18.4,-2.5) {\includegraphics[height=4.3cm]{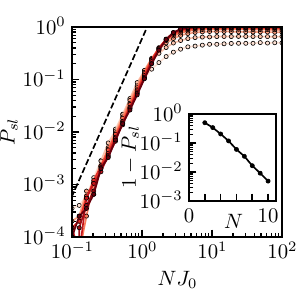}};

\node[rotate=0] at (5.1,-0.4) {\small \textbf{Rule} $\boldsymbol{0}$};
\node[rotate=0] at (9.7,-0.4) {\small \textbf{Rule} $\boldsymbol{1}$};
\node[rotate=0] at (14.3,-0.4) {\small \textbf{Rule} $\boldsymbol{1'}$};
\node[rotate=0] at (18.9,-0.4) {\small \textbf{Rule} $\boldsymbol{2}$};

\node[rotate=0] at (3.87,-1.07) {\small (a)};
\node[rotate=0] at (8.47,-1.07) {\small (b)};
\node[rotate=0] at (13.07,-1.07) {\small (c)};
\node[rotate=0] at (17.67,-1.07) {\small (d)};

\node[rotate=0] at (4.6,-6.75) {\includegraphics[height=4.3cm]{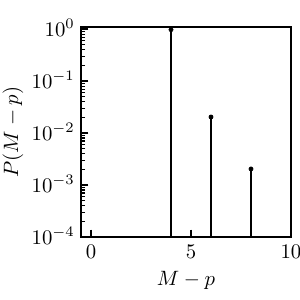}};
\node[rotate=0] at (9.2,-6.75) {\includegraphics[height=4.3cm]{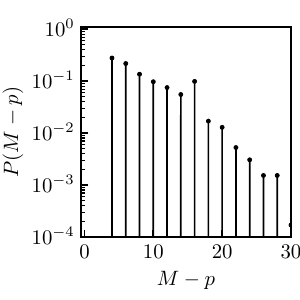}};
\node[rotate=0] at (13.8,-6.75) {\includegraphics[height=4.3cm]{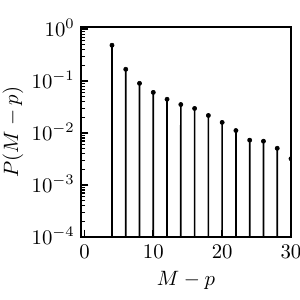}};
\node[rotate=0] at (18.4,-6.75) {\includegraphics[height=4.3cm]{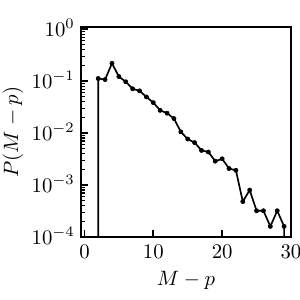}};

\fill[white] (4.68,-8.79) rectangle (5.57,-8.45);
\node[rotate=0] at (5.13,-8.6) {\small $L$};

\fill[white] (9.28,-8.79) rectangle (10.17,-8.45);
\node[rotate=0] at (9.73,-8.6) {\small $L$};

\fill[white] (13.88,-8.79) rectangle (14.77,-8.45);
\node[rotate=0] at (14.33,-8.6) {\small $L$};

\fill[white] (18.48,-8.79) rectangle (19.37,-8.45);
\node[rotate=0] at (18.93,-8.6) {\small $L$};

\fill[white] (2.45,-7.19) rectangle (2.87,-5.83);
\node[rotate=90] at (2.64,-6.51) {\small $P(L)$};

\fill[white] (7.05,-7.19) rectangle (7.47,-5.83);
\node[rotate=90] at (7.24,-6.51) {\small $P(L)$};

\fill[white] (11.65,-7.19) rectangle (12.07,-5.83);
\node[rotate=90] at (11.84,-6.51) {\small $P(L)$};

\fill[white] (16.25,-7.19) rectangle (16.67,-5.83);
\node[rotate=90] at (16.44,-6.51) {\small $P(L)$};

\node[rotate=0] at (6.22,-5.33) {\small (e)};
\node[rotate=0] at (10.82,-5.33) {\small (f)};
\node[rotate=0] at (15.42,-5.33) {\small (g)};
\node[rotate=0] at (20.02,-5.33) {\small (h)};

\end{tikzpicture}
\vspace*{-0.9cm}
\caption{\small{\textbf{Self-loop statistics for arbitrarily-coupled hysterons and different race condition rules.} (a-d) Probability $P_{sl}$ of at least one self-loops occurring for any $H$ as a function of $NJ_0$, for different $N \in \left[ 2, 3, \dots, 10 \right]$, color-coded from light to dark red as $N$ increases; the black dashed lines represent the slope $3$; (inset) probability to be self-loop free $1 - P_{sl}$ as a function of $N$ in the large coupling limit ($NJ_0 = 10^2$). (e-h) Self-loops size distributions for fixed $NJ_0 = 20$ and $N = 8$. (a/e) Rule $0$; (b/f) rule $1$; (c/g) rule $1'$; (d/h) rule $2$.}}
\label{fig:arbitrary_differentRules}
\end{figure}

Here, we examine different race condition rules for hysterons with random asymmetric interactions. First, in the case where all unstable hysterons flip simultaneously (rule $2$), we find that all self-loop sizes are allowed starting from $L=2$ (Fig. \ref{fig:arbitrary_differentRules}-h).
In contrast, all race conditions rules which involve flipping elements one by one (rules $0$, $1$, and $1'$) lead to self-loops whose sizes are always even, starting from $L=4$ (Figs. \ref{fig:arbitrary_differentRules}-e to g). This parity is expected given all the hysterons involved in a self-loop must first flip and then unflip to revisit a state.

Let us first show that self-loops of size $L=2$ are forbidden when hysterons flip one-by-one, because they correspond to loops involving a single hysterons with negative span. Let us consider that state $S$ is unstable at a given value of the drive $H$ because hysteron $i$ is unstable, e.g. $H > H_i^{+}(S)$ (without loss of generality). The system enters a self-loop of size $2$ if hysteron $i$ remains unstable after the snap, i.e. $H < H_i^{-}(S)$, which yields:
\begin{equation} \label{eq:condition_self-loop_2}
h_{i}^{-} - \sum_{j \neq i} c_{ij} s_{j} > h_{i}^{+} - \sum_{j \neq i} c_{ij} s_{j}.
\end{equation}
All hysterons $j \neq i$ being unchanged, Eq. (\ref{eq:condition_self-loop_2}) implies $h_{i}^{-} > h_{i}^{+}$, which corresponds to hysteron $i$ having a negative span, i.e. $\sigma_i < 0$, which is forbidden (excluded from our sampling).

Interestingly, for all race condition rules, the probability $P_{sl}$ of finding at least one self-loop increases as a power law for $NJ_0 \ll 1$, and saturates toward a constant value with $J_0$ for $NJ_0 \gg 1$ (Figs. \ref{fig:arbitrary_differentRules}-a to d). Similar dependencies have been observed for other properties of interacting hysterons \cite{van2021profusion, lindeman2021multiple, lindeman2025generalizing}. Except for rule $0$, the large-$J_0$ plateau value increases monotonically with $N$, and in all cases the data suggest that $P_{sl} \rightarrow 1$ when $N$ increases. Moreover, in all cases, the size of self-loops seems to be exponentially distributed, modulo the specific constraints on the self-loop sizes discussed above.

\subsection{Well-behaved ensembles}

\begin{figure}[b!]
\hspace*{-0.25cm}
\begin{tikzpicture}

\node[rotate=0] at (4.6,-2.5) {\includegraphics[height=4.3cm]{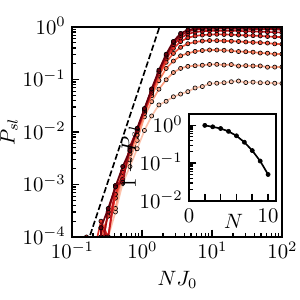}};
\node[rotate=0] at (9.2,-2.5) {\includegraphics[height=4.3cm]{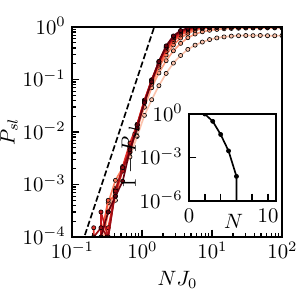}};
\node[rotate=0] at (13.8,-2.5) {\includegraphics[height=4.3cm]{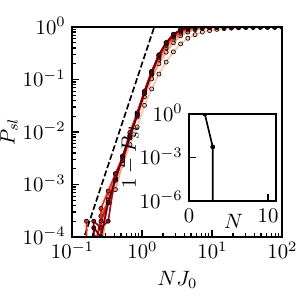}};

\node[rotate=0] at (5.1,-0.4) {\small \textbf{Symmetric}};
\node[rotate=0] at (9.7,-0.4) {\small \textbf{Constant-columns}};
\node[rotate=0] at (14.3,-0.4) {\small \textbf{Constant-rows}};

\node[rotate=0] at (3.85,-1.05) {\small (a)};
\node[rotate=0] at (8.45,-1.05) {\small (b)};
\node[rotate=0] at (13.05,-1.05) {\small (c)};

\end{tikzpicture}
\vspace*{-0.5cm}
\caption{\small{\textbf{Self-loop statistics for well-behaved models of coupled hysterons}, where race conditions are resolved with rule $2$. Statistical measures for self-loops scale when
plotted as function of $NJ_0$ and dominate for $NJ_0 \gg 1$ ($10^5$ samples; color from light to dark as $N$ increases from $2$ to $10$).
Probability $P_{sl}$ of finding at least one self-loop for any value of $H$ (dashed line indicates slope $4$).
Inset: The probability to be self-loop free, $1-P_{sl}$, decays to zero with $N$ for large couplings ($NJ_0 = 10^2$). (a) Symmetric interactions; (b) constant-columns interactions; (c) constant-rows interactions.}}
\label{fig:well_behaved_rule2}
\end{figure}

Here, we focus on the well-behaved models introduced in the main text, namely symmetric, constant-columns, constant-rows, and ferromagnetic interactions. We have shown that self-loops are forbidden for all these classes when hysterons flip one by one (rules $0$, $1$ and $1'$). Here, we analyze the probability of finding self-loops and the self-loop size distributions when race conditions are resolved with rule $2$.

Let us first focus on symmetric, constant-columns, and constant-rows interactions. We find that all self-loops have size $L=2$, independently of the model (not shown here). The presence of $2$-cycles is a well-known feature for symmetrically-coupled spins where all unstable spins flip simultaneously \cite{hopfield1982neural, nutzel1993subtle, eissfeller1994mean}, which is generally called synchronous or parallel update in the context of spin glasses and neural networks. Second, the probability $P_{sl}$ of finding at least one $L=2$ self-loop for any $H$ increases as a power law for $NJ_0 \ll 1$, and saturates toward a constant value with $J_0$ for $NJ_0 \gg 1$ (Figs. \ref{fig:well_behaved_rule2}), which asymptotes to $1$ as $N$ increases (Figs. \ref{fig:well_behaved_rule2}-insets)

Finally, for purely ferromagnetic interactions, we find no self-loops when race conditions are resolved with rule $2$, which suggests that the $2$-cycles emerging with rule $2$ are more related to the presence of antiferromagnetic interactions than to non-symmetric ones, which is very different than in the case where hysterons flip one by one.

\section{The combinatorics and structure of self-loops}
\label{sec:selfLoop_structure}

Transition graphs have emerged as a powerful framework for capturing and studying sequential responses \cite{paulsen2019minimal,mungan2019networks,van2021profusion,teunisse2024transition}. In these, states $S$ are represented by nodes, and their transitions under zero-temperature, quasistatic driving with a global field $H$, form directed edges (Fig. 1-c of the main text, and Figs. \ref{fig:cycles_ensemble}). The range of stability of state $S$ is given by its upper and lower switching fields $H^{\pm}(S)$ (which follow from the extrema of $H^{\pm}_i (S)$, see main text), and an up or down transition is initiated when $H > H^{+}(S)$ or $H < H^{-}(S)$.
Here, we use this framework to represent the structure of self-loops, focusing on the repeating cycle of their unstable states.

\subsection{Structure of self-loops}

Here we discuss how to count the number of fundamental self-loop structures of size $L$ and involving $n_e$ elements, \( M(n_{e}, L) \), irrespective of whether they are realizable in a certain interaction ensemble.
We first determine all potential self-loops using a naive procedure based on
the combinatorics of flip sequences, which generates cycles of length $L$ on a ($n_e$)-hypercube. However, many of these are related under two trivial symmetries---relabeling and timeshifting (see below)---and in the second step we determine the fundamental self-loops, i.e., those that can not be mapped onto each other by trivial symmetries.

\begin{figure}[b!]
\hspace*{-0.2cm}
\begin{tikzpicture}

\node[rotate=0] at (-0.1,0.0) {\includegraphics[height=3.0cm]{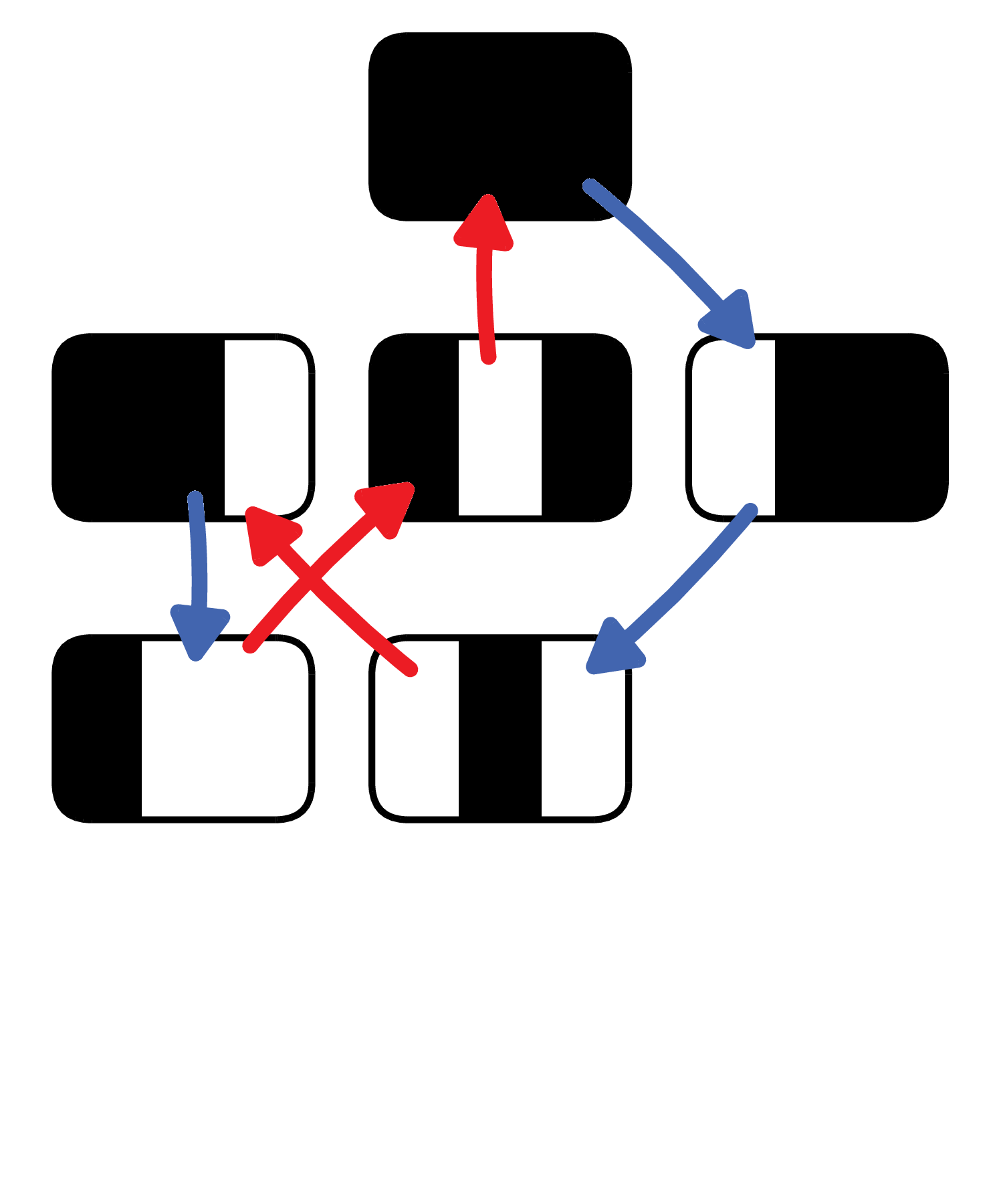}};
\node[rotate=0] at (5.1,0.0) {\includegraphics[height=3.0cm]{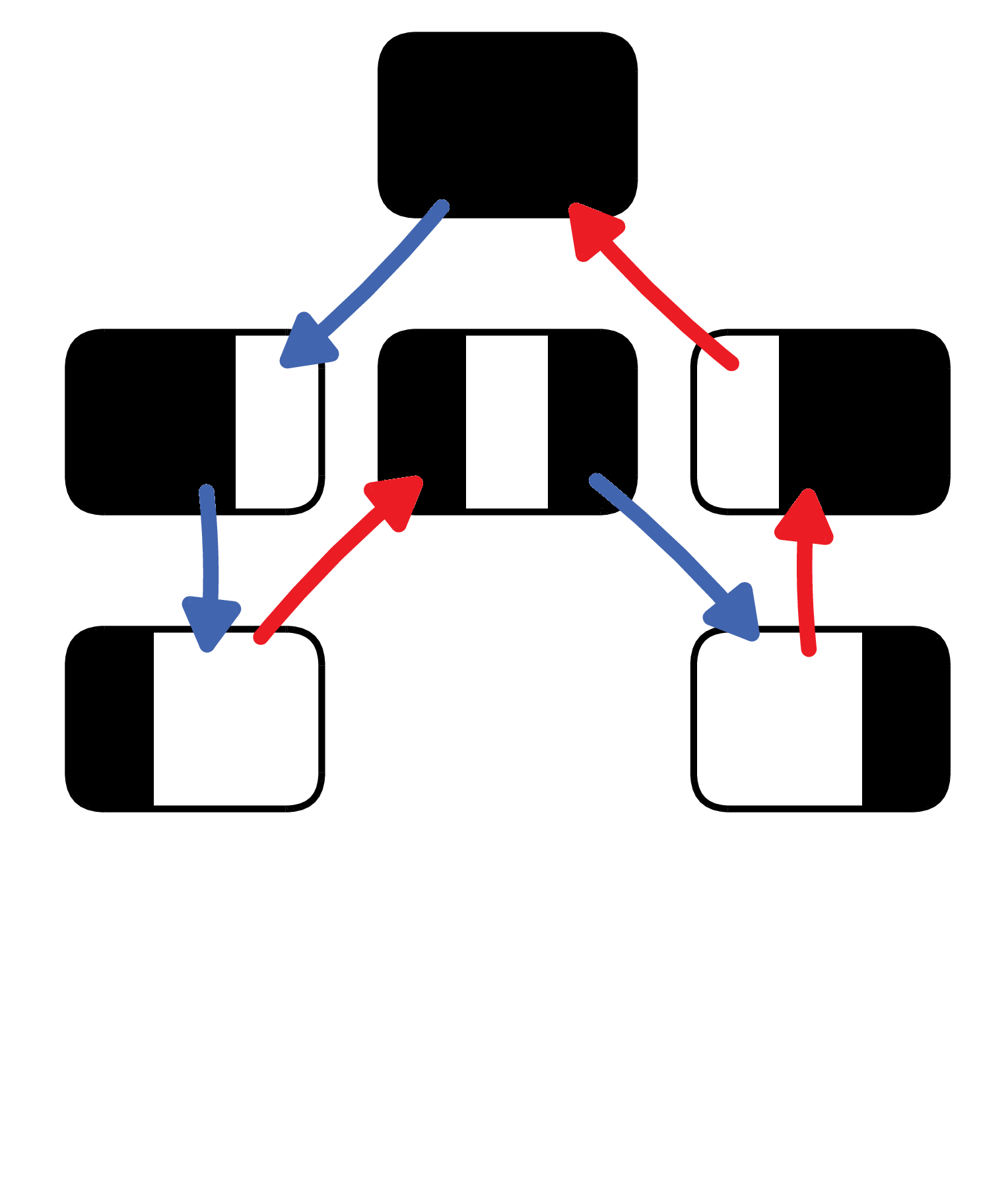}};

\node[rotate=0] at (-0.1,1.18) {\small \scalebox{0.64}{$\color{white}\boldsymbol{+}\!\color{white}\boldsymbol{+}\!\color{white}\boldsymbol{+}$}};

\node[rotate=0] at (-0.1,0.43) {\small \scalebox{0.64}{$\color{white}\boldsymbol{+}\!\color{black}\boldsymbol{-}\!\color{white}\boldsymbol{+}$}};
\node[rotate=0] at (0.68,0.43) {\small \scalebox{0.64}{$\color{black}\boldsymbol{-}\!\color{white}\boldsymbol{+}\!\color{white}\boldsymbol{+}$}};
\node[rotate=0] at (-0.895,0.43) {\small \scalebox{0.64}{$\color{white}\boldsymbol{+}\!\color{white}\boldsymbol{+}\!\color{black}\boldsymbol{-}$}};

\node[rotate=0] at (-0.1,-0.325) {\small \scalebox{0.64}{$\color{black}\boldsymbol{-}\!\color{white}\boldsymbol{+}\!\color{black}\boldsymbol{-}$}};
\node[rotate=0] at (-0.900,-0.325) {\small \scalebox{0.64}{$\color{white}\boldsymbol{+}\!\color{black}\boldsymbol{-}\!\color{black}\boldsymbol{-}$}};

\node[rotate=0] at (5.1,1.18) {\small \scalebox{0.64}{$\color{white}\boldsymbol{+}\!\color{white}\boldsymbol{+}\!\color{white}\boldsymbol{+}$}};

\node[rotate=0] at (5.1,0.43) {\small \scalebox{0.64}{$\color{white}\boldsymbol{+}\!\color{black}\boldsymbol{-}\!\color{white}\boldsymbol{+}$}};
\node[rotate=0] at (5.895,0.43) {\small \scalebox{0.64}{$\color{black}\boldsymbol{-}\!\color{white}\boldsymbol{+}\!\color{white}\boldsymbol{+}$}};
\node[rotate=0] at (4.315,0.43) {\small \scalebox{0.64}{$\color{white}\boldsymbol{+}\!\color{white}\boldsymbol{+}\!\color{black}\boldsymbol{-}$}};

\node[rotate=0] at (4.305,-0.325) {\small \scalebox{0.64}{$\color{white}\boldsymbol{+}\!\color{black}\boldsymbol{-}\!\color{black}\boldsymbol{-}$}};
\node[rotate=0] at (5.905,-0.325) {\small \scalebox{0.64}{$\color{black}\boldsymbol{-}\!\color{black}\boldsymbol{-}\!\color{white}\boldsymbol{+}$}};

\draw[<->] (1.6,0.2) -- (3.4,0.2);
\node[rotate=0] at (2.5,0.55) {\small relabeling};

\node[rotate=0] at (-0.1,-1.0) {\small $(2,1,0,2,0,1)$};
\node[rotate=0] at (-0.1,-1.45) {\small $S^0 = (+1,-1,-1)$};

\node[rotate=0] at (5.1,-1.0) {\small $(1,0,2,1,2,0)$};
\node[rotate=0] at (5.1,-1.45) {\small $S^0 = (-1,-1,+1)$};

\end{tikzpicture}
\vspace*{-0.35cm}
\caption{\small{\textbf{Relabeling symmetry.} Each self-loop is defined by the sequence of flips and the starting state $S^0$. Two self-loops are equivalent if one can be obtained from the other by a relabeling of the hysterons.}}
\label{fig:margot1}
\end{figure}

\begin{figure}[t!]
\hspace*{-0.2cm}
\begin{tikzpicture}

\draw[dashed] (-1.4,-0.33) -- (1.3,-0.33);
\node[rotate=0] at (0.9,-0.14) {\scriptsize $m = -1$};

\draw[dashed] (3.8,0.43) -- (7.3,0.43);
\node[rotate=0] at (6.9,0.63) {\scriptsize $m = +1$};

\draw[dashed] (9.7,0.43) -- (13.2,0.43);
\node[rotate=0] at (12.8,0.63) {\scriptsize $m = +1$};

\node[rotate=0] at (-0.1,0.0) {\includegraphics[height=3.0cm]{self_loop_6_bis_3.png}};
\node[rotate=0] at (5.1,0.0) {\includegraphics[height=3.0cm]{self_loop_6_bis_3.png}};
\node[rotate=0] at (11.0,0.0) {\includegraphics[height=3.0cm]{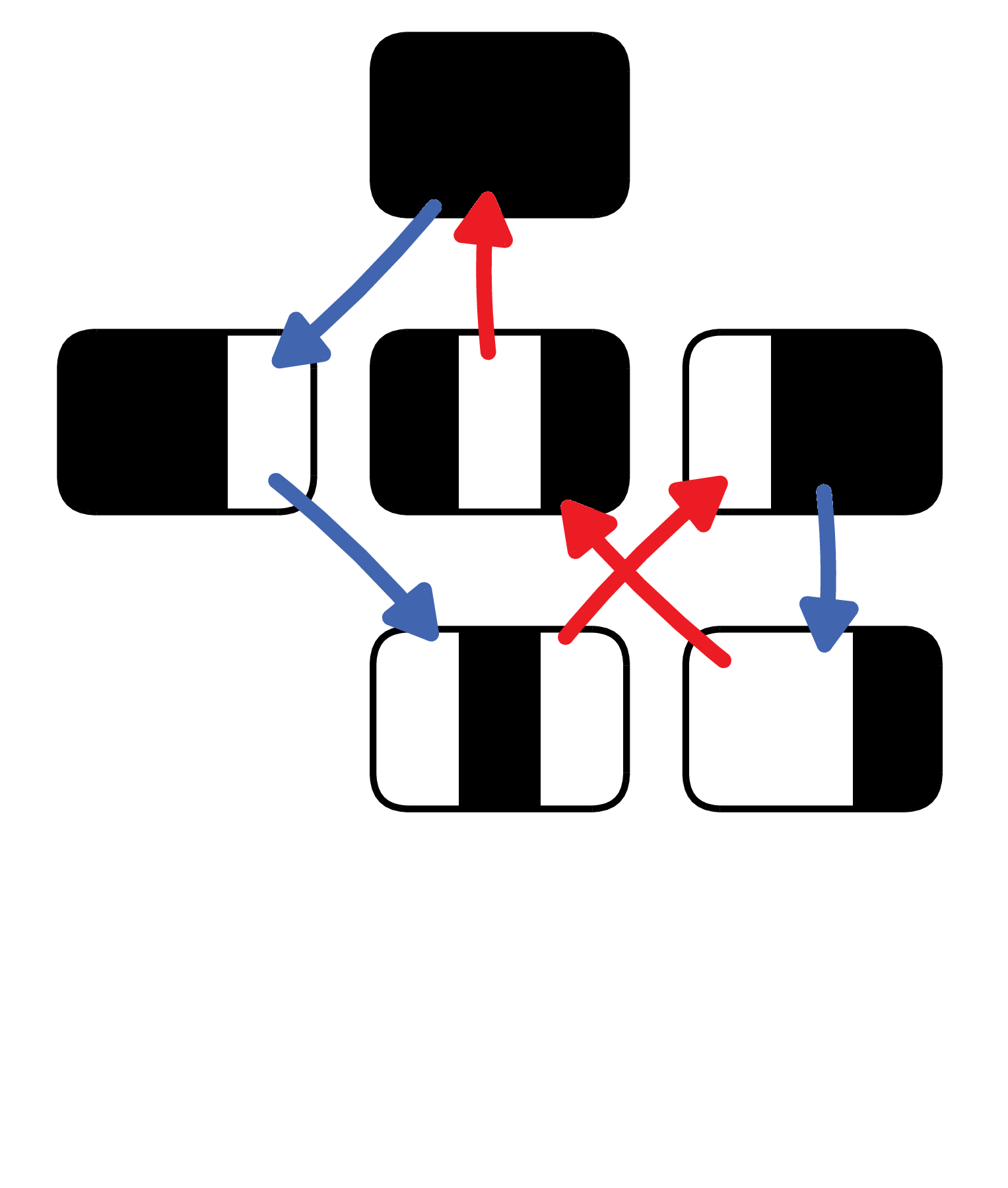}};

\node[rotate=0] at (-0.1,1.18) {\small \scalebox{0.64}{$\color{white}\boldsymbol{+}\!\color{white}\boldsymbol{+}\!\color{white}\boldsymbol{+}$}};

\node[rotate=0] at (-0.1,0.43) {\small \scalebox{0.64}{$\color{white}\boldsymbol{+}\!\color{black}\boldsymbol{-}\!\color{white}\boldsymbol{+}$}};
\node[rotate=0] at (0.68,0.43) {\small \scalebox{0.64}{$\color{black}\boldsymbol{-}\!\color{white}\boldsymbol{+}\!\color{white}\boldsymbol{+}$}};
\node[rotate=0] at (-0.895,0.43) {\small \scalebox{0.64}{$\color{white}\boldsymbol{+}\!\color{white}\boldsymbol{+}\!\color{black}\boldsymbol{-}$}};

\node[rotate=0] at (-0.1,-0.325) {\small \scalebox{0.64}{$\color{black}\boldsymbol{-}\!\color{white}\boldsymbol{+}\!\color{black}\boldsymbol{-}$}};
\node[rotate=0] at (-0.900,-0.325) {\small \scalebox{0.64}{$\color{white}\boldsymbol{+}\!\color{black}\boldsymbol{-}\!\color{black}\boldsymbol{-}$}};

\node[rotate=0] at (5.1,1.18) {\small \scalebox{0.64}{$\color{white}\boldsymbol{+}\!\color{white}\boldsymbol{+}\!\color{white}\boldsymbol{+}$}};

\node[rotate=0] at (5.1,0.43) {\small \scalebox{0.64}{$\color{white}\boldsymbol{+}\!\color{black}\boldsymbol{-}\!\color{white}\boldsymbol{+}$}};
\node[rotate=0] at (5.88,0.43) {\small \scalebox{0.64}{$\color{black}\boldsymbol{-}\!\color{white}\boldsymbol{+}\!\color{white}\boldsymbol{+}$}};
\node[rotate=0] at (4.305,0.43) {\small \scalebox{0.64}{$\color{white}\boldsymbol{+}\!\color{white}\boldsymbol{+}\!\color{black}\boldsymbol{-}$}};

\node[rotate=0] at (5.1,-0.325) {\small \scalebox{0.64}{$\color{black}\boldsymbol{-}\!\color{white}\boldsymbol{+}\!\color{black}\boldsymbol{-}$}};
\node[rotate=0] at (4.3,-0.325) {\small \scalebox{0.64}{$\color{white}\boldsymbol{+}\!\color{black}\boldsymbol{-}\!\color{black}\boldsymbol{-}$}};

\node[rotate=0] at (11.0,1.18) {\small \scalebox{0.64}{$\color{white}\boldsymbol{+}\!\color{white}\boldsymbol{+}\!\color{white}\boldsymbol{+}$}};

\node[rotate=0] at (11.0,0.43) {\small \scalebox{0.64}{$\color{white}\boldsymbol{+}\!\color{black}\boldsymbol{-}\!\color{white}\boldsymbol{+}$}};
\node[rotate=0] at (11.78,0.43) {\small \scalebox{0.64}{$\color{black}\boldsymbol{-}\!\color{white}\boldsymbol{+}\!\color{white}\boldsymbol{+}$}};
\node[rotate=0] at (10.205,0.43) {\small \scalebox{0.64}{$\color{white}\boldsymbol{+}\!\color{white}\boldsymbol{+}\!\color{black}\boldsymbol{-}$}};

\node[rotate=0] at (11.0,-0.325) {\small \scalebox{0.64}{$\color{black}\boldsymbol{-}\!\color{white}\boldsymbol{+}\!\color{black}\boldsymbol{-}$}};
\node[rotate=0] at (11.78,-0.325) {\small \scalebox{0.64}{$\color{black}\boldsymbol{-}\!\color{black}\boldsymbol{-}\!\color{white}\boldsymbol{+}$}};

\draw[dotted] (-0.9,-0.32) circle (0.4);
\draw[dotted] (5.09,0.44) circle (0.4);
\draw[dotted] (10.19,0.44) circle (0.4);

\draw[<->] (1.6,0.2) -- (3.4,0.2);
\node[rotate=0] at (2.5,0.55) {\small time shift};

\draw[<->] (7.5,0.2) -- (9.3,0.2);
\node[rotate=0] at (8.4,0.55) {\small convention};
\node[rotate=0] at (8.4,0.85) {\small labeling};

\node[rotate=0] at (-0.1,-1.0) {\small $(2,1,0,2,0,1)$};
\node[rotate=0] at (-0.1,-1.45) {\small $S^0 = (+1,-1,-1)$};

\node[rotate=0] at (5.1,-1.0) {\small $(1,0,2,0,1,2)$};
\node[rotate=0] at (5.1,-1.45) {\small $S^0 = (+1,-1,+1)$};

\node[rotate=0] at (11.0,-1.0) {\small $(2,1,0,1,2,0)$};
\node[rotate=0] at (11.0,-1.45) {\small $S^0 = (+1,+1,-1)$};

\end{tikzpicture}
\vspace*{-0.35cm}
\caption{\small{\textbf{Time shift symmetry between two self-loops.} Each self-loop is defined by the sequence of flips and the starting state $S^0$. Two self-loops are the same if one can be obtained from the other by arbitrarily shifting the starting state. This  ambiguity on the starting state is partially lifted by decomposing self-loops by the magnetization $m$ of their starting state, and imposing the magnetization cannot go below $m$.}}
\label{fig:margot2}
\end{figure}

\begin{figure}[t!]
\hspace*{-0.2cm}
\begin{tikzpicture}

\node[rotate=0] at (-0.1,0.0) {\includegraphics[height=3.0cm]{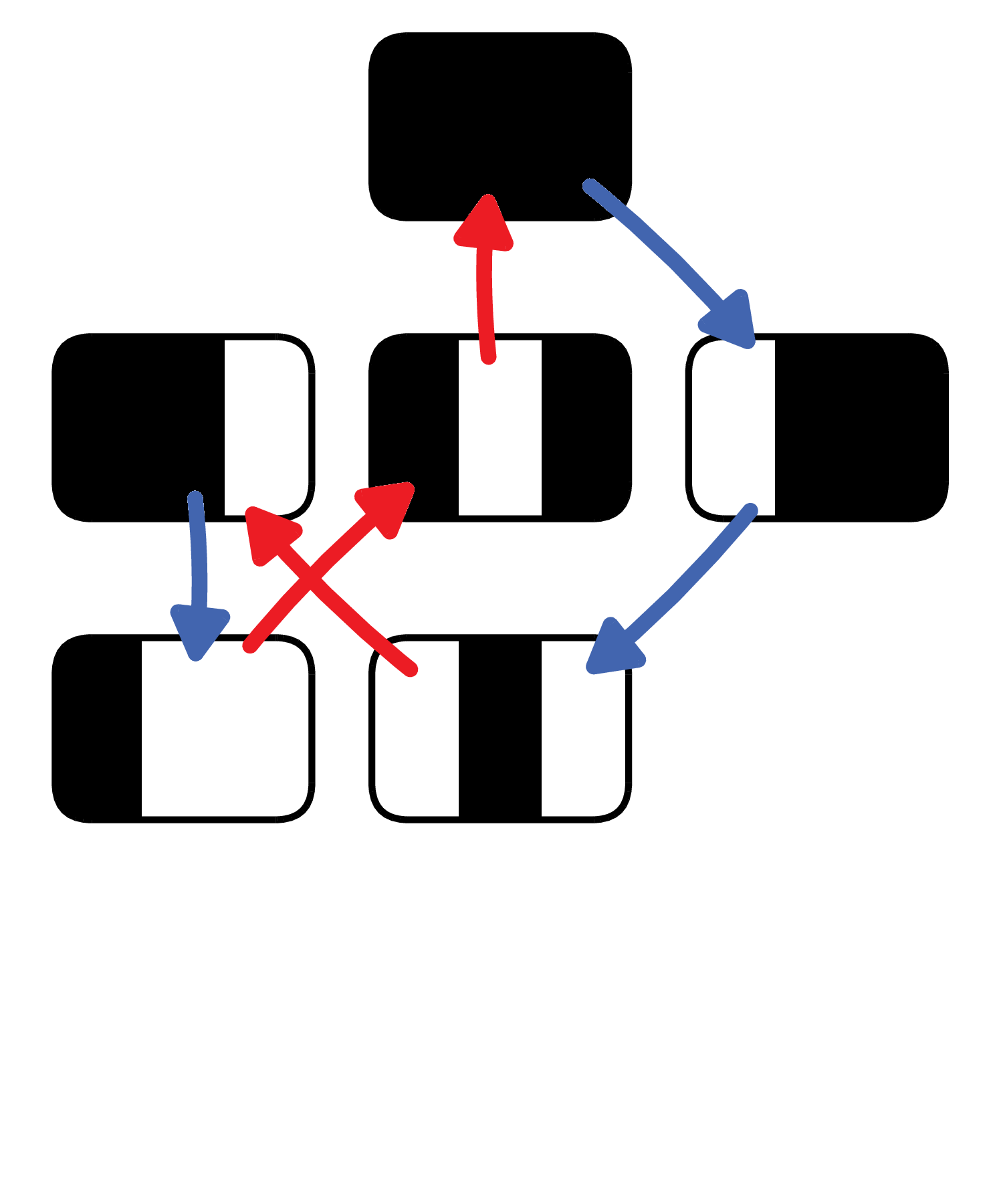}};
\node[rotate=0] at (5.1,0.0) {\includegraphics[height=3.0cm]{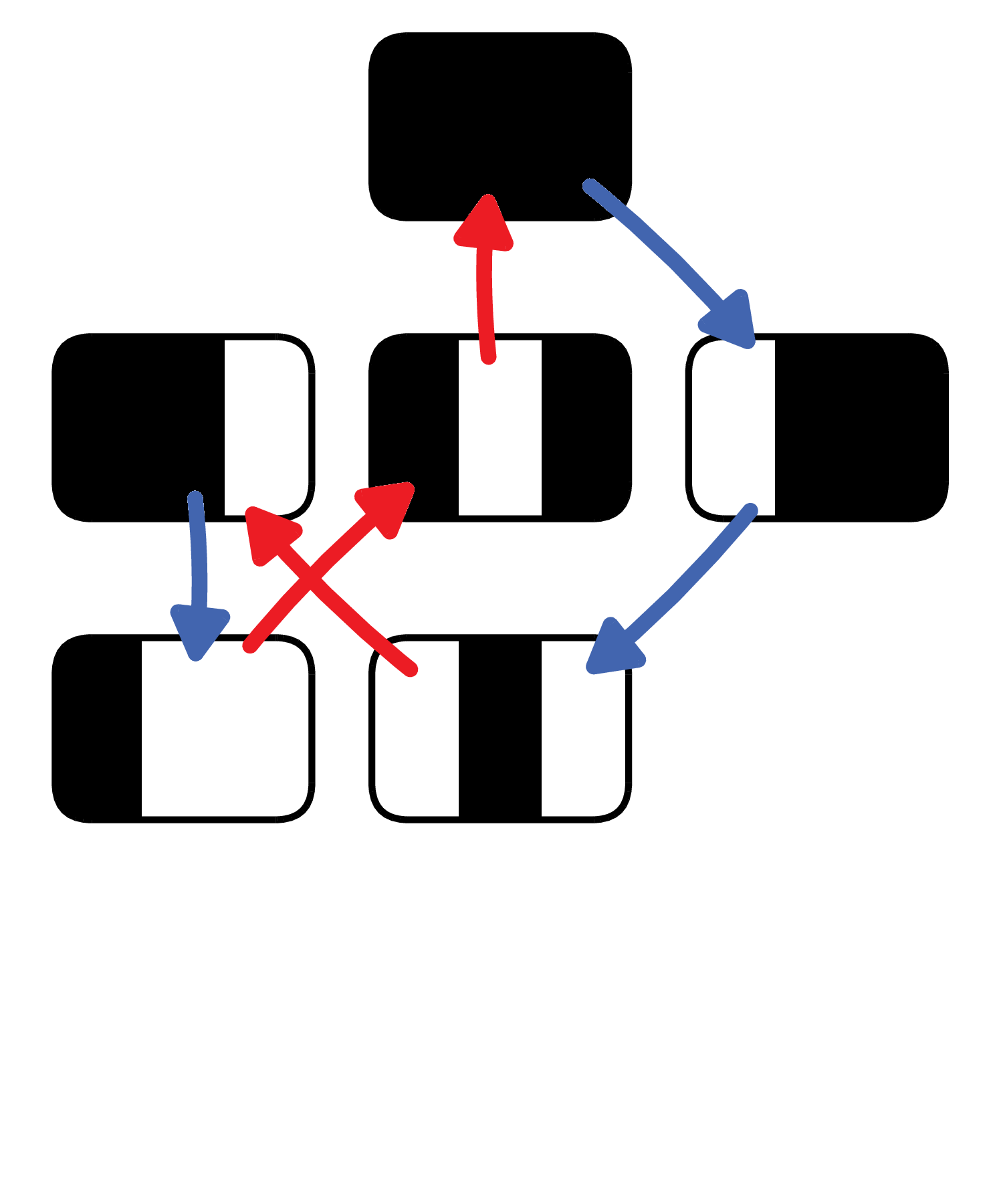}};
\node[rotate=0] at (10.3,0.0) {\includegraphics[height=3.0cm]{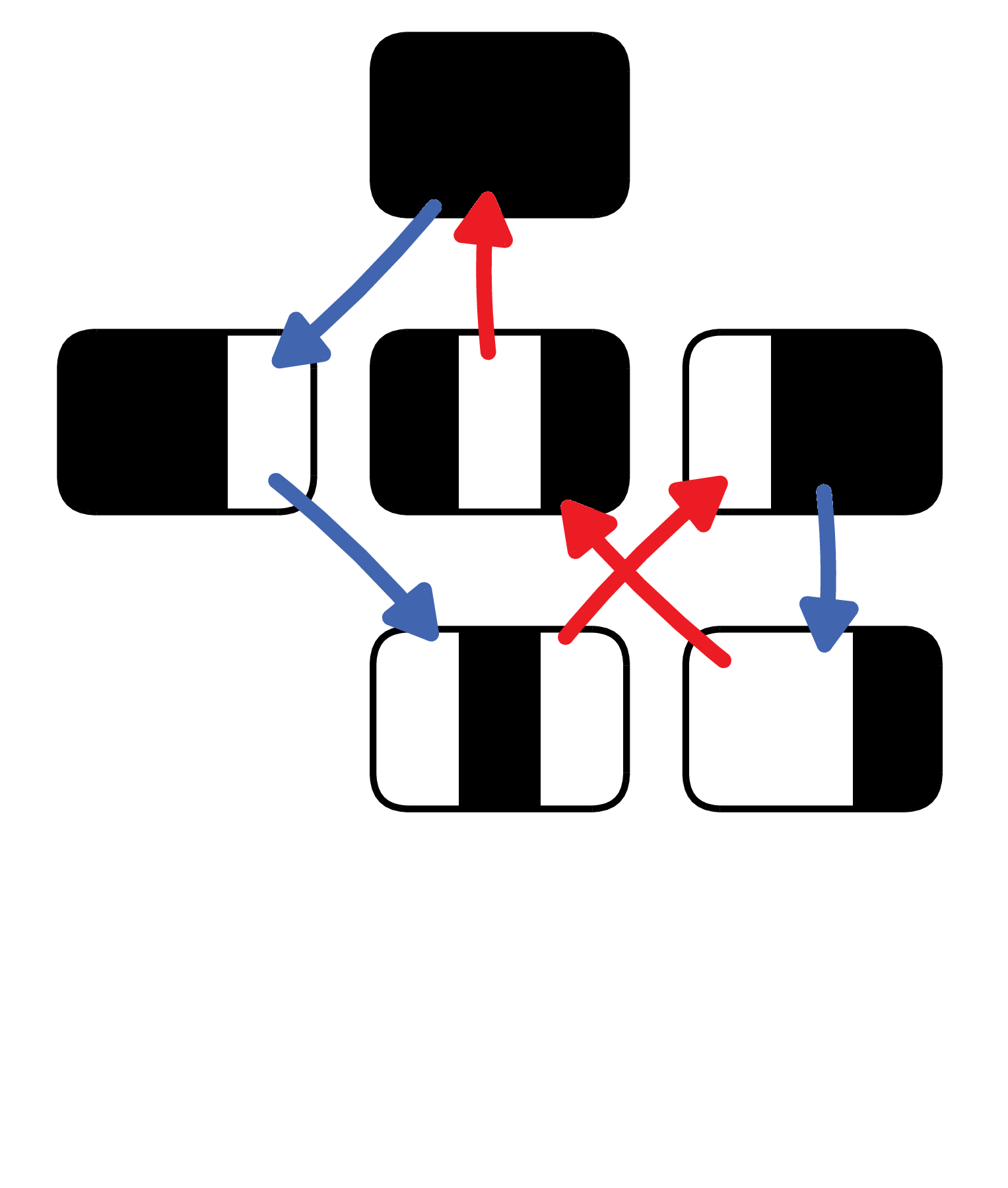}};

\node[rotate=0] at (-0.1,1.18) {\small \scalebox{0.64}{$\color{white}\boldsymbol{+}\!\color{white}\boldsymbol{+}\!\color{white}\boldsymbol{+}$}};

\node[rotate=0] at (-0.1,0.43) {\small \scalebox{0.64}{$\color{white}\boldsymbol{+}\!\color{black}\boldsymbol{-}\!\color{white}\boldsymbol{+}$}};
\node[rotate=0] at (0.68,0.43) {\small \scalebox{0.64}{$\color{black}\boldsymbol{-}\!\color{white}\boldsymbol{+}\!\color{white}\boldsymbol{+}$}};
\node[rotate=0] at (-0.895,0.43) {\small \scalebox{0.64}{$\color{white}\boldsymbol{+}\!\color{white}\boldsymbol{+}\!\color{black}\boldsymbol{-}$}};

\node[rotate=0] at (-0.1,-0.325) {\small \scalebox{0.64}{$\color{black}\boldsymbol{-}\!\color{white}\boldsymbol{+}\!\color{black}\boldsymbol{-}$}};
\node[rotate=0] at (-0.900,-0.325) {\small \scalebox{0.64}{$\color{white}\boldsymbol{+}\!\color{black}\boldsymbol{-}\!\color{black}\boldsymbol{-}$}};

\node[rotate=0] at (5.1,1.18) {\small \scalebox{0.64}{$\color{white}\boldsymbol{+}\!\color{white}\boldsymbol{+}\!\color{white}\boldsymbol{+}$}};

\node[rotate=0] at (5.1,0.43) {\small \scalebox{0.64}{$\color{white}\boldsymbol{+}\!\color{black}\boldsymbol{-}\!\color{white}\boldsymbol{+}$}};
\node[rotate=0] at (5.88,0.43) {\small \scalebox{0.64}{$\color{black}\boldsymbol{-}\!\color{white}\boldsymbol{+}\!\color{white}\boldsymbol{+}$}};
\node[rotate=0] at (4.305,0.43) {\small \scalebox{0.64}{$\color{white}\boldsymbol{+}\!\color{white}\boldsymbol{+}\!\color{black}\boldsymbol{-}$}};

\node[rotate=0] at (5.1,-0.325) {\small \scalebox{0.64}{$\color{black}\boldsymbol{-}\!\color{white}\boldsymbol{+}\!\color{black}\boldsymbol{-}$}};
\node[rotate=0] at (4.3,-0.325) {\small \scalebox{0.64}{$\color{white}\boldsymbol{+}\!\color{black}\boldsymbol{-}\!\color{black}\boldsymbol{-}$}};

\node[rotate=0] at (10.3,1.18) {\small \scalebox{0.64}{$\color{white}\boldsymbol{+}\!\color{white}\boldsymbol{+}\!\color{white}\boldsymbol{+}$}};

\node[rotate=0] at (10.3,0.43) {\small \scalebox{0.64}{$\color{white}\boldsymbol{+}\!\color{black}\boldsymbol{-}\!\color{white}\boldsymbol{+}$}};
\node[rotate=0] at (11.08,0.43) {\small \scalebox{0.64}{$\color{black}\boldsymbol{-}\!\color{white}\boldsymbol{+}\!\color{white}\boldsymbol{+}$}};
\node[rotate=0] at (9.505,0.43) {\small \scalebox{0.64}{$\color{white}\boldsymbol{+}\!\color{white}\boldsymbol{+}\!\color{black}\boldsymbol{-}$}};

\node[rotate=0] at (10.3,-0.325) {\small \scalebox{0.64}{$\color{black}\boldsymbol{-}\!\color{white}\boldsymbol{+}\!\color{black}\boldsymbol{-}$}};
\node[rotate=0] at (11.08,-0.325) {\small \scalebox{0.64}{$\color{black}\boldsymbol{-}\!\color{black}\boldsymbol{-}\!\color{white}\boldsymbol{+}$}};

\draw[dotted] (-0.9,-0.32) circle (0.4);
\draw[dotted] (5.09,-0.32) circle (0.4);
\draw[dotted] (10.99,-0.32) circle (0.4);

\draw[<->] (1.6,0.2) -- (3.4,0.2);
\node[rotate=0] at (2.5,0.55) {\small time shift};

\draw[<->] (6.8,0.2) -- (8.6,0.2);
\node[rotate=0] at (7.7,0.55) {\small convention};
\node[rotate=0] at (7.7,0.85) {\small labeling};

\node[rotate=0] at (-0.1,-1.0) {\small $(2,1,0,2,0,1)$};
\node[rotate=0] at (-0.1,-1.45) {\small $S^0 = (+1,-1,-1)$};

\draw[red] (-1.5,-1.7) rectangle (1.3,-0.75);
\node[rotate=0] at (-0.1,-2.02) {\small lowest rank};

\node[rotate=0] at (5.1,-1.0) {\small $(0,1,2,1,0,2)$};
\node[rotate=0] at (5.1,-1.45) {\small $S^0 = (-1,+1,-1)$};

\node[rotate=0] at (10.3,-1.0) {\small $(2,1,0,1,2,0)$};
\node[rotate=0] at (10.3,-1.45) {\small $S^0 = (-1,+1,-1)$};

\end{tikzpicture}
\vspace*{-0.3cm}
\caption{\small{\textbf{Time shift symmetry between two self-loops.} To remove the final ambiguity on the starting state when there exist multiple
possible starting states with the same magnetization, we compare the ranking of self-loops (see text), and pick the loop with the smallest ranking.}}
\label{fig:margot3}
\end{figure}

\begin{figure}[b!]
\hspace*{-0.2cm}
\begin{tikzpicture}



\node[rotate=0] at (8.5,-1.85) {\includegraphics[height=2.5cm]{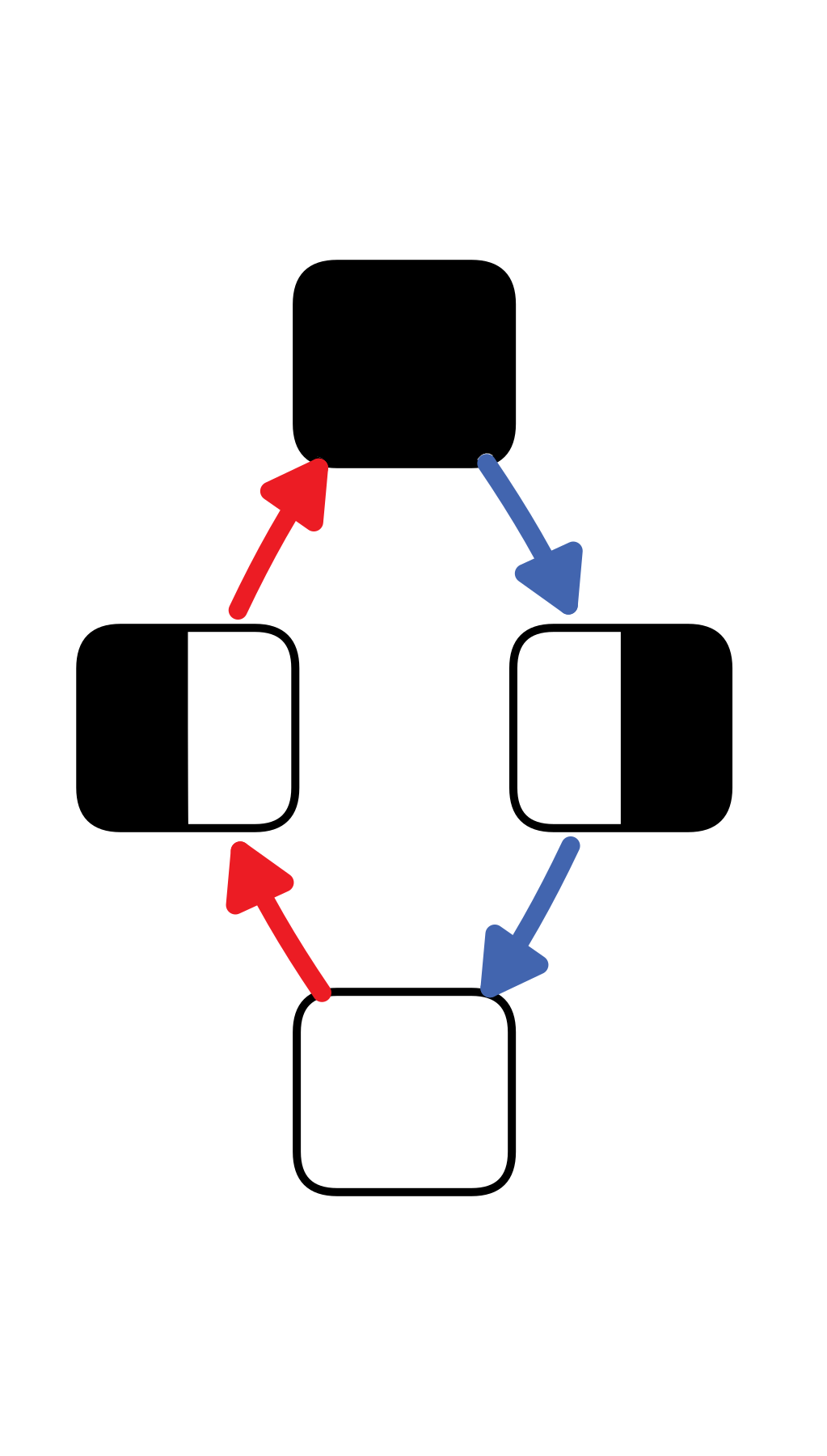}};

\node[rotate=0] at (8.48,-2.48) {\small \scalebox{0.58}{$\color{black}\boldsymbol{-}\color{black}\boldsymbol{-}$}};
\node[rotate=0] at (8.48,-1.225) {\small \scalebox{0.58}{$\color{white}\boldsymbol{+}\color{white}\boldsymbol{+}$}};

\node[rotate=0] at (8.11,-1.85) {\small \scalebox{0.58}{$\color{white}\boldsymbol{+}\color{black}\boldsymbol{-}$}};
\node[rotate=0] at (8.86,-1.85) {\small \scalebox{0.58}{$\color{black}\boldsymbol{-}\color{white}\boldsymbol{+}$}};

\draw[] (7.3,-2.9) rectangle (9.7,-0.8);
\node[rotate=0] at (7.6,-1.1) {\small (a)};


\node[rotate=0] at (2.8,-4.64) {\includegraphics[height=2.1cm]{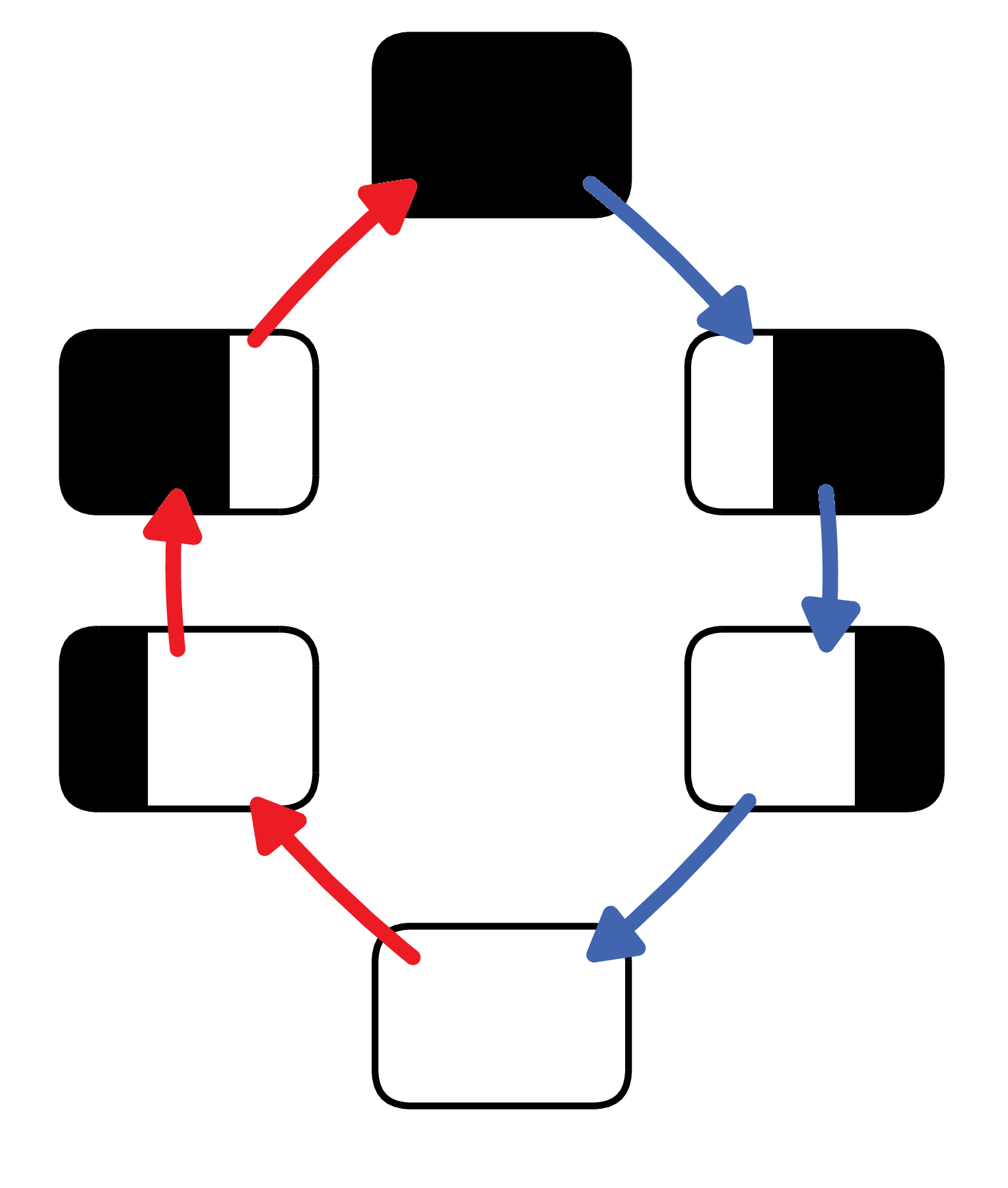}};

\node[rotate=0] at (2.795,-5.39) {\small \scalebox{0.45}{$\color{black}\boldsymbol{-}\!\color{black}\boldsymbol{-}\!\color{black}\boldsymbol{-}$}};

\node[rotate=0] at (2.795,-3.805) {\small \scalebox{0.45}{$\color{white}\boldsymbol{+}\!\color{white}\boldsymbol{+}\!\color{white}\boldsymbol{+}$}};

\node[rotate=0] at (2.24,-4.86) {\small \scalebox{0.45}{$\color{white}\boldsymbol{+}\!\color{black}\boldsymbol{-}\!\color{black}\boldsymbol{-}$}};
\node[rotate=0] at (2.24,-4.34) {\small \scalebox{0.45}{$\color{white}\boldsymbol{+}\!\color{white}\boldsymbol{+}\!\color{black}\boldsymbol{-}$}};

\node[rotate=0] at (3.35,-4.34) {\small \scalebox{0.45}{$\color{black}\boldsymbol{-}\!\color{white}\boldsymbol{+}\!\color{white}\boldsymbol{+}$}};
\node[rotate=0] at (3.35,-4.86) {\small \scalebox{0.45}{$\color{black}\boldsymbol{-}\!\color{black}\boldsymbol{-}\!\color{white}\boldsymbol{+}$}};

\node[rotate=0] at (5.3,-4.64) {\includegraphics[height=2.1cm]{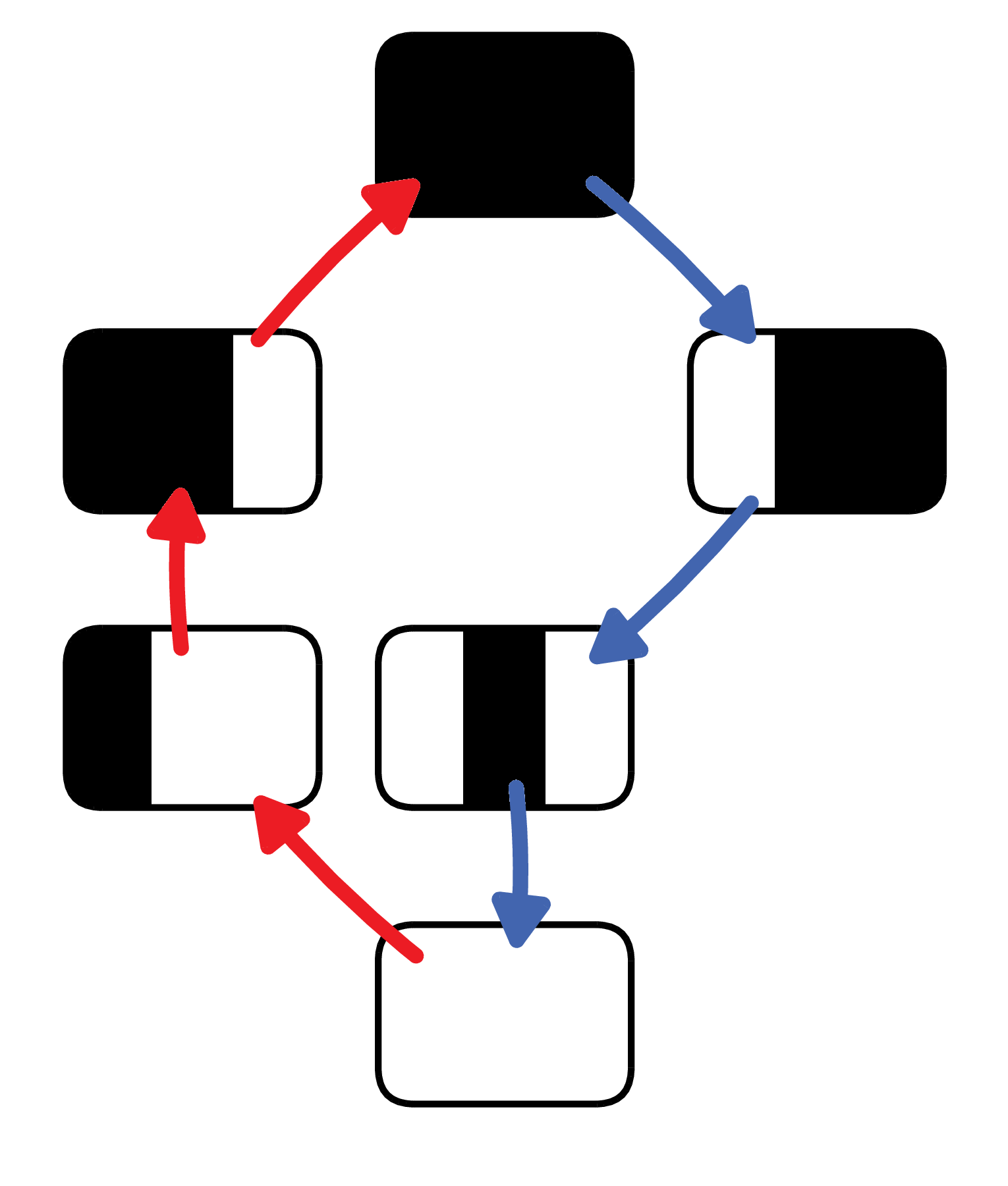}};

\node[rotate=0] at (5.295,-5.39) {\small \scalebox{0.45}{$\color{black}\boldsymbol{-}\!\color{black}\boldsymbol{-}\!\color{black}\boldsymbol{-}$}};

\node[rotate=0] at (5.295,-3.805) {\small \scalebox{0.45}{$\color{white}\boldsymbol{+}\!\color{white}\boldsymbol{+}\!\color{white}\boldsymbol{+}$}};
\node[rotate=0] at (5.295,-4.86) {\small \scalebox{0.45}{$\color{black}\boldsymbol{-}\!\color{white}\boldsymbol{+}\!\color{black}\boldsymbol{-}$}};

\node[rotate=0] at (4.74,-4.86) {\small \scalebox{0.45}{$\color{white}\boldsymbol{+}\!\color{black}\boldsymbol{-}\!\color{black}\boldsymbol{-}$}};
\node[rotate=0] at (4.74,-4.34) {\small \scalebox{0.45}{$\color{white}\boldsymbol{+}\!\color{white}\boldsymbol{+}\!\color{black}\boldsymbol{-}$}};

\node[rotate=0] at (5.85,-4.34) {\small \scalebox{0.45}{$\color{black}\boldsymbol{-}\!\color{white}\boldsymbol{+}\!\color{white}\boldsymbol{+}$}};

\node[rotate=0] at (7.8,-4.64) {\includegraphics[height=2.1cm]{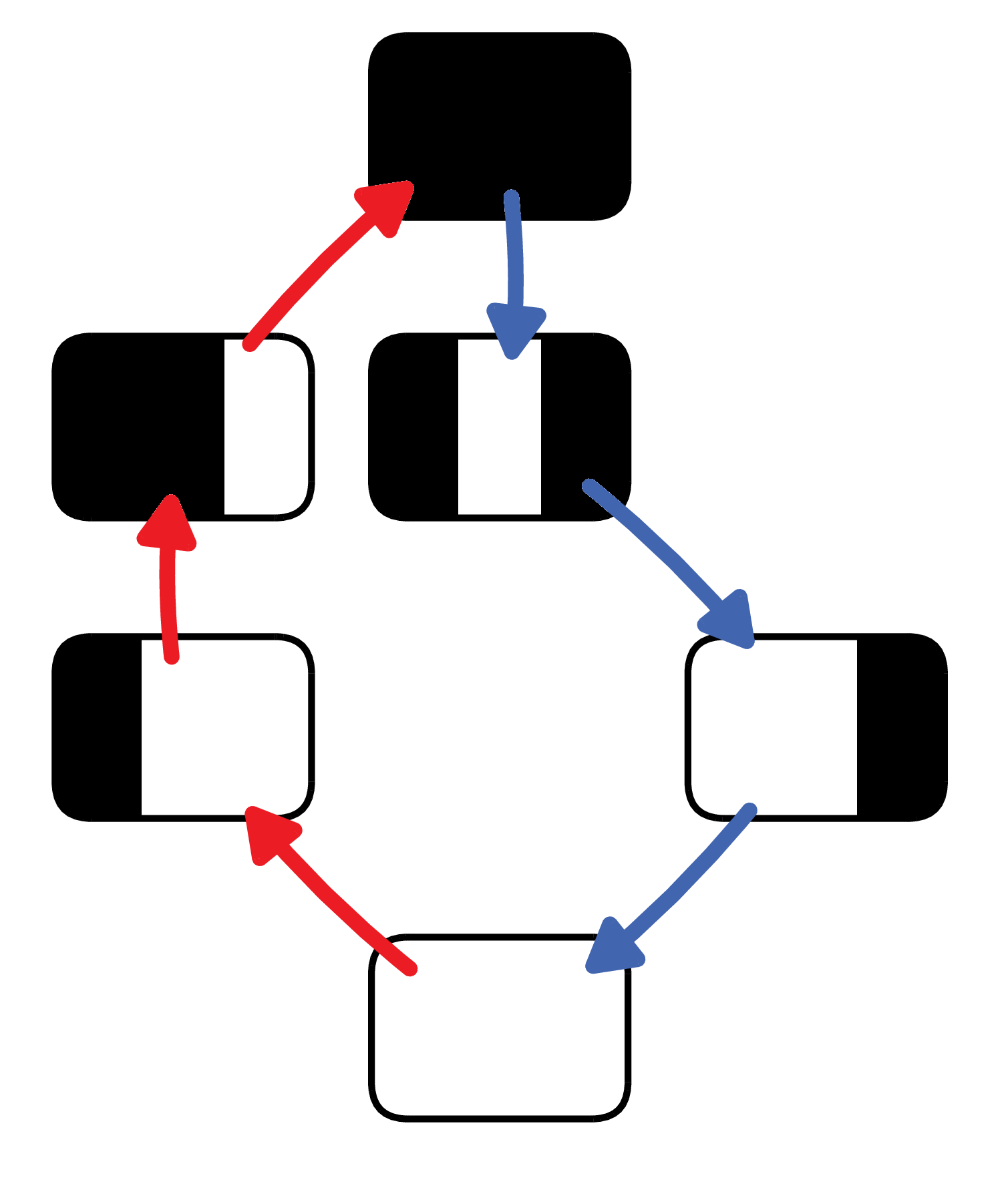}};

\node[rotate=0] at (7.795,-5.39) {\small \scalebox{0.45}{$\color{black}\boldsymbol{-}\!\color{black}\boldsymbol{-}\!\color{black}\boldsymbol{-}$}};

\node[rotate=0] at (7.795,-3.805) {\small \scalebox{0.45}{$\color{white}\boldsymbol{+}\!\color{white}\boldsymbol{+}\!\color{white}\boldsymbol{+}$}};

\node[rotate=0] at (7.795,-4.34) {\small \scalebox{0.45}{$\color{white}\boldsymbol{+}\!\color{black}\boldsymbol{-}\!\color{white}\boldsymbol{+}$}};

\node[rotate=0] at (7.24,-4.86) {\small \scalebox{0.45}{$\color{white}\boldsymbol{+}\!\color{black}\boldsymbol{-}\!\color{black}\boldsymbol{-}$}};
\node[rotate=0] at (7.24,-4.34) {\small \scalebox{0.45}{$\color{white}\boldsymbol{+}\!\color{white}\boldsymbol{+}\!\color{black}\boldsymbol{-}$}};

\node[rotate=0] at (8.35,-4.86) {\small \scalebox{0.45}{$\color{black}\boldsymbol{-}\!\color{black}\boldsymbol{-}\!\color{white}\boldsymbol{+}$}};

\node[rotate=0] at (10.3,-4.64) {\includegraphics[height=2.1cm]{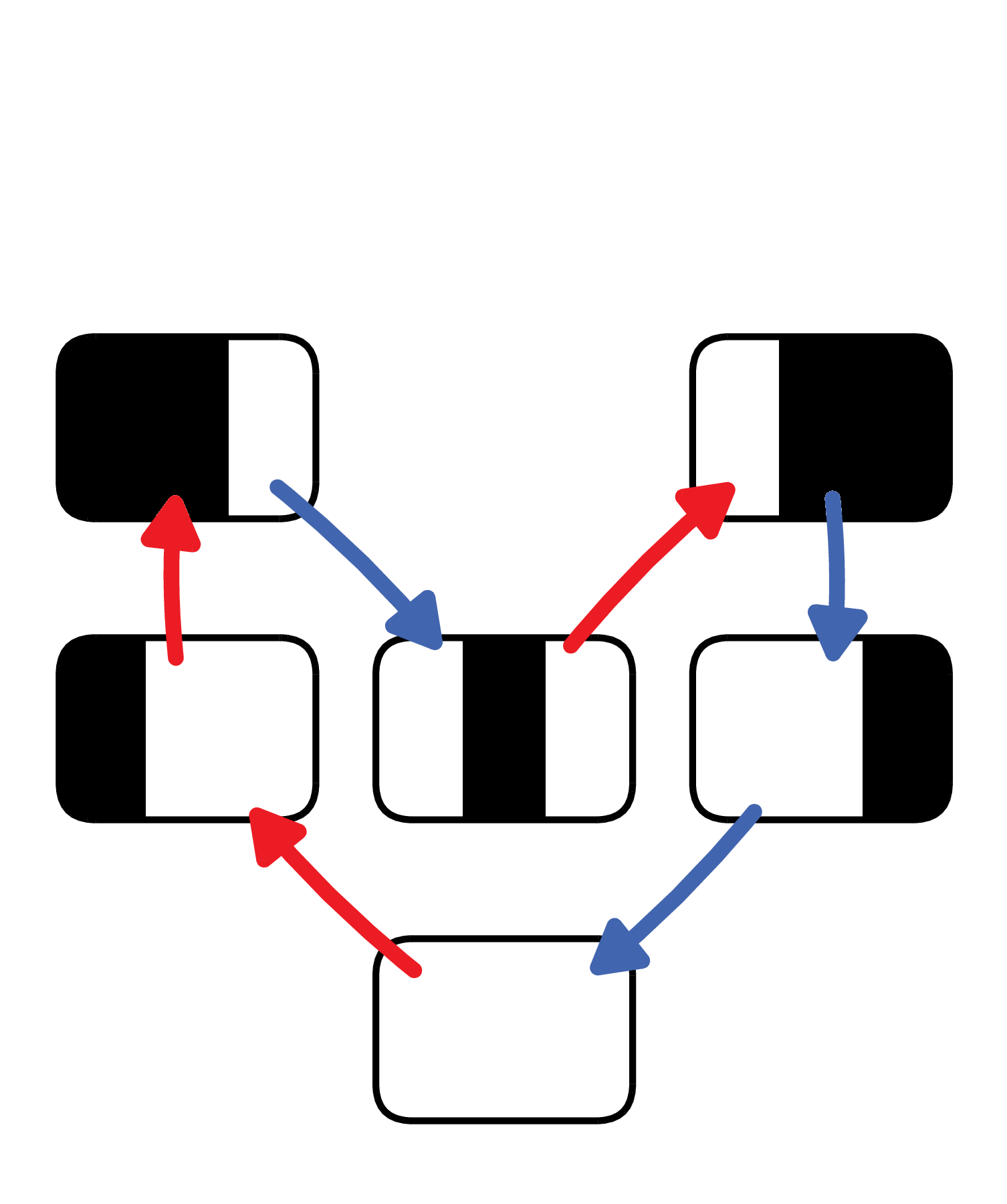}};

\node[rotate=0] at (10.295,-5.39) {\small \scalebox{0.45}{$\color{black}\boldsymbol{-}\!\color{black}\boldsymbol{-}\!\color{black}\boldsymbol{-}$}};

\node[rotate=0] at (10.295,-4.86) {\small \scalebox{0.45}{$\color{black}\boldsymbol{-}\!\color{white}\boldsymbol{+}\!\color{black}\boldsymbol{-}$}};

\node[rotate=0] at (9.74,-4.86) {\small \scalebox{0.45}{$\color{white}\boldsymbol{+}\!\color{black}\boldsymbol{-}\!\color{black}\boldsymbol{-}$}};
\node[rotate=0] at (9.74,-4.34) {\small \scalebox{0.45}{$\color{white}\boldsymbol{+}\!\color{white}\boldsymbol{+}\!\color{black}\boldsymbol{-}$}};

\node[rotate=0] at (10.85,-4.34) {\small \scalebox{0.45}{$\color{black}\boldsymbol{-}\!\color{white}\boldsymbol{+}\!\color{white}\boldsymbol{+}$}};
\node[rotate=0] at (10.85,-4.86) {\small \scalebox{0.45}{$\color{black}\boldsymbol{-}\!\color{black}\boldsymbol{-}\!\color{white}\boldsymbol{+}$}};

\node[rotate=0] at (12.8,-4.64) {\includegraphics[height=2.1cm]{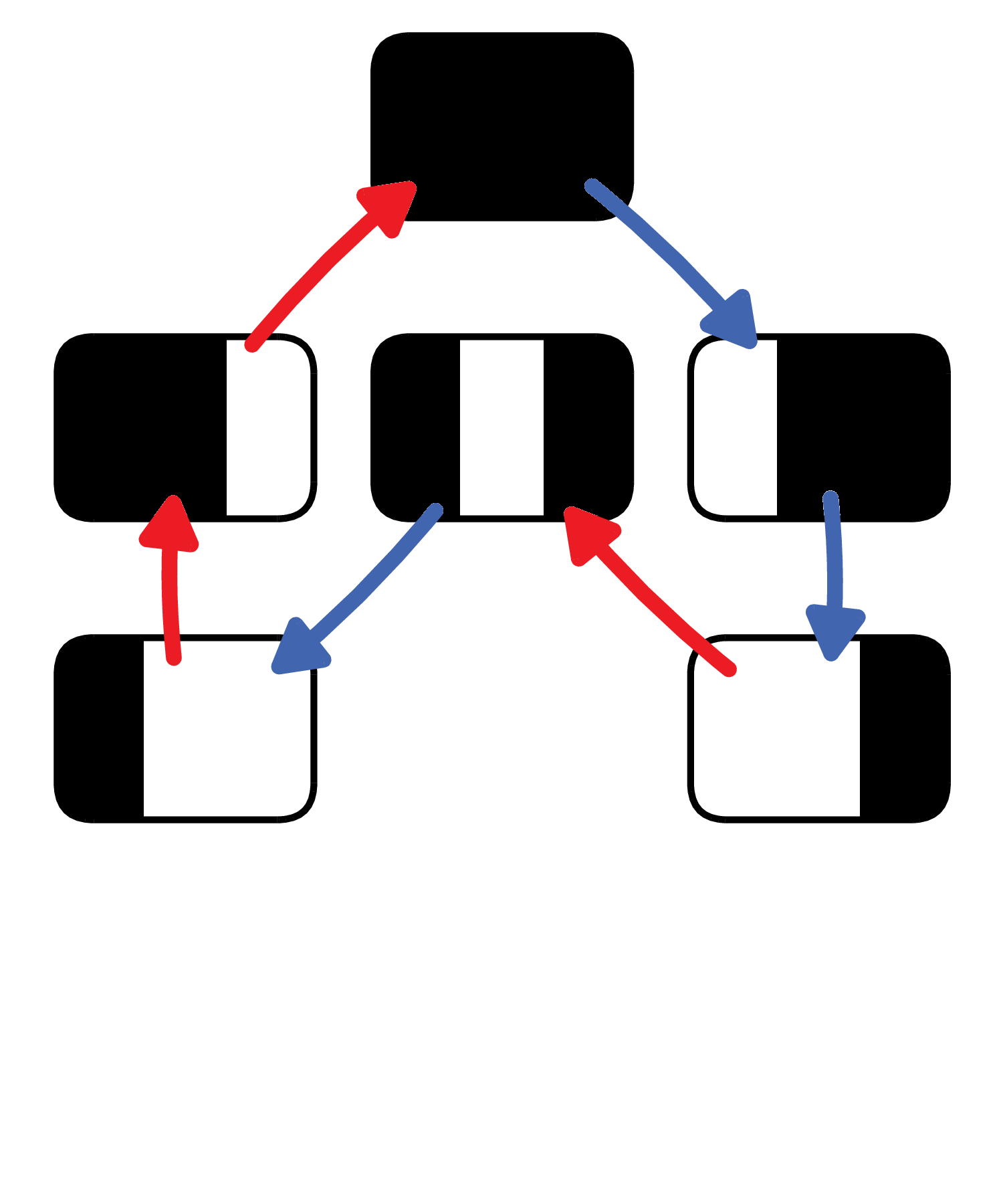}};

\node[rotate=0] at (12.795,-4.34) {\small \scalebox{0.45}{$\color{white}\boldsymbol{+}\!\color{black}\boldsymbol{-}\!\color{white}\boldsymbol{+}$}};

\node[rotate=0] at (12.795,-3.805) {\small \scalebox{0.45}{$\color{white}\boldsymbol{+}\!\color{white}\boldsymbol{+}\!\color{white}\boldsymbol{+}$}};

\node[rotate=0] at (12.24,-4.86) {\small \scalebox{0.45}{$\color{white}\boldsymbol{+}\!\color{black}\boldsymbol{-}\!\color{black}\boldsymbol{-}$}};
\node[rotate=0] at (12.24,-4.34) {\small \scalebox{0.45}{$\color{white}\boldsymbol{+}\!\color{white}\boldsymbol{+}\!\color{black}\boldsymbol{-}$}};

\node[rotate=0] at (13.35,-4.34) {\small \scalebox{0.45}{$\color{black}\boldsymbol{-}\!\color{white}\boldsymbol{+}\!\color{white}\boldsymbol{+}$}};
\node[rotate=0] at (13.35,-4.86) {\small \scalebox{0.45}{$\color{black}\boldsymbol{-}\!\color{black}\boldsymbol{-}\!\color{white}\boldsymbol{+}$}};

\node[rotate=0] at (15.3,-4.64) {\includegraphics[height=2.1cm]{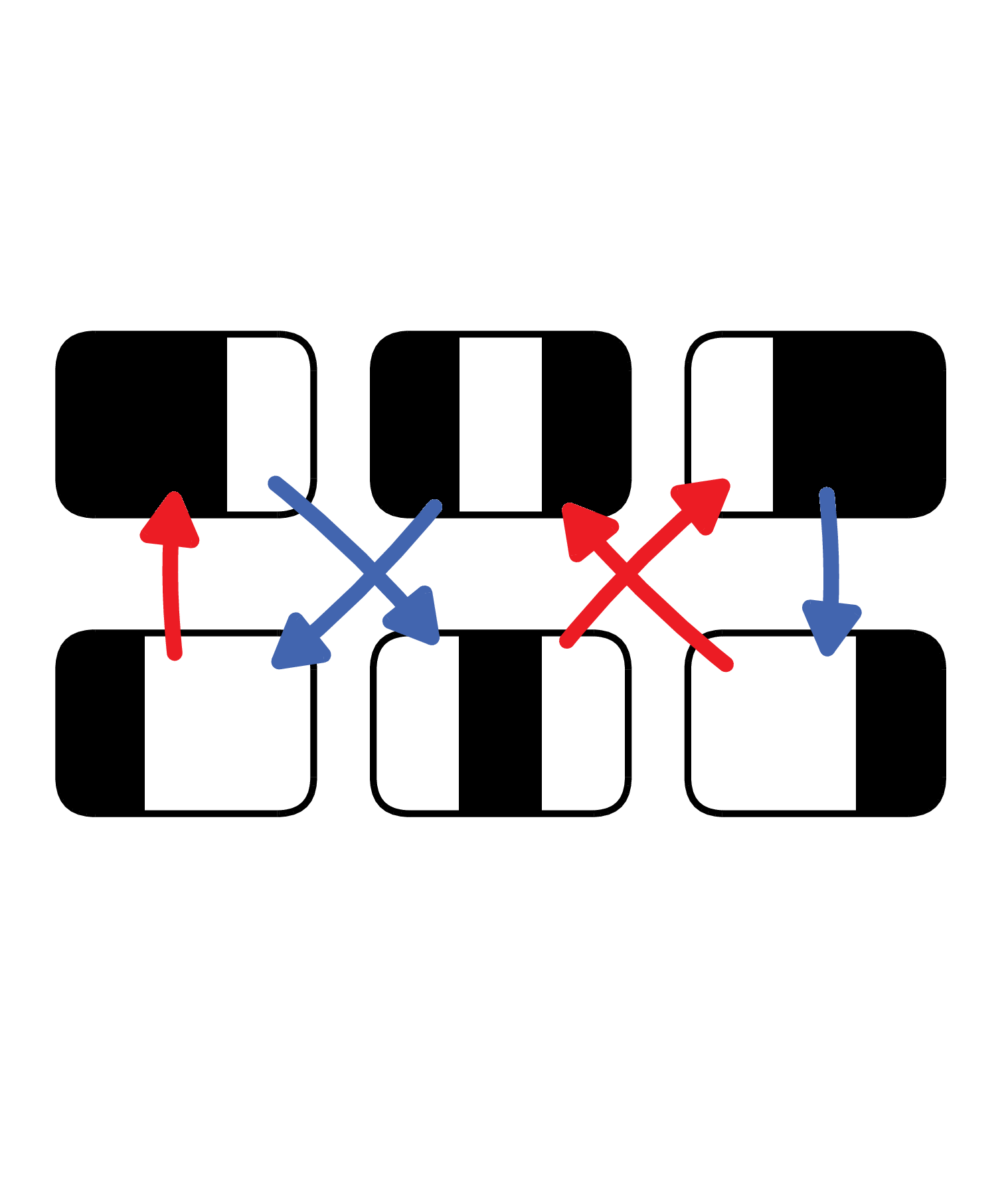}};

\node[rotate=0] at (15.295,-4.34) {\small \scalebox{0.45}{$\color{white}\boldsymbol{+}\!\color{black}\boldsymbol{-}\!\color{white}\boldsymbol{+}$}};

\node[rotate=0] at (15.295,-4.86) {\small \scalebox{0.45}{$\color{black}\boldsymbol{-}\!\color{white}\boldsymbol{+}\!\color{black}\boldsymbol{-}$}};

\node[rotate=0] at (14.74,-4.86) {\small \scalebox{0.45}{$\color{white}\boldsymbol{+}\!\color{black}\boldsymbol{-}\!\color{black}\boldsymbol{-}$}};
\node[rotate=0] at (14.74,-4.34) {\small \scalebox{0.45}{$\color{white}\boldsymbol{+}\!\color{white}\boldsymbol{+}\!\color{black}\boldsymbol{-}$}};

\node[rotate=0] at (15.85,-4.34) {\small \scalebox{0.45}{$\color{black}\boldsymbol{-}\!\color{white}\boldsymbol{+}\!\color{white}\boldsymbol{+}$}};
\node[rotate=0] at (15.85,-4.86) {\small \scalebox{0.45}{$\color{black}\boldsymbol{-}\!\color{black}\boldsymbol{-}\!\color{white}\boldsymbol{+}$}};

\draw[] (1.3,-5.7) rectangle (16.8,-3.5);
\node[rotate=0] at (1.6,-3.8) {\small (b)};


\node[rotate=0] at (2.55,-7.75) {\includegraphics[height=2.5cm]{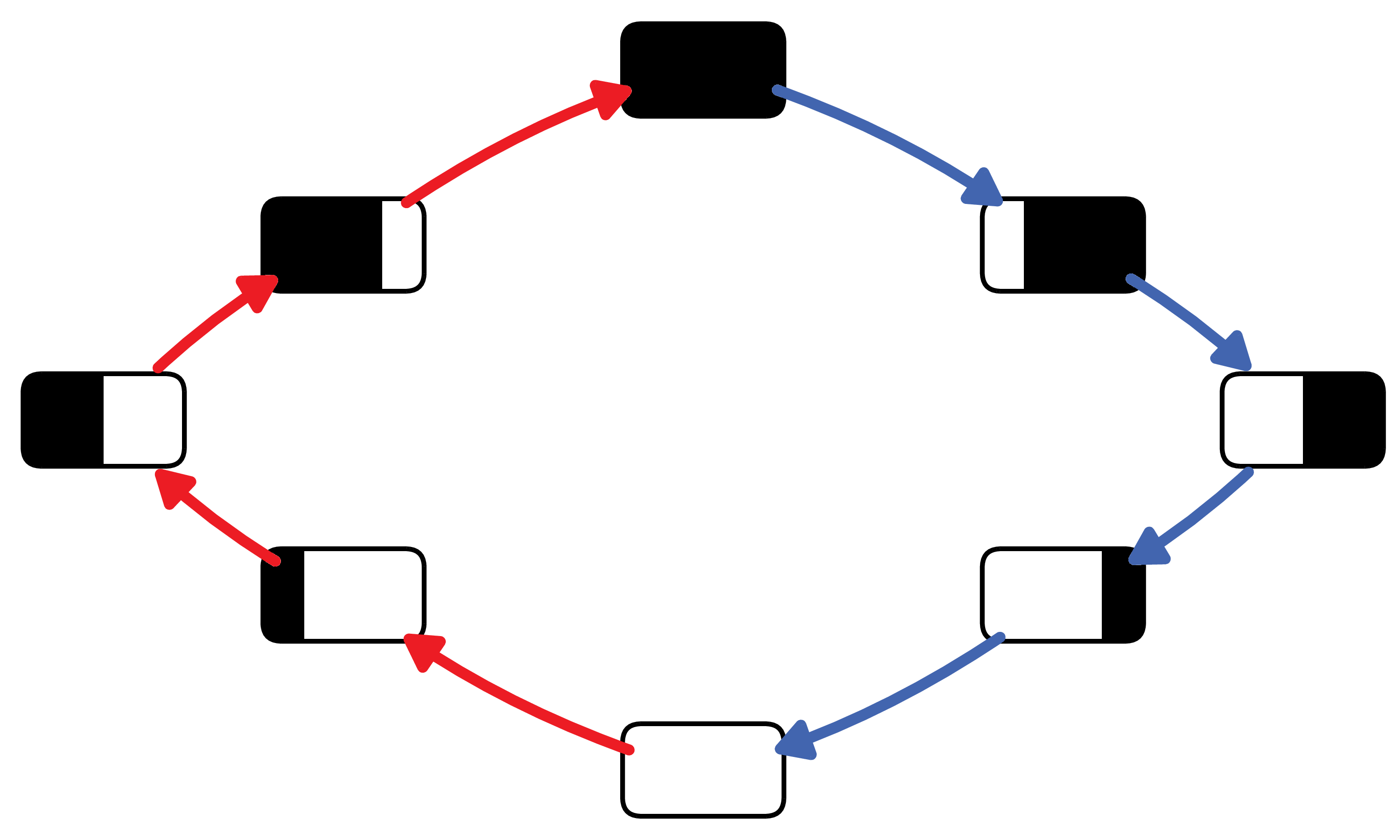}};

\node[rotate=0] at (2.56,-8.79) {\small \scalebox{0.39}{$\boldsymbol{-}\mkern-4mu\boldsymbol{-}\mkern-4mu\boldsymbol{-}\mkern-0mu\boldsymbol{-}$}};
\node[rotate=0] at (2.56,-6.705) {\small \scalebox{0.39}{$\color{white}\boldsymbol{+}\mkern-4mu\boldsymbol{+}\mkern-4mu\boldsymbol{+}\mkern-0mu\boldsymbol{+}$}};

\node[rotate=0] at (1.49,-7.225) {\small \scalebox{0.39}{$\color{white}\boldsymbol{+}\mkern-4mu\boldsymbol{+}\mkern-4mu\boldsymbol{+}\mkern-0mu\color{black}\boldsymbol{-}$}};
\node[rotate=0] at (1.485,-8.27) {\small \scalebox{0.39}{$\color{white}\boldsymbol{+}\mkern-4mu\color{black}\boldsymbol{-}\mkern-4mu\boldsymbol{-}\mkern-0mu\boldsymbol{-}$}};

\node[rotate=0] at (3.63,-7.225) {\small \scalebox{0.39}{$\color{black}\boldsymbol{-}\mkern-4mu\color{white}\boldsymbol{+}\mkern-4mu\boldsymbol{+}\mkern-0mu\boldsymbol{+}$}};
\node[rotate=0] at (3.635,-8.27) {\small \scalebox{0.39}{$\color{black}\boldsymbol{-}\mkern-4mu\color{black}\boldsymbol{-}\mkern-4mu\boldsymbol{-}\mkern-0mu\color{white}\boldsymbol{+}$}};

\node[rotate=0] at (0.77,-7.75) {\small \scalebox{0.39}{$\color{white}\boldsymbol{+}\mkern-4mu\boldsymbol{+}\mkern-4mu\color{black}\boldsymbol{-}\mkern-0mu\boldsymbol{-}$}};

\node[rotate=0] at (4.345,-7.75) {\small \scalebox{0.39}{$\color{black}\boldsymbol{-}\mkern-4mu\boldsymbol{-}\mkern-4mu\color{white}\boldsymbol{+}\mkern-0mu\boldsymbol{+}$}};

\node[rotate=0] at (6.85,-7.75) {\includegraphics[height=2.5cm]{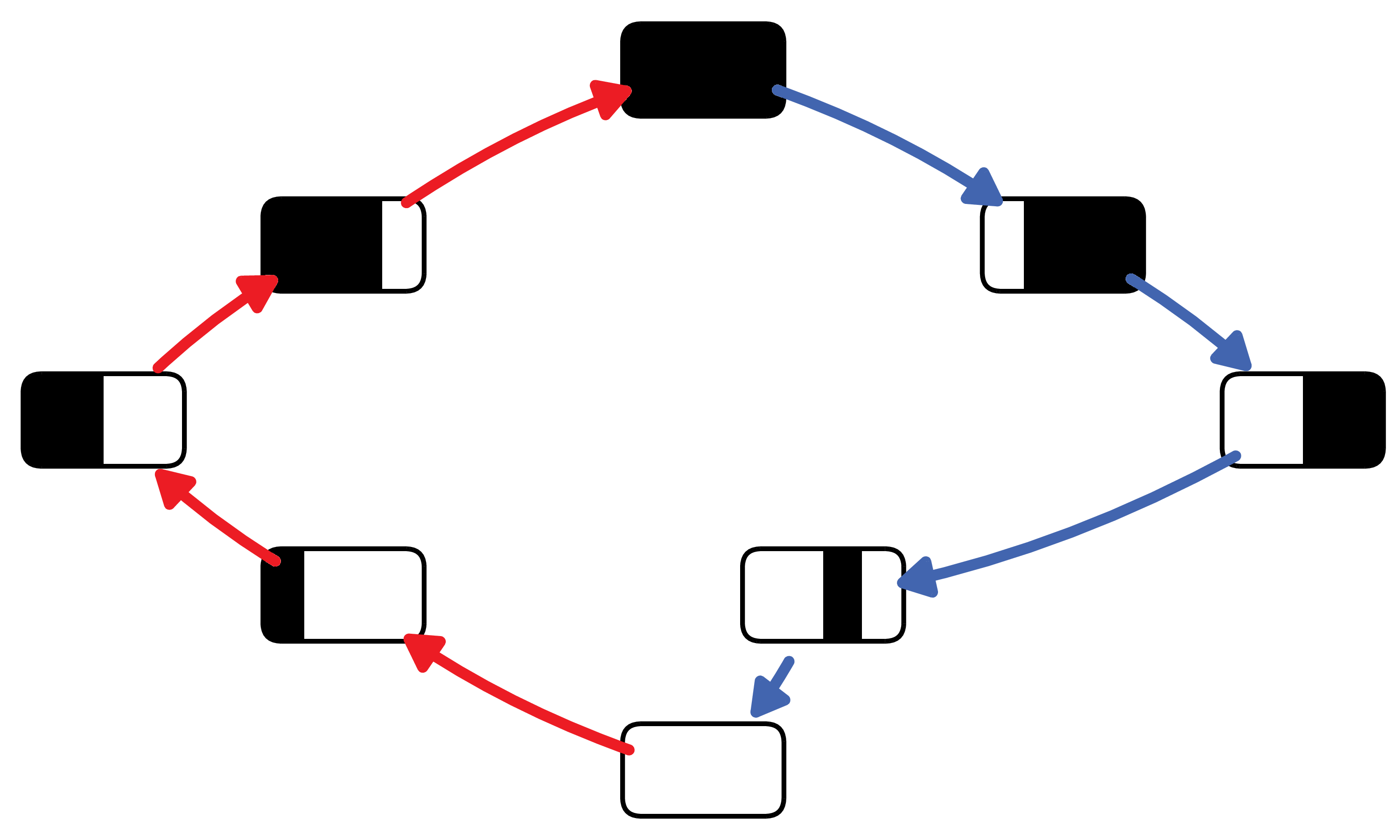}};

\node[rotate=0] at (6.86,-8.79) {\small \scalebox{0.39}{$\boldsymbol{-}\mkern-4mu\boldsymbol{-}\mkern-4mu\boldsymbol{-}\mkern-0mu\boldsymbol{-}$}};
\node[rotate=0] at (6.86,-6.705) {\small \scalebox{0.39}{$\color{white}\boldsymbol{+}\mkern-4mu\boldsymbol{+}\mkern-4mu\boldsymbol{+}\mkern-0mu\boldsymbol{+}$}};

\node[rotate=0] at (5.79,-7.225) {\small \scalebox{0.39}{$\color{white}\boldsymbol{+}\mkern-4mu\boldsymbol{+}\mkern-4mu\boldsymbol{+}\mkern-0mu\color{black}\boldsymbol{-}$}};
\node[rotate=0] at (5.785,-8.27) {\small \scalebox{0.39}{$\color{white}\boldsymbol{+}\mkern-4mu\color{black}\boldsymbol{-}\mkern-4mu\boldsymbol{-}\mkern-0mu\boldsymbol{-}$}};

\node[rotate=0] at (7.93,-7.225) {\small \scalebox{0.39}{$\color{black}\boldsymbol{-}\mkern-4mu\color{white}\boldsymbol{+}\mkern-4mu\boldsymbol{+}\mkern-0mu\boldsymbol{+}$}};

\node[rotate=0] at (7.22,-8.27) {\small \scalebox{0.39}{$\color{black}\boldsymbol{-}\mkern-4mu\color{black}\boldsymbol{-}\mkern-4mu\color{white}\boldsymbol{+}\mkern-0mu\color{black}\boldsymbol{-}$}};

\node[rotate=0] at (5.07,-7.75) {\small \scalebox{0.39}{$\color{white}\boldsymbol{+}\mkern-4mu\boldsymbol{+}\mkern-4mu\color{black}\boldsymbol{-}\mkern-0mu\boldsymbol{-}$}};

\node[rotate=0] at (8.645,-7.75) {\small \scalebox{0.39}{$\color{black}\boldsymbol{-}\mkern-4mu\boldsymbol{-}\mkern-4mu\color{white}\boldsymbol{+}\mkern-0mu\boldsymbol{+}$}};

\node[rotate=0] at (11.2,-7.75) {\includegraphics[height=2.5cm]{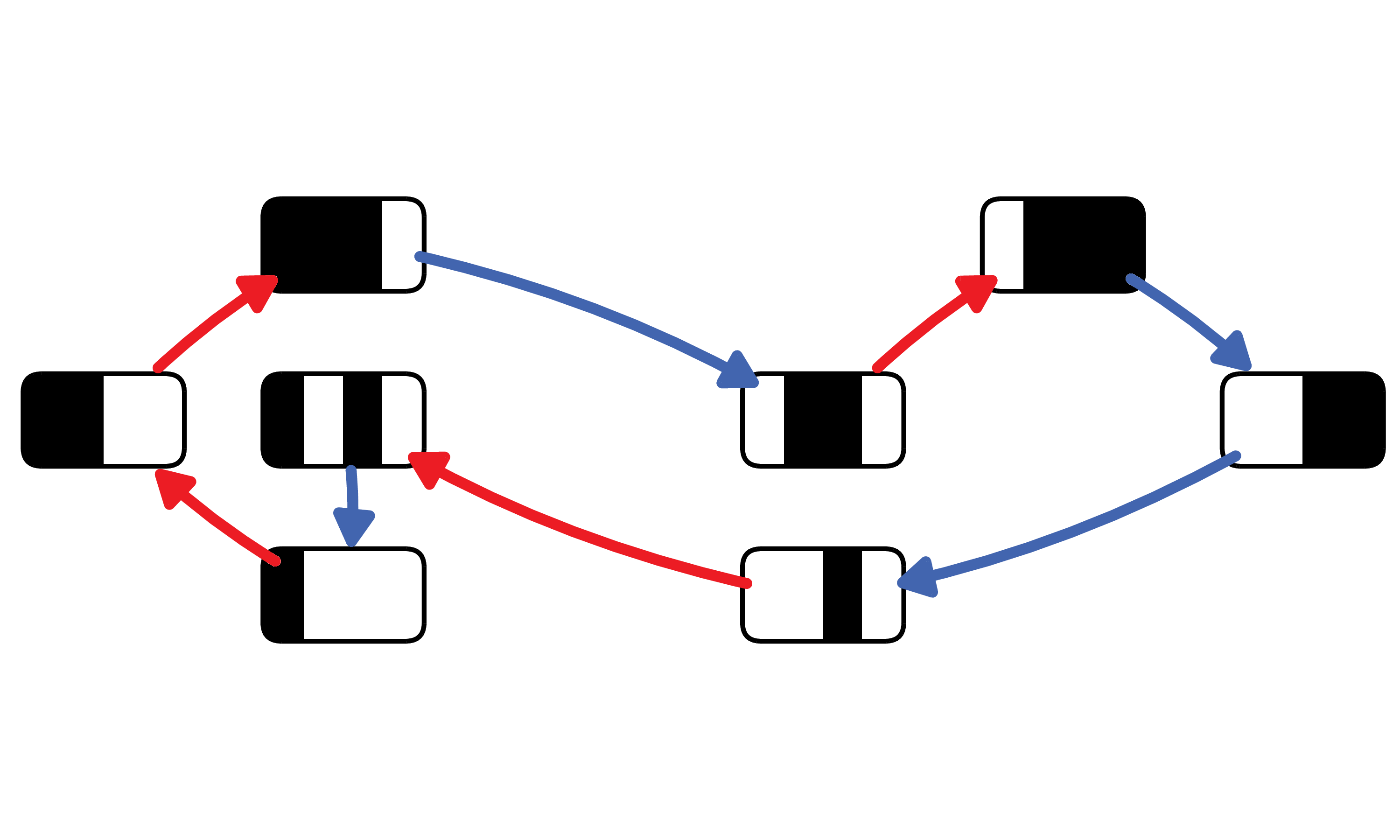}};

\node[rotate=0] at (10.14,-7.225) {\small \scalebox{0.39}{$\color{white}\boldsymbol{+}\mkern-4mu\boldsymbol{+}\mkern-4mu\boldsymbol{+}\mkern-0mu\color{black}\boldsymbol{-}$}};
\node[rotate=0] at (10.135,-7.75) {\small \scalebox{0.39}{$\color{white}\boldsymbol{+}\mkern-4mu\color{black}\boldsymbol{-}\mkern-4mu\color{white}\boldsymbol{+}\mkern-0mu\color{black}\boldsymbol{-}$}};
\node[rotate=0] at (10.135,-8.27) {\small \scalebox{0.39}{$\color{white}\boldsymbol{+}\mkern-4mu\color{black}\boldsymbol{-}\mkern-4mu\boldsymbol{-}\mkern-0mu\boldsymbol{-}$}};

\node[rotate=0] at (12.28,-7.225) {\small \scalebox{0.39}{$\color{black}\boldsymbol{-}\mkern-4mu\color{white}\boldsymbol{+}\mkern-4mu\boldsymbol{+}\mkern-0mu\boldsymbol{+}$}};

\node[rotate=0] at (11.57,-8.27) {\small \scalebox{0.39}{$\color{black}\boldsymbol{-}\mkern-4mu\color{black}\boldsymbol{-}\mkern-4mu\color{white}\boldsymbol{+}\mkern-0mu\color{black}\boldsymbol{-}$}};
\node[rotate=0] at (11.57,-7.75) {\small \scalebox{0.39}{$\boldsymbol{-}\mkern-4mu\color{white}\boldsymbol{+}\mkern-4mu\boldsymbol{+}\mkern-0mu\color{black}\boldsymbol{-}$}};

\node[rotate=0] at (9.42,-7.75) {\small \scalebox{0.39}{$\color{white}\boldsymbol{+}\mkern-4mu\boldsymbol{+}\mkern-4mu\color{black}\boldsymbol{-}\mkern-0mu\boldsymbol{-}$}};

\node[rotate=0] at (12.995,-7.75) {\small \scalebox{0.39}{$\color{black}\boldsymbol{-}\mkern-4mu\boldsymbol{-}\mkern-4mu\color{white}\boldsymbol{+}\mkern-0mu\boldsymbol{+}$}};

\node[rotate=0] at (15.55,-7.75) {\includegraphics[height=2.5cm]{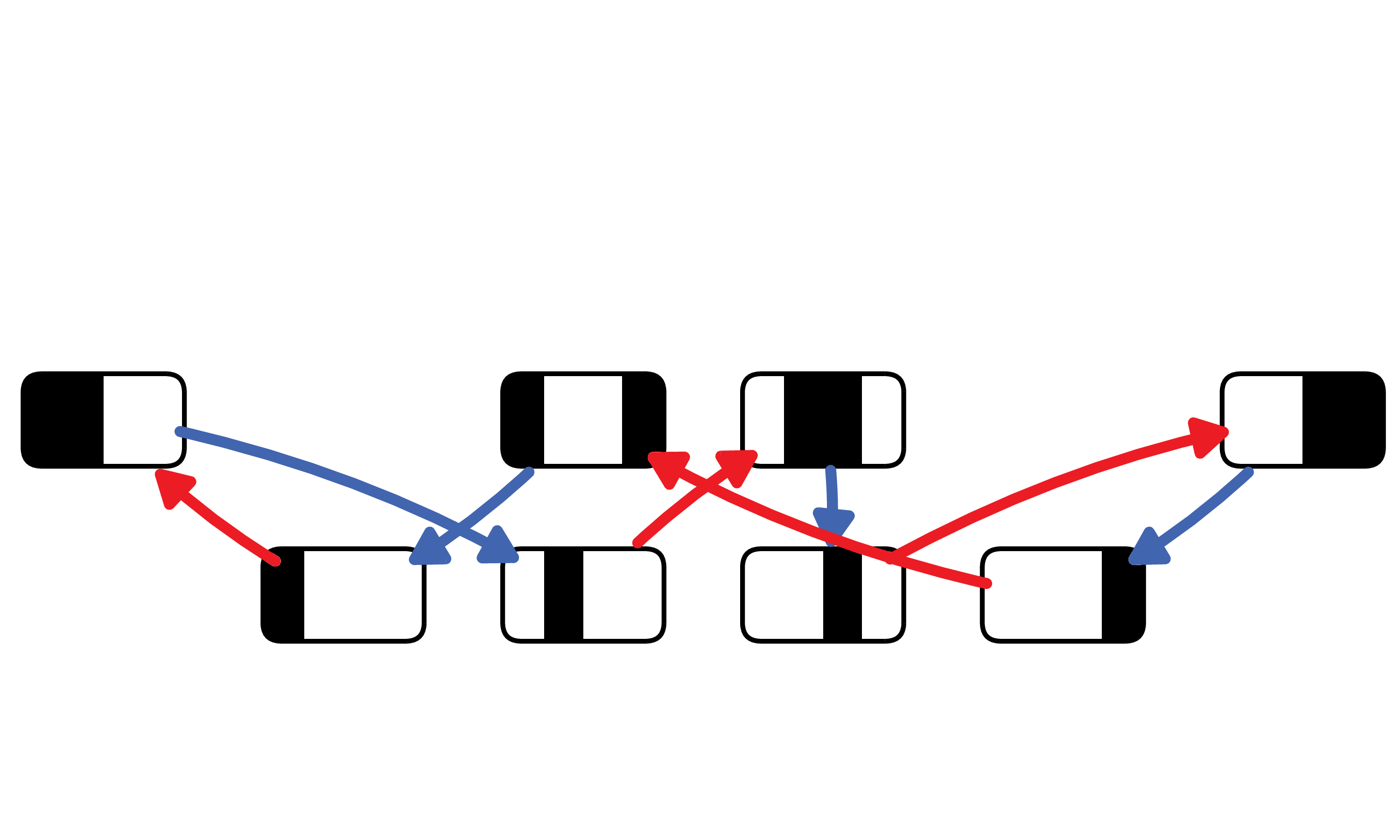}};

\node[rotate=0] at (14.485,-8.27) {\small \scalebox{0.39}{$\color{white}\boldsymbol{+}\mkern-4mu\color{black}\boldsymbol{-}\mkern-4mu\boldsymbol{-}\mkern-0mu\boldsymbol{-}$}};

\node[rotate=0] at (15.92,-8.27) {\small \scalebox{0.39}{$\color{black}\boldsymbol{-}\mkern-4mu\color{black}\boldsymbol{-}\mkern-4mu\color{white}\boldsymbol{+}\mkern-0mu\color{black}\boldsymbol{-}$}};
\node[rotate=0] at (15.92,-7.75) {\small \scalebox{0.39}{$\boldsymbol{-}\mkern-4mu\color{white}\boldsymbol{+}\mkern-4mu\boldsymbol{+}\mkern-0mu\color{black}\boldsymbol{-}$}};

\node[rotate=0] at (16.635,-8.27) {\small \scalebox{0.39}{$\color{black}\boldsymbol{-}\mkern-4mu\color{black}\boldsymbol{-}\mkern-4mu\color{black}\boldsymbol{-}\mkern-0mu\color{white}\boldsymbol{+}$}};

\node[rotate=0] at (15.20,-8.27) {\small \scalebox{0.39}{$\color{black}\boldsymbol{-}\mkern-4mu\color{white}\boldsymbol{+}\mkern-4mu\color{black}\boldsymbol{-}\mkern-0mu\color{black}\boldsymbol{-}$}};
\node[rotate=0] at (15.20,-7.75) {\small \scalebox{0.39}{$\color{white}\boldsymbol{+}\mkern-4mu\color{black}\boldsymbol{-}\mkern-4mu\boldsymbol{-}\mkern-0mu\color{white}\boldsymbol{+}$}};

\node[rotate=0] at (13.77,-7.75) {\small \scalebox{0.39}{$\color{white}\boldsymbol{+}\mkern-4mu\boldsymbol{+}\mkern-4mu\color{black}\boldsymbol{-}\mkern-0mu\boldsymbol{-}$}};

\node[rotate=0] at (17.345,-7.75) {\small \scalebox{0.39}{$\color{black}\boldsymbol{-}\mkern-4mu\boldsymbol{-}\mkern-4mu\color{white}\boldsymbol{+}\mkern-0mu\boldsymbol{+}$}};

\draw[] (0.3,-9.13) rectangle (\textwidth+0.3,-6.4);
\node[rotate=0] at (0.6,-6.7) {\small (c)};

\end{tikzpicture}
\vspace*{-0.2cm}
\caption{\small{\textbf{Ensembles of fundamental self-loops of different sizes.} Blue(/red) arrows represent up(/down) transitions. (a) Size $L=4$ ($1$ self-loop). (b) Size $L=6$ ($6$ self-loops). (c) Size $L=8$ ($58$ self-loops, not all shown).}}
\label{fig:cycles_ensemble}
\end{figure}

To illustrate this, consider \( M(3, 6) \). A naive count works as follows: there are $24$ self-loops of length $L=6$ starting from, e.g.,  state $S = (-1-1-1)$, yielding a total of $2^3 \times 24 = 192$ potential self-loops. To find the fundamental self-loops, we discuss two symmetries on their structure, i.e., sequence of states. First, we consider self-loops equivalent if they can be mapped onto each other by relabeling of the hysterons, which reduces the number of loops by a factor $n_e!$  (Fig.~\ref{fig:margot1}). Note, however, that each of these corresponds to a separate polytopes in parameter space. Second, since any state in a self-loop can be taken as the starting state, we consider self-loops equivalent if they can be mapped on each other by a 'timeshift' (Fig.~\ref{fig:margot2}-\ref{fig:margot3}). We note that all loops related by shifts lead to the same polytope in parameter space. For the example of $n_e=3, L=6$, the relabeling symmetry reduces the $192$ potential self-loops to $32$, and the timeshift symmetry lowers it further to six distinct fundamental self-loops (Fig.~\ref{fig:cycles_ensemble}).

We now describe our algorithm for calculating \( M(n_{e}, L) \). We specify each self-loop by its starting state $S^0$ and the sequence of element flips (Fig.~\ref{fig:margot1}). The relabeling symmetry implies that the sequence of flips $(2, 1, 0, 2, 0, 1)$ from state $S^0=(+1-1-1)$ gives rise to the same self-loop as the sequence $(1, 0, 2, 1, 2, 0)$ from state $S^0=(-1-1+1)$. We break this permutation symmetry by indexing hysterons according to the order in which they are flipped, and only keeping loops where the different hysteron flips occur in descending order, such as $(2, 1, 0, \dots)$ and $(2, 1, 2, 0, \dots)$. Using this convention, we only keep the first sequence $(2, 1, 0, 2, 0, 1)$, and overall this convention reduces the number of generated self-loops by a factor $n_e!$.

Second, we deal with the timeshift symmetry. For example, the sequence $(2, 1, 0, 2, 0, 1)$ from state $(+1-1-1)$ can also be 'shifted' by one, such that we obtain the sequence $(1, 0, 2, 0, 1, 2)$ starting from state $(+1-1+1)$; with the relabeling convention described above, these cycle maps to $(2, 1, 0, 1, 2, 0)$ starting from state $(+1+1-1)$ (Fig.~\ref{fig:margot2}). For computational efficiency, we break up the check for timeshift symmetry in two parts.

First, when generating self-loops, we order these by their minimum magnetization $m:=\sum_i s_i$:
\begin{equation}
	M(n_e, L) = \sum_{m=-n_e}^{n_e} M_m(n_e, L)
\end{equation}
where $M_m (n_e, L)$ is the number of self-loops which starts and ends at the same magnetisation $m$, and does not go below $m$. Hence, we first generate all loops starting at state $(-1,-1,-1,\dots)$ with magnetization $m=-n_e$; then we start at states with  magnetization $m=-n_e+2$, but whenever such potential loop reaches a lower magnetization, we dismiss it. This approach filters out loops such as $(2, 1, 0, 1, 2, 0)$ starting from state $(+1+1-1)$. Iterating over all magnetizations is computationally effective, e.g.,  for $n_e = 3, L=6$, this procedure reduces the number of potential self-loops from $32$ to $7$.

The labeling and magnetization conventions limit the number of self-loop structures, but do not fully eliminate all duplicates.
For our example, both loops $(2, 1, 0, 2, 0, 1)$ from state $(+1-1-1)$ and $(2, 1, 0, 1, 2, 0)$ from state $(-1+1-1)$ would be included, even though there are related by shifting and relabeling (Fig.~\ref{fig:margot3}).
To select which of these to keep, we assign to each of these loops a ranking composed of the initial state and the sequence of flipped hysterons: in the above example these rankings would be $100210102$ and $010210120$ (where we replaced all the phases $-1$ by $0$ for notational convenience). For equivalent loops, we only include the loop with the lowest ranking. This allows us to identify one more pair of duplicates for $n_e = 3, L=6$ -- namely, the two loops described above -- bringing the final number of ($n_e=3,L=6$) fundamental self-loops down to $6$, as shown in the main text.

\begin{table}[p]
\small
\vspace*{0.35cm}
\begin{center}
\renewcommand{\arraystretch}{1.5}
\begin{tabular}{|p{.08\textwidth}|p{.08\textwidth}|p{.08\textwidth}|p{.08\textwidth}|p{.08\textwidth}|}
\hline
 & $n_{e}=2$ & $n_{e}=3$ & $n_{e}=4$ & $n_{e}=5$ \\
\hline
$L=4$ & $1$ & -- & -- & -- \\
\hline
$L=6$ & -- & $6$ & -- & -- \\
\hline
$L=8$ & -- & $2$ & $56$ & -- \\
\hline
$L=10$ & -- & -- & $176$ & $796$ \\
\hline
$L=12$ & -- & -- & $420$ & $9028$ \\
\hline
$L=14$ & -- & -- & $448$ & $76640$ \\
\hline
$L=16$ & -- & -- & $112$ & $535584$ \\
\hline
$L=18$ & -- & -- & -- & x \\
\hline
$L=20$ & -- & -- & -- & x \\
\hline
$L=22$ & -- & -- & -- & x \\
\hline
$L=24$ & -- & -- & -- & x \\
\hline
$L=26$ & -- & -- & -- & x \\
\hline
$L=28$ & -- & -- & -- & x \\
\hline
$L=30$ & -- & -- & -- & x \\
\hline
$L=32$ & -- & -- & -- & $15109096$ \\
\hline
\end{tabular}
\end{center}
\begin{flushleft}
\end{flushleft}
\vspace{-0.6cm}
\caption[table caption]{\small{\textbf{Number fundamental self-loop structures which can be drawn}. The left column (resp. top row) indicates the size $L$ of self-loops (resp. the number of hysterons involved in the loop $n_{e}$); the number inside each box is the number of different self-loop structures that can be drawn (not all of them are realizable with pairwise interactions).}}
\label{table:self_loops_summary}
\end{table}

\begin{table}[p]
\small
\vspace*{0.35cm}
\begin{center}
\begin{minipage}{0.48\textwidth}
\renewcommand{\arraystretch}{1.5}
\begin{tabular}{|p{.16\textwidth}|p{.16\textwidth}|p{.16\textwidth}|p{.16\textwidth}|p{.16\textwidth}|}
\hline
 & $n_{e}=2$ & $n_{e}=3$ & $n_{e}=4$ & $n_{e}=5$ \\
\hline
$L=4$ & $1$ & -- & -- & -- \\
\hline
$L=6$ & -- & $6$ & -- & -- \\
\hline
$L=8$ & -- & $0$ & $56$ & -- \\
\hline
$L=10$ & -- & -- & $114$ & $796$ \\
\hline
$L=12$ & -- & -- & $145$ & x \\
\hline
$L=14$ & -- & -- & $48$ & x \\
\hline
$L=16$ & -- & -- & $4$ & x \\
\hline
$L=18$ & -- & -- & -- & x \\
\hline
$L=20$ & -- & -- & -- & x \\
\hline
$L=22$ & -- & -- & -- & x \\
\hline
$L=24$ & -- & -- & -- & x \\
\hline
$L=26$ & -- & -- & -- & x \\
\hline
$L=28$ & -- & -- & -- & x \\
\hline
$L=30$ & -- & -- & -- & x \\
\hline
$L=32$ & -- & -- & -- & x \\
\hline
\end{tabular}
\end{minipage}
\hfill
\begin{minipage}{0.48\textwidth}
\renewcommand{\arraystretch}{1.5}
\begin{tabular}{|p{.16\textwidth}|p{.16\textwidth}|p{.16\textwidth}|p{.16\textwidth}|p{.16\textwidth}|}
\hline
 & $n_{e}=2$ & $n_{e}=3$ & $n_{e}=4$ & $n_{e}=5$ \\
\hline
$L=4$ & $0$ & -- & -- & -- \\
\hline
$L=6$ & -- & $2$ & -- & -- \\
\hline
$L=8$ & -- & $0$ & $24$ & -- \\
\hline
$L=10$ & -- & -- & $4$ & $376$ \\
\hline
$L=12$ & -- & -- & $1$ & x \\
\hline
$L=14$ & -- & -- & $0$ & x \\
\hline
$L=16$ & -- & -- & $0$ & x \\
\hline
$L=18$ & -- & -- & -- & x \\
\hline
$L=20$ & -- & -- & -- & x \\
\hline
$L=22$ & -- & -- & -- & x \\
\hline
$L=24$ & -- & -- & -- & x \\
\hline
$L=26$ & -- & -- & -- & x \\
\hline
$L=28$ & -- & -- & -- & x \\
\hline
$L=30$ & -- & -- & -- & x \\
\hline
$L=32$ & -- & -- & -- & x \\
\hline
\end{tabular}
\end{minipage}
\end{center}
\begin{flushleft}
\end{flushleft}
\vspace{-0.6cm}
\caption[table caption]{\small{\textbf{Number of fundamental self-loop structures which are realizable}. The left column (resp. top row) indicates the size $L$ of self-loops (resp. the number of hysterons involved in the loop $n_{e}$); the number inside each box is the number of different self-loop structures which are realizable (left), and which are realizable when restricting to weak asymmetry (right).}}
\label{table:self_loops_realizable_summary}
\end{table}

We implement this procedure using a suite of numerical tools developed earlier \cite{teunisse2024transition}, and summarize the results in Table \ref{table:self_loops_summary} and Fig. \ref{fig:cycles_ensemble}. Our results show a rapid
proliferation of the number of fundamental self-loops when $L$ and $n_e$ increase.

\subsection{Realizability of self-loops}

Now that we have established the structure of the fundamental self-loops as a function of $(n_e,L)$, we determine their realizability. For every fundamental self-loop we can construct a set of linear inequalities of the hysteron parameters $(h_i^{\pm},c_{ij})$  \cite{van2021profusion,lindeman2025generalizing, teunisse2024transition}, and solve them using linear programming. We adapt the methods introduced in \cite{teunisse2024transition} to deal with self-loops and race conditions, and eliminate the role of the value of the driving parameter.

The framework in \cite{teunisse2024transition} did not explicitly discuss parameter conditions to realize self-loops, but instead focused on non-loop transitions, for which separate design inequalities for the initial, intermediate and final state of an avalanche were derived. As self-loops have no final state, it suffices to construct the initial and intermediate inequalities. Moreover, we are
not interested in whether a self-loop arises for a specific value of the driving $H$, but only whether a self-loop is realizable for {\em any} driving. We account for this by eliminating the driving from the set of linear inequalities. For example, if we have the set of design inequalities:

\begin{equation}
\begin{aligned}
H > H^+(S^0),\\
H < H^-(S^1),\\
H > H^+(S^2),\\
H < H^-(S^3),\\
H > H^+(S^4),\\
H < H^-(S^5),\\
\end{aligned}
\end{equation}
which realizes a self-loop of length $L=6$ and driving $H$, we can eliminate the driving $H$ to obtain inequalities that enforce that there is a range of the driving $H$ where a self-loop occurs:
\begin{equation}
\begin{aligned}
 H^-(S^1) > H^+(S^0),\\
 H^-(S^1) > H^+(S^2),\\
 H^-(S^1) >  H^+(S^4),\\
 H^-(S^3) > H^+(S^0),\\
 H^-(S^3) > H^+(S^2),\\
 H^-(S^3) >  H^+(S^4),\\
H^-(S^5) > H^+(S^0),\\
 H^-(S^5) > H^+(S^2),\\
 H^-(S^5) >  H^+(S^4).\\
\end{aligned}
\end{equation}

While previously, we considered transitions with race conditions ill-defined in \cite{teunisse2024transition}, we can easily implement our resolution of the race conditions by flipping the most unstable hysteron. We note that as a consequence we do not need to keep track of all switching fields $H_i^\pm(S)$, but only of the state switching fields $H^\pm(S)$.

We finally note subtlety for states that are 'unconditionally unstable' -- i.e., for which  $H^-(S) > H^+(S)$. While these states pose no issue if $H < H^+(S)$ or $H > H^- (S)$, the range $H^+(S)< H < H^-(S)$ is problematic, as in this case one cannot judge whether a hysteron flips up or down from the ordering of the switching fields alone. As a proper accounting for these cases would significantly increase the complexity of the inequalities, we choose to take a more conservative approach, where a self-loop is counted as non-realizable if it is contingent upon this exceptional case. Note that this in contrast with section \ref{sec:conditions_self_loops}, where such cases were taken into account to find the regions where self-loops emerge.


The results of our realizability checks
for the general case and weakly asymmetric case
are shown in Tables \ref{table:self_loops_realizable_summary}.


\section{Response for strictly self-loop-free ensembles} \label{sec:large_avalanches}

The absence of self-loops in the different well-behaved models allows to explore statistics of avalanches and of the response to cyclic drive for large $N$ and arbitrary $J_0$. We focus below on race condition rule $1$, where the most unstable element flips first, and on ensembles of hysterons with distributed spans ($\sigma_i$ flatly sampled from $\left[ 0, 0.5 \right]$).

\subsection{Avalanche sizes}

\begin{figure}[b!]
\hspace*{-0.25cm}
\begin{tikzpicture}

\node[] at (0.5,2.2) {\small \textbf{Symmetric} };
\node[] at (5.1,2.2) {\small \textbf{Constant-columns} };
\node[] at (9.7,2.2) {\small \textbf{Constant-rows} };

\node[rotate=0] at (0.0,0.0) {\includegraphics[height=4.3cm]{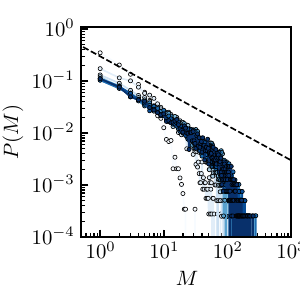}};

\fill[white] (-2.14,0.21) rectangle (-1.81,0.53);
\node[rotate=90] at (-1.98,0.36) {\small $A$};
\fill[white] (0.36,-1.99) rectangle (0.69,-1.67);
\node[] at (0.52,-1.84) {\small $A$};

\draw[->] (0.2,-0.3) -- (1.55,-0.4);
\node[rotate=0] at (1.8,-0.4) {\small $J_0$};

\node[rotate=0] at (4.6,0.0) {\includegraphics[height=4.3cm]{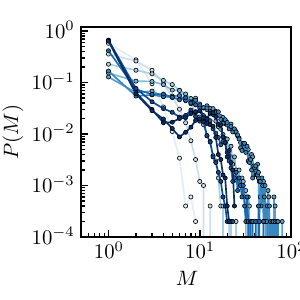}};

\fill[white] (2.46,0.21) rectangle (2.79,0.53);
\node[rotate=90] at (2.62,0.36) {\small $A$};
\fill[white] (4.96,-1.99) rectangle (5.29,-1.67);
\node[] at (5.12,-1.84) {\small $A$};

\node[rotate=0] at (9.2,0.0) {\includegraphics[height=4.3cm]{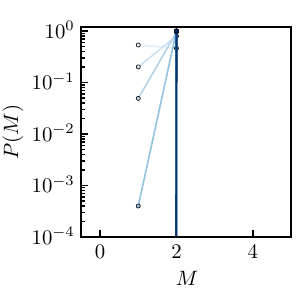}};

\fill[white] (7.06,0.21) rectangle (7.39,0.53);
\node[rotate=90] at (7.22,0.36) {\small $A$};
\fill[white] (9.56,-1.99) rectangle (9.89,-1.67);
\node[] at (9.72,-1.84) {\small $A$};

\node[rotate=0] at (0.0,-4.2) {\includegraphics[height=4.3cm]{meanAvalancheSize_N_reciprocal.pdf}};

\fill[white] (-2.23,-4.1) rectangle (-1.6,-3.78);
\node[rotate=90] at (-1.77,-3.95) {\small $A$};

\draw[->] (0.6,-4.8) -- (0.9,-3.2);
\node[rotate=0] at (0.95,-2.95) {\small $N$};

\node[rotate=0] at (4.6,-4.2) {\includegraphics[height=4.3cm]{meanAvalancheSize_N_serial.pdf}};

\fill[white] (2.37,-4.1) rectangle (3.0,-3.78);
\node[rotate=90] at (2.83,-3.95) {\small $A$};
\fill[white] (3.73,-3.27) rectangle (4.02,-2.95);
\node[rotate=90] at (3.85,-3.12) {\small $A$};

\draw[->] (5.25,-4.7) -- (6.2,-3.6);
\node[rotate=0] at (6.4,-3.4) {\small $N$};

\node[rotate=0] at (9.2,-4.2) {\includegraphics[height=4.3cm]{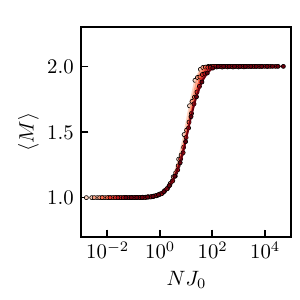}};

\fill[white] (7.3,-4.1) rectangle (7.6,-3.78);
\node[rotate=90] at (7.43,-3.95) {\small $A$};

\node[rotate=0] at (1.6,1.4) {\small (a)};
\node[rotate=0] at (6.2,1.4) {\small (b)};
\node[rotate=0] at (10.8,1.4) {\small (c)};

\node[rotate=0] at (1.6,-2.8) {\small (d)};
\node[rotate=0] at (6.2,-2.8) {\small (e)};
\node[rotate=0] at (10.8,-2.8) {\small (f)};

\end{tikzpicture}
\vspace*{-0.5cm}
\caption{\small{\textbf{Large avalanches for well-behaved models}. (a-c) Avalanche size distributions for different $J_0 \in \left[ 10^{-2}, 10^{2}\right]$; color-coded from light to dark blue as $J_0$ increases; fixed $N = 1024$. (d-f) Mean avalanche size $\langle A \rangle$ as a function of $NJ_0$, for different $N \in \left[ 16, 32, 64, 128, 256, 512\right]$, color-coded from light to dark red as $N$ increases. (a/d) Symmetric interactions; (b/e) constant-columns interactions; (c/f) constant-rows interactions.}}
\label{fig:well_behaved_avalanches}
\end{figure}

We consider $N = 1024$ hysterons and place ourselves at $H = 0$. Starting from a random initial state, the system is
stabilized by flipping unstable hysterons one by one until a stable state $S^0$ is found (which is guaranteed in the absence of self-loops). We simulate $S^0 \rightarrow S^1$ over $5 \times 10^3$ samples, and record the avalanche sizes $A$ -- the number of flips before the system settles in a stable state.
For all ensembles, when $NJ_0 \ll 1$, only Preisach-like non-avalanche transitions ($A=1$) are found. For symmetric interactions,
the larger $J_0$, the broader the avalanche size distribution and the larger the mean avalanche size, with a crossover around $NJ_0 \simeq 1$ (Figs. \ref{fig:well_behaved_avalanches}-a and d). For $NJ_0 \gg 1$, avalanche sizes are power-law distributed with a cutoff growing with $J_0$ and $N$, and saturating below system size. For constant-columns interactions, we find a similar transition
scenario at $NJ_0 \simeq 1$ (Figs. \ref{fig:well_behaved_avalanches}-b and e). However, for $J_0 \simeq 1$, the mean avalanche size reaches a maximum, which increases with system size (Fig. \ref{fig:well_behaved_avalanches}-e, inset), and then decreases abruptly, and increases again for larger $J_0$. Finally, for constant-rows interactions, numerical simulations confirm that transitions can only be Preisach-like nonavalanche transitions or avalanches of size $2$ that do not alter the magnetization $m = \sum_i s_i$ (Figs. \ref{fig:well_behaved_avalanches}-c and f), with a crossover between
the two around $NJ_0 \simeq 1$.

\subsection{Response to cyclic drive}

Let us now consider cyclic drive conditions - when the input $U$ is swept between $U_{\rm min}$ and $U_{\rm max}$. We focus on two aspects of the response: the number of driving cycles $\tau$ taken to reach a periodic orbit, and the period $T$ of the orbit relative to the driving cycle. In the Preisach model (limit of zero interactions), $\tau \leq 1$ and $T = 1$, which can be understood by noting that each (independent) hysteron requires at most one cycle before it reaches a periodic orbit. Only a few results exist for finite interactions between hysterons. For small systems, it was shown that the transients $\tau$ and periodicity $T$ are distributed exponentially \cite{lindeman2021multiple}.

\begin{figure}[t!]
\hspace*{-0.25cm}
\begin{tikzpicture}

\node[] at (0.5,2.2) {\small \textbf{Symmetric} };
\node[] at (5.1,2.2) {\small \textbf{Constant-columns} };
\node[] at (9.7,2.2) {\small \textbf{Constant-rows} };

\node[rotate=0] at (0.0,0.0) {\includegraphics[height=4.3cm]{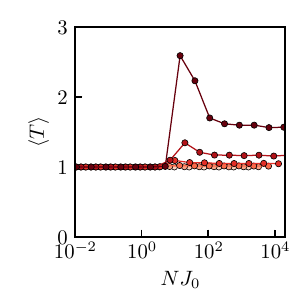}};

\draw[->] (0.4,-0.45) -- (1.05,0.7);
\node[rotate=0] at (1.25,0.9) {\small $N$};

\node[rotate=0] at (4.6,0.0) {\includegraphics[height=4.3cm]{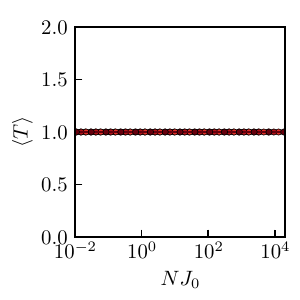}};

\node[rotate=0] at (9.2,0.0) {\includegraphics[height=4.3cm]{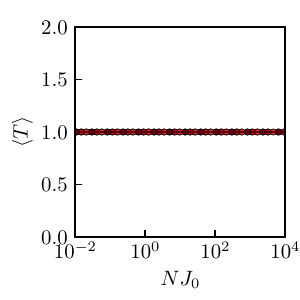}};

\node[rotate=0] at (0.0,-4.2) {\includegraphics[height=4.3cm]{transientLength_reciprocal.pdf}};

\draw[->] (0.45,-5.3) -- (1.3,-2.95);
\node[rotate=0] at (1.4,-2.7) {\small $N$};

\node[rotate=0] at (4.6,-4.2) {\includegraphics[height=4.3cm]{transientLength_serial.pdf}};

\draw[->] (5.5,-5.1) -- (6.2,-3.15);
\node[rotate=0] at (6.3,-2.9) {\small $N$};

\node[rotate=0] at (9.2,-4.2) {\includegraphics[height=4.3cm]{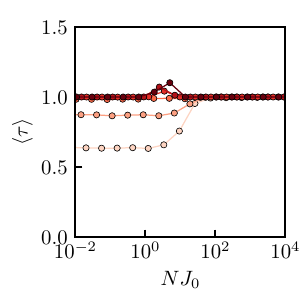}};

\node[rotate=0] at (1.6,1.4) {\small (a)};
\node[rotate=0] at (6.2,1.4) {\small (b)};
\node[rotate=0] at (10.8,1.4) {\small (c)};

\node[rotate=0] at (-0.7,-2.8) {\small (d)};
\node[rotate=0] at (3.9,-2.8) {\small (e)};
\node[rotate=0] at (8.5,-2.8) {\small (f)};

\end{tikzpicture}
\vspace*{-0.5cm}
\caption{\small{\textbf{Transients and periodicity during cyclic drive for the well-behaved models}, for different $N \in \left[ 16, 32, 64, 128, 256, 512 \right]$, color coded from light to dark red as $N$ increases. (a-c) Ensemble averaged transient length $\langle \tau \rangle$ as a function of $J_0$. (d-f) Ensemble averaged periodicity $\langle T \rangle$ as a function of $J_0$. (a/d) Symmetric interactions; (b/e) constant-columns interactions; (c-f) constant-rows interactions.}}
\label{fig:well_behaved_response}
\end{figure}

Here, we consider $N = 512$ hysterons and place ourselves at $U = 0$. Starting from a random initial condition, the system is stabilized by flipping unstable hysterons one by one until a stable state is found. Finally, the drive $U$ is swept between $0$ and $U_{\rm max}$, where $U_{\rm max}$ is the drive amplitude leading to a magnetization $m = \sum_i s_i$ equal or larger than $N/2$ ($U_{\rm max}$ is determined during the first drive cycle and is kept fixed for the rest of the simulation). For the symmetrically-coupled case (Figs. \ref{fig:well_behaved_response}-a and d), when $NJ_0 \ll 1$, both $\langle \tau \rangle$ and $\langle T \rangle$ are equal to $1$, and when $NJ_0 \gg 1$, both $\langle \tau \rangle$ and $\langle T \rangle$ are greater than $1$ and constant with $J_0$. Remarkably, both $\langle \tau \rangle$ and $\langle T \rangle$ reach a maximum for intermediate $J_0 \simeq 1/N$ \cite{van2021profusion}. For constant-columns and constant-rows interactions, we systematically find $T = 1$ (Figs. \ref{fig:well_behaved_response}-b and c): all orbits have the same period as the drive. This is expected for constant-columns interactions given the absence of scrambling \cite{liu2024controlled}. Also, $\langle \tau \rangle = 1$ for both $NJ_0 \ll 1$ and $NJ_0 \gg 1$ (Figs. \ref{fig:well_behaved_response}-e and f). For constant-columns interactions, $\langle \tau \rangle$ reaches a maximum for intermediate $J_0$ at the same value corresponding to the maximum of $\langle A \rangle$, i.e. $J_0 \simeq 1$. In contrast, for constant-rows interactions, the maximum of $\langle \tau \rangle$ is reached for $J_0 \simeq 1/N$.

\end{document}